\documentclass[submission,Phys]{SciPost}
\pdfoutput=1
\usepackage{amsmath,amssymb,mathtools,xspace,units}
\usepackage{booktabs,multirow,graphicx,tabularx,slashed}
\usepackage{color,xcolor}
\usepackage[normalem]{ulem}
\usepackage{enumitem}
\usepackage{feynmp}
\usepackage{braket,stackrel}
\usepackage{tikz}
\usepackage{floatrow}
\usepackage{subfig}
\usetikzlibrary{arrows}
\usetikzlibrary{shapes.geometric, arrows}

\graphicspath{{./Figures/}{./}}
\makeatletter
\@ifundefined{pdfoutput}{}{\DeclareGraphicsRule{*}{mps}{*}{}}
\makeatother

\makeatletter
\DeclareRobustCommand*{\bfseries}{%
   \not@math@alphabet\bfseries\mathbf
   \fontseries\bfdefault\selectfont
   \boldmath
}
\makeatother

\setitemize{itemsep=2pt,topsep=2pt,parsep=0pt,partopsep=0pt,leftmargin=*}
\setenumerate{itemsep=0pt,topsep=2pt,parsep=0pt,partopsep=0pt,labelindent=3pt,leftmargin=*}

\definecolor{Gcolor}{HTML}{3b528b}
\definecolor{Dcolor}{HTML}{e41a1c}

\tikzstyle{generator} = [rectangle, rounded corners, minimum width=3cm, minimum height=1cm,text centered, draw=Gcolor]
\tikzstyle{discriminator} = [rectangle, rounded corners, minimum width=3cm, minimum height=1cm,text centered, draw=Dcolor]
\tikzstyle{io} = [circle, trapezium left angle=70, trapezium right angle=110, minimum width=1cm, minimum height=1cm, text centered, draw=black]

\tikzstyle{process} = [rectangle, minimum width=1cm, minimum height=1cm, text centered, draw=black]
\tikzstyle{decision} = [rectangle, minimum width=1cm, minimum height=1cm, text centered, draw=black]

\tikzstyle{arrow} = [thick,->,>=stealth]
\usepackage{xcolor}

\newcommand{\rd}{\ensuremath{\mathrm{d}}}
\newcommand{\fpWt}{f_{\phi Q}^{(3)}}
\newcommand{\fpWs}{f_{\phi Q}^{(1)}}

\def\g{\gamma}

\def\hc{\text{h.c.}}

\def\qm{Q}

\newcommand\one{\leavevmode\hbox{\small1\normalsize\kern-.33em1}}

\newcommand{\p}{\partial}

\newcommand{\lag}{\mathcal{L}}

\newcommand{\ope}{\mathcal{O}}
\renewcommand{\O}{\mathcal{O}}

\newcommand{\qqquad}{\qquad \qquad}

\newcommand{\msbar}{\overline{\text{MS}}}

\def\slashchar#1{\setbox0=\hbox{$#1$}           
   \dimen0=\wd0                                 
   \setbox1=\hbox{/} \dimen1=\wd1               
   \ifdim\dimen0>\dimen1                        
      \rlap{\hbox to \dimen0{\hfil/\hfil}}      
      #1                                        
   \else                                        
      \rlap{\hbox to \dimen1{\hfil$#1$\hfil}}   
      /                                         
   \fi}

\newcommand{\sfitter}{\textsc{SFitter}}

\newcommand{\Loop}{\ensuremath{\ell}}
\newcommand{\EFT}{\ensuremath{\text{EFT}}\xspace}
\newcommand{\UV}{\ensuremath{\text{UV}}\xspace}

\DeclareMathOperator{\Tr}{Tr}

\newcommand{\gt}{\tilde{g}}
\newcommand{\tV}{\ensuremath{\widetilde{V}}}
\newcommand{\tW}{\ensuremath{\widetilde{W}}}
\newcommand{\tmv}{\ensuremath{\tilde{m}_V}}

\begin{document}

\begin{center}{\Large \textbf{
From Models to SMEFT and Back?
}}\end{center}

\begin{center}
Ilaria Brivio\textsuperscript{1},
Sebastian Bruggisser\textsuperscript{1},
Emma Geoffray\textsuperscript{1},
Wolfgang Kilian\textsuperscript{2},\\ 
Michael Kr\"amer\textsuperscript{3},
Michel Luchmann\textsuperscript{1},
Tilman Plehn\textsuperscript{1}, and 
Benjamin Summ\textsuperscript{3,4}
\end{center}

\begin{center}
{\bf 1} Institut f\"ur Theoretische Physik, Universit\"at Heidelberg, Germany\\
{\bf 2}  Department of Physics, University of Siegen,  Germany\\
{\bf 3} Institut f\"ur Theoretische Teilchenphysik und Kosmologie, RWTH Aachen University, Germany\\
{\bf 4} Institut f\"ur Theoretische Physik und Astrophysik, Universit\"at W\"urzburg, Germany\\
geoffray@thphys.uni-heidelberg.de
\end{center}

\tikzstyle{int}=[thick,draw, minimum size=2em]

\section*{Abstract}
{\bf We present a global analysis of the Higgs and electroweak sector,
  in the SMEFT framework and matched to a UV-completion. As the UV-model
  we use the triplet extension of the electroweak gauge sector. The
  matching is performed at one loop, employing functional methods. In
  the SFitter analysis, we pay particular attention to theory
  uncertainties arising from the matching. Our results highlight the
  complementarity between SMEFT and model-specific analyses.
}

\vspace{10pt}
\noindent\rule{\textwidth}{1pt}
\tableofcontents\thispagestyle{fancy}
\noindent\rule{\textwidth}{1pt}

\begin{fmffile}{feynman}
\section{Introduction}
\label{sec:intro}

The Higgs discovery~\cite{Aad:2012tfa,Chatrchyan:2012ufa} and many
measurements of the Higgs Lagrangian~\cite{Dawson:2018dcd} indicate
that the Standard Model with its single, weakly interacting Higgs
boson might well be the correct effective theory around the
electroweak scale. However, the Standard Model is extremely unlikely
to be the full story. Many theoretical considerations, including
electroweak baryogenesis, dark matter, or neutrino mass generation,
point to an extended electroweak or scalar sector. To avoid a bias
through a specific, pre-selected signal hypothesis, modern LHC searches
for beyond the Standard Model (BSM) physics are often conducted in the
Standard Model effective theory (SMEFT)~\cite{Brivio:2017vri}. Because
of its vast operator landscape, the corresponding experimental
searches~\cite{Aad:2019mbh,Sirunyan:2018koj} and global
analyses~\cite{Biekotter:2018rhp,Ellis:2018gqa,Almeida:2018cld,Kraml:2019sis,vanBeek:2019evb,Dawson:2020oco,Ellis:2020unq}
provide a comprehensive probe of rates and kinematic patterns in LHC
processes.

One of the complications of SMEFT analyses of LHC data is that the
effective theory truncated at dimension six has a limited validity
range, and that LHC measurements span a large energy range. Moreover,
even if we assume the SMEFT to be generally valid, it is not clear how
much information on a full BSM model is lost when we confront it with
LHC data via a truncated SMEFT Lagrangian rather than the original
full model. Combining these questions, it is instructive to consider
concrete, albeit simplified, BSM models and examine the limits
extracted through a SMEFT interpretation matched to these models in
comparison with the constraints obtained from direct
searches~\cite{Brehmer:2015rna,Biekotter:2016ecg,Dawson:2020oco}.

The naive expectation behind SMEFT analyses is that we can use the
complete, correlated information on the Wilson coefficients from a
global analysis and derive limits on any BSM model through matching.
However, if the BSM scale is not sufficiently well-separated from the
electroweak (EW) scale, an interpretation based on the SMEFT
Lagrangian truncated at dimension six will likely give inaccurate
results~\cite{Kilian:2015opv,Freitas:2016iwx}.  The theory
uncertainties related to the matching to full models are usually not
accounted for in global analyses, which instead take their Lagrangian
as a fixed interpretation framework. In general, limits derived on BSM
models through a SMEFT framework using the same data and with all
uncertainties accounted for will differ from limits derived on the
full model directly, where the former can be significantly weaker or
stronger than the latter.

This work aims at exploring the complementarity of the two analysis
strategies and at highlighting general aspects that emerge when the
SMEFT results are related to a concrete BSM scenario. We address this question for a global analysis of
electroweak, di-boson and Higgs measurements, matching the relevant
Wilson coefficients to the UV-model at one loop, using functional
matching methods.  We use the \sfitter\ framework and include a proper
estimate of a new and non-negligible theory uncertainty from the
variation of the matching scale.  As a UV-model we use a
triplet-extended gauge
sector~\cite{Low:2009di,delAguila:2010mx,Pappadopulo:2014qza,Biekoetter:2014jwa,Brehmer:2015rna}
a standard scenario when it comes to motivating the SMEFT approach to
the Higgs and electroweak sector. Such a triplet model can be linked
for instance to the weakly coupled gauge group $SU(3) \times SU(2)
\times SU(2) \times U(1)$~\cite{Barger:1980ti} or deconstructed extra
dimensions~\cite{ArkaniHamed:2001nc}.

The paper is organized as follows: in Sec.~\ref{sec:basics} we review
the basics of functional one-loop matching, we define the gauge
triplet model under study, and we provide details about the
\sfitter\ setup.  In Sec.~\ref{sec:toy} we discuss the decoupling
limit of the new heavy states and the relevance of the matching scale
choice. The impact of these two aspects on the global analysis is
illustrated via simplified fits.  In Sec.~\ref{sec:global} we present
the results of a global fit to the full vector triplet model, based on
the dimension-6 SMEFT Lagrangian, and compare our results with limits
obtained from direct searches.  We conclude in
Sec.~\ref{sec:conclusions}.

\section{Basics}
\label{sec:basics}

In this section, we briefly review the one-loop matching procedure, the
UV-model, as well as the \sfitter\ setup. Experienced readers are welcome to
skip this section.

\subsection{One-loop matching: generic approach}
\label{sec:basics_match}

The methods of constructing and matching effective-field
theories~\cite{Appelquist:1974tg,Weinberg:1975gm} have been in use for
more than four
decades~\cite{Miller:1982ij,Gaillard:1985uh,Chan:1986jq,Cheyette:1987qz}.
Generic expressions for the low-energy effective action of a gauge
theory at the one-loop order were derived in the
80s~\cite{Ball:1988xg}.  More recently, the approach has been further
explored, particularly within the context of
SMEFT~\cite{Henning:2014wua,Henning:2016lyp,Drozd:2015rsp,Ellis:2017jns,Fuentes-Martin:2016uol,Zhang:2016pja,Kramer:2019fwz,Ellis:2020ivx,Angelescu:2020yzf}.

We consider a UV model which can be defined in terms of light fields
$\psi$ and heavy fields $\Psi$, and which supports a perturbative
expansion based on a local Lagrangian.  Heavy fields are characterized
by the condition that the support of their spectral functions vanishes
below a certain threshold.  We may identify the threshold with a mass
$M$, typically the lightest mass that belongs to the heavy spectrum.
The remaining fields are understood as light fields.

The UV model is expressible in terms of an effective action
$\Gamma_\UV[\psi,\Psi]$, the generating functional of its one-particle
irreducible (1PI) vertex functions.  If fields of spin higher than
$1/2$ are involved, or if global symmetries are present, it is
constrained to be a solution of a Slavnov-Taylor identity.  By
assumption, $\Gamma_\UV$ is calculable in a loop expansion from a
local Lagrangian $\lag_\UV(\psi(x),\Psi(x))$ with a finite number of
fields and parameters.  The parameters depend on the choice of a
regularization and renormalization scheme and are redefined order by
order by suitable renormalization conditions.  This includes resolving
inherent ambiguities associated with field reparameterizations, such
as wave-function renormalization and terms vanishing by equations of
motion.

The EFT is likewise expressible in terms of an effective action
$\Gamma_\EFT[\psi]$, a functional of the light fields only.  Again, we
assume that a perturbative loop expansion is possible, and that it can
be computed from a local Lagrangian $\lag_\EFT(\psi(x))$.  The number
of parameters of $\lag_\EFT$ is intended to be finite, but it
increases without bounds with the accuracy that we want to implement
via matching conditions.  To keep the EFT parameter set manageable, we
have to define an organizing principle which amounts to a series of
approximations, and a prescription to truncate this series at a
certain order.

To find the EFT Lagrangian iteratively, one introduces the
one-light-particle irreducible (1LPI) effective action
$\Gamma_{\text{L,UV}}[\psi]$.  Formally, this is a double Legendre
transform of $\Gamma_\UV[\psi,\Psi]$; in practice, it amounts to
absorbing a maximal set of independent heavy-field propagators in the
skeleton expansion of S-matrix elements.  This results in redefined
light-field effective vertices. By contrast, the light-field
propagators are kept explicit.  In general they still carry a mixture
of light and residual heavy degrees of freedom, depending on the
precise definition of the original UV model.  Like the original
effective action, $\Gamma_{\text{L,UV}}[\psi]$ depends on
conventions regarding renormalization and handling the equations of
motion.  In terms of this entity, the matching condition reads
\begin{align}
  \label{eq:matching-master}
  \Gamma_{\text{L,UV}}[\psi] = \Gamma_\EFT[\psi]
  + \Delta\Gamma[\psi] \; .
\end{align}
The matching error $\Delta\Gamma[\psi]$ describes a set of vertex-function
corrections $\Delta\Gamma_i(x)$ that are not calculable from a local
Lagrangian involving light fields only.  The matching procedure succeeds if,
in momentum space, all contributions to this error are sufficiently
power-suppressed at low energy,
\begin{align}
  \label{eq:DeltaGamma}
  \Delta\Gamma_i(p) < c|p|^k \; ,
\end{align}
where $p$ is any light-particle mass or momentum component.

At the tree level, the 1LPI effective
action~$\Gamma^{(0)}_{\text{L,UV}}[\psi]$ of the UV model can be
derived by simple variable changes, applying the equations of motion.
Unless the $\psi$ multiplets are incomplete under a symmetry, the
result satisfies the tree-level Slavnov-Taylor identity with only
light fields taken into account.  The tree-level effective action
$S_\EFT[\psi]=\Gamma^{(0)}_\EFT[\psi]$ is evaluated, to arbitrary
order, by means of a momentum-space Taylor expansion of the 1LPI
effective action on the l.h.s.\ of Eq.\eqref{eq:matching-master}.  In
this expansion, residual heavy degrees of freedom are naturally
removed from the tree-level light-field propagators.  The latter
assume their canonical tree-level form while any extra terms are
shifted to the interaction part of $S_\EFT[\psi]$.
 
The operator content of the tree-level effective action $S_\EFT[\psi]$
can be determined independently by algebraic methods.  Their
coefficients are fixed by a term-by-term comparison with the vertices
of $\Gamma^{(0)}_{\text{L,UV}}$.  The symmetries are preserved in this
expansion if covariant derivatives are used consistently.  At one
loop, new contributions to the UV effective action arise which are
generically non-local, and can be formally summarized as
\begin{align}
  \Gamma^{1\Loop}_\UV[\psi,\Psi]
  &=
  i c_s \Tr \log \left(-
  \frac{\delta^2 S_\UV[\psi,\Psi]}{\delta^2(\psi,\Psi)}\right) \; ,
\end{align}
where the trace is integrated over all field components at all
space-time points and $c_s$ accounts for the statistics of the fields
that are integrated over.  This evaluates to the sum of all one-loop
Feynman graphs with external fields attached.  In expressions of this
kind, the external field insertions act as bookkeeping devices, or
background
fields~\cite{DeWitt:1967ub,Bardeen:1969md,tHooft:1975uxh,DeWitt:1980jv,Boulware:1980av,Abbott:1980hw,Abbott:1981ke}. This
allows for employing gauges and conventions that distinguish between
internal and external lines, a generic feature of working with 1PI
vertex functions.  The trace is in general UV divergent and requires
the application of a regularization scheme and the addition of local
counterterms, such as dimensional regularization and minimal
subtraction.

To match the UV model to the EFT at the one-loop order, we have to
evaluate Eq.\eqref{eq:matching-master} again.  Initially,
\begin{align}
  \label{eq:matching-1loop}
  \Delta\Gamma^{1\Loop}[\psi]
  =
  \left.
  i c_s \Tr \left[
    \log \left(-\frac{\delta^2 S_\UV[\psi,\Psi]}{\delta^2(\psi,\Psi)}\right)
    - \log\left(-\frac{\delta^2 S^{(0)}_\EFT[\psi]}{\delta^2\psi}\right)
  \right]\right|_{\Psi=0} \;,
\end{align}
where the formal trace includes the integral over all one-loop
diagrams which are 1LPI and do not contain open external $\Psi$ lines.
Because
$S^{(0)}_\EFT=\Gamma^{(0)}_\EFT=\Gamma^{(0)}_{\text{L,UV}}+\ope(|p|^k)$,
the difference is well-behaved in the IR.  Loops of canonical light
propagators only would exactly cancel between the two terms, but since
the light-field propagators need not coincide between the two
Lagrangians, we have to be careful to take all terms into account.  In
any case, due to the IR cancellation the one-loop functional
Eq.\eqref{eq:matching-1loop} again admits a Taylor expansion up to the
order of the previous tree-level truncation.  The result can be
expressed as a finite set of local terms that modify the coefficients
of terms which are already present in the generic effective Lagrangian
of the tree-level EFT.  They are absorbed in $S_\EFT[\psi]$,
\begin{align}
  \label{eq:matching-ren-1loop}
  S_\EFT^{1\Loop}[\psi]
  =
  - \Delta\Gamma^{1\Loop}[\psi]\Big|_{\text{local, truncated}} \; ,
\end{align}
and disappear from Eq.\eqref{eq:matching-1loop}.  In effect, the
remainder still contains all non-local parts of the matching error but
satisfies Eq.\eqref{eq:DeltaGamma}, to one-loop order.

By the same reasoning, the difference in Eq.\eqref{eq:matching-1loop}
is not well-behaved but divergent in the UV, and therefore requires
regularization and renormalization.  The renormalization conditions
are given by the matching conditions themselves and thus indirectly refer
to the renormalization conditions of the UV model.  All free
parameters of the EFT are fixed, order by order, in terms of the
original parameters of the UV model.  Nevertheless, a practical scheme
such as dimensional regularization with minimal subtraction may
introduce an intermediate renormalization which depends on an
arbitrary scale $\mu_R$.  The implications of this additional mass
scale will be discussed in detail below.

In analogy with the tree-level matching procedure, in order to
manifestly preserve the symmetries of the theory one should
consistently work with covariant derivatives in the one-loop matching
calculation, as discussed in the following subsection.  However, due
to the presence of UV divergences in the matching conditions the
Slavnov-Taylor identity need not be compatible with a local Taylor
expansion of the one-loop vertex functions, and the separation of the
UV effective action into a gauge-invariant low-energy effective action
and a remainder like in Eq.\eqref{eq:matching-master} may
fail~\cite{Ball:1988xg,Preskill:1990fr,Cata:2020crs}.  In the current
paper, we assume that such an obstruction does not critically affect our
argument.

\subsection{One-loop matching: implementation}
\label{sec:basics_technical}

Instead of constructing the difference in Eq.\eqref{eq:matching-1loop}
in terms of Feynman graphs explicitly, the subtraction may be
accounted for in the integrand by employing the method of
regions~\cite{Beneke:1997zp,Smirnov:2002pj,Jantzen:2011nz}.  The
matching correction (Eqs.\eqref{eq:matching-1loop}
and~\eqref{eq:matching-ren-1loop}) is replaced by
\begin{align}
  S^{1\Loop}_\EFT[\psi] = i c_s \left. \Tr
  \log \left(-\frac{\delta^2
    S_\UV[\Psi,\psi]}{\delta(\Psi,\psi)^2}\right)\right|_\text{hard}.
  \label{eq:general-one-loop-matching}
\end{align}
The label `hard' has to be understood in the following way: the
functional trace is computed in momentum space.  Two different regions
are of interest in the matching, the hard and the soft region. If $q$
denotes the typical size of a loop momentum, the hard region is
defined by $q\sim M \gg m$, whereas the soft region is defined by $q
\sim m \ll M$. Here $m$ stands for the typical mass scale of the light sector.
As discussed above, only the hard region is relevant
while in the soft region the matching integral is well behaved. It has
been shown that the tree-level induced EFT contribution to the
matching cancels the soft region contributions from the UV-theory in
the difference in
Eq.\eqref{eq:matching-1loop}~\cite{Zhang:2016pja,Summ:2021aek}. Therefore,
the integrands of the loop integrals in
Eq.\eqref{eq:general-one-loop-matching} are expanded only in the hard
region. The evaluation of the functional trace then reduces to
computing integrals of the form
\begin{align}
    \int \frac{\rd^d q}{(2\pi)^d} \frac{q^{\mu_1} \dots q^{\mu_{2 n_c}}}{(q^2-M^2_{i_1})^{n_1}\dots(q^2-M^2_{i_m})^{n_m} (q^2)^{n_0}}.
\end{align}
Here, all masses $M_{i_1},\dots,M_{i_m}$ are of the order of $M$. This
implies that the dependence of Eq.\eqref{eq:general-one-loop-matching}
on any external momentum or mass $|p|$ is analytic, and no logarithms
of the form $\log ( m/|p|)$ or $\log ( |p|/M )$ can appear. The only
logarithm possible is $\log (M/\mu_R )$, and to avoid large logarithms
in the relation between EFT and UV parameters we need to choose $\mu_R
\sim M$.

Apart from the prescription `hard', the second derivatives of the
UV-action evaluated at the background field configurations appear in
the matching. To derive a universal result these derivatives are split
into a part that contains the gauge-kinetic term of the field and its
mass term, generating the propagator of the field, and a pure
interaction contribution that appears in the final result. For the
field $\psi$ this latter piece is given by
\begin{align}
X_{\psi \psi} = - \frac{\delta ^2 S_\text{UV,int.}}{\delta \psi^2} \; ,
\label{eq:general-X-matrix}
\end{align}
where only the interaction part of the action excluding the
interactions with gauge bosons through the covariant derivative
appears. The interactions with the gauge bosons are included in the
propagator part of the functional derivative, which allows for an
evaluation in which only gauge covariant objects appear at every step
and the final result is manifestly gauge invariant. The price to be
paid for this manifest gauge covariance is that every occurrence of a
covariant derivative has to be shifted by a loop momentum in the
evaluation of the functional trace in
Eq.\eqref{eq:general-one-loop-matching}. We therefore have to
parameterize Eq.\eqref{eq:general-X-matrix} as
\begin{align}
X_{\psi \psi} = U_{\psi \psi} + i D_\mu Z^\mu_{\psi \psi} + i Z^{\dagger \mu}_{\psi \psi} D_\mu + \dots,
\label{Eq: X-parameterization}
\end{align}
where $D_\mu$ is the covariant derivative of the UV-model. The
quantities $U_{\psi \psi}, Z^\mu_{\psi \psi}$ and $Z^{\dagger
  \mu}_{\psi \psi}$ only depend on covariant derivatives through
commutators whereas the explicit covariant derivatives appearing in
Eq.\eqref{Eq: X-parameterization} are so-called open covariant
derivatives that act on everything to their right. The ellipsis
denotes terms with further open covariant derivatives. Importantly,
contributions with one open covariant derivative arise at dimension
six whenever there is a scalar field charged under the gauge group and
therefore they contribute to the matching through the presence of the
Higgs field. Consequently, for our matching computations we use an
extension of the results of Ref.~\cite{Kramer:2019fwz}, adding gauge
bosons and the heavy resonance of our model. Since the gauge boson
fluctuations appear in loops they have to be gauge fixed. This gauge
fixing does not disturb the manifest gauge invariance at the level of
the background fields and the gauge-fixing parameter can be chosen at
convenience. Choosing Feynman gauge allows for easy incorporation of
these operators into the results of Ref.~\cite{Kramer:2019fwz}, since
we can treat gauge bosons like scalar fields with an extra index. Care
has to be taken to account for the overall sign in the propagator. For
the resonance this choice is not available since it does not have a
gauge-fixing term and some operators with up to two open covariant
derivatives have to be computed for the matching.

\subsection{Triplet model}
\label{sec:basics_model}

The UV model we study in this paper is a gauge-triplet
extension of the Standard
Model~\cite{Low:2009di,delAguila:2010mx,Pappadopulo:2014qza,Biekoetter:2014jwa,Brehmer:2015rna}.
In the unbroken electroweak phase, the Lagrangian reads
\begin{align}
  \lag = \lag_\text{SM}
  &- \frac{1}{4} \tV^{\mu \nu A} \tV_{\mu \nu}^A
  - \frac{\gt_M}{2} \; \tV^{\mu \nu A} \tW_{\mu \nu}^A 
  + \frac{\tilde{m}_V^2}{2} \tV^{\mu A} \tV_{\mu}^A \notag \\
  &+  \sum_f \tilde{g}_f \; \tV^{\mu A} J_{\mu}^{fA}
  + \tilde{g}_H \; \tV^{\mu A} J_\mu^{HA}
  + \frac{\tilde{g}_{VH}}{2} \; |\phi|^2 \tV^{\mu A} \tV_{\mu}^A \; ,
\label{eq:lagrangian1}
\end{align}
where $\tV_\mu^A$ is a new, massive vector field transforming as a
triplet of $SU(2)_L$, $\tW_\mu ^A$ are the SM weak gauge bosons, and
$\phi$ is the SM Higgs doublet.  The kinetic term of the vector field
includes a covariant derivative,
\begin{align}
\label{eq:cov_D_triplet}
  \tV_{\mu \nu}^A = \widetilde{D}_\mu \tV_\nu^A -  \widetilde{D}_\nu \tV_\mu^A
  \quad \text{with} \quad
  \widetilde{D}_\mu \tV_\nu^A = \partial_\mu \tV_\nu^A - g_2 f^{ABC} \tW^B_\mu \tV^C_\nu  \; .
\end{align}
where $A,B,C$ are $SU(2)_L$ indices and the covariant derivative carries a tilde to indicate that it contains the fields $\tW_\mu ^A$.
The currents coupling the heavy vector to the SM-fields are given by
\begin{align}
  J_{\mu}^{l A} =\bar l_i \g_\mu t^A l_j\, \delta^{ij} \,,
  \qqquad
 J_{\mu}^{q A} = \bar q_i \g_\mu t^A q_j\, \delta^{ij} \,,
 \qqquad
 J_\mu^{H A} = \phi^\dag i\overleftrightarrow{D}_{\mu}^A \phi \; ,
\end{align}
with $l$, $q$ being the SM lepton and quark doublets, $t^A=\sigma^A/2$
the $SU(2)$ generators and $\sigma^A$ the Pauli matrices. $i,j$ are
flavor indices and the Lagrangian is defined in a flavor-symmetric
limit. In the Higgs current, $(\phi^\dag
i\overleftrightarrow{D}_{\mu}^A \phi) = i\phi^\dag t^A(D_\mu\phi) - i
(D_\mu \phi^\dag) t^A\phi$.  As pointed out in \cite{Summ:2021aek},
the theory cannot be quantized in a self-consistent way for $\gt_{VH}
< 0$.

The gauge mixing described by the triplet model is familiar from the
general case of extra-$U(1)$ bosons~\cite{Bauer:2018onh}. A special
feature is the explicit $\tV$-mass term, which would have to be
generated by some kind of symmetry breaking and likely involve
additional fields; we ignore these additional fields, for instance in
their effect on $\gt_M$. The Higgs doublet $\phi$ is yet to develop a
VEV, which means we are working in the unbroken electroweak
phase. Underlying this choice is the assumption that a SMEFT expansion
for the EFT exists. This is the case unless there are additional
sources of electroweak symmetry breaking, or a heavy particle obtains
all of its mass from the Higgs VEV~\cite{Cohen:2020xca}.  Even in the
weakly coupled UV-completion of the triplet model there are no
additional sources of electroweak symmetry breaking, because the
additional scalar breaks $SU(2)\times SU(2)$ to $SU(2)_L$ and leaves
the electroweak symmetry completely intact.

To remove the kinetic mixing, we can re-define the SM-gauge field
as~\cite{delAguila:2010mx,Pappadopulo:2014qza}
\begin{align}
  W^{\mu \nu A}
  = \tW^{\mu \nu A} + \gt_M \tV^{\mu \nu A} 
  = \p^\mu ( \tW^{\nu A} + \gt_M \tV^{\nu A} )
     - \p^\nu ( \tW^{\mu A} + \gt_M \tV^{\mu A} )
     + \cdots
     \label{eq:sm-redef}
\end{align}
For the triplet field we only allow for a re-scaling, $\tV^{\mu A} =
\alpha V^{\mu A}$, so that the triplet mass does not get transferred
into the SM-gauge sector. The triplet mass also fixes the phase of the
real vector field $\tV_\mu^A$, such that $\alpha$ has to be
real. Requiring a canonical normalization of the new kinetic term
$V^{\mu \nu A} V_{\mu \nu}^A$ we find $\alpha^2 = 1/(1 - \gt_M^2)$.
This relation requires $\gt_M \neq \pm 1$, to ensure a valid model
with a propagating heavy vector. Furthermore, as we will see in
Sec.~\ref{sec:toy}, we need to require $|\gt_M|<1$ for the squared
pole mass of the resonance to be positive.  The final form of the
gauge field re-definition in Eq.\eqref{eq:sm-redef} becomes
\begin{align}
  \tW^{\mu A} =  W^{\mu A} -\frac{\tilde{g}_M}{\sqrt{1-\tilde{g}_M^2}} V^{\mu A} 
  \qquad \text{and} \qquad
  \tV^{\mu A} = \frac{1}{\sqrt{1-\tilde{g}_M^2}} V^{\mu A} \; ,
  \label{eq:field-redef}
\end{align}
and brings the Lagrangian into the form
\begin{align}
  \lag = \lag_\text{SM}
  &- \frac{1}{4}V^{\mu \nu A} V_{\mu \nu}^A
  + \frac{m_V^2}{2}V^{\mu A}V_{\mu}^A \notag \\
  &+ \sum_f g_f \; V^{\mu A} J_{\mu}^{fA} 
  + g_H V^{\mu A} J_\mu^{HA} 
  + \frac{g_{VH}}{2} \; |H|^2V^{\mu A}V_{\mu}^A \notag \\
  &+ \frac{g_{3V}}{2} f^{ABC} \; V^{\mu A}  V^{\nu B}  V_{\mu \nu}^C 
  - \frac{g_{2VW}}{2} f^{ABC} \; V^{\mu B}  V^{\nu C}  W_{\mu \nu}^A  \; ,
\label{eq:lagrangian2}
\end{align}
which has the same structure as Eq.\eqref{eq:lagrangian1}, but
additional triple and quartic gauge couplings between the weak and
triplet sectors.  The Lagrangian parameters are related through
\begin{alignat}{7}
 m_V^2 &= \frac{\tilde m_V^2}{1-\gt_M^2} \; ,
 &\qquad 
 g_H &= \frac{\gt_H+g_2 \gt_M}{\sqrt{1-\gt_M^2}} \; ,
 &\qquad 
 g_f &= \frac{\gt_f +g_2 \gt_M}{\sqrt{1-\gt_M^2}}\; ,
 \notag \\
 g_{VH} &= \frac{2\gt_{VH}+g_2^2 \gt_M^2+2g_2\gt_H\gt_M}{2(1-\gt_M^2)}\; ,
 &\quad
 g_{3V} &= -\frac{2 g_2 \gt_M}{(1-\gt_M^2)^{1/2}}\; ,
 &\qquad 
 g_{2VW} &= \frac{g_2 \gt_M^2}{1-\gt_M^2} \; ,
\label{eq:couplings}
\end{alignat}
where $g_2$ denotes the $SU(2)_L$ gauge coupling. The heavy vector
triplet couples to the weak gauge bosons not only via the
$g-$couplings in Eq.\eqref{eq:lagrangian2}, but also through the
non-abelian component of the covariant derivative
Eq.\eqref{eq:cov_D_triplet}, that leads to interaction terms of the
form $(\partial V) VW$ and $VVWW$. These interactions are weighted by
the weak gauge coupling, and therefore are present even if
$g_i\,(\gt_i)\equiv 0$.

\subsection{SFitter setup}
\label{sec:sfitter}

The \sfitter\ framework~\cite{Lafaye:2004cn} has been long employed
for global analyses of LHC measurements in the context of Higgs
couplings and
EFTs~\cite{Klute:2012pu,Corbett:2015ksa,Corbett:2015mqf,Butter:2016cvz,Biekotter:2018rhp,Brivio:2019ius},
including a comprehensive study of an analysis in terms of Higgs
couplings and its UV-completion~\cite{Lopez-Val:2013yba}.  The
approach is unique in that it allows a comprehensive treatment of
uncertainties: \sfitter\ uses a likelihood set up that includes a
broad set of statistical, systematic, and theory
uncertainties. Statistical and most systematic ones are described by a
Poisson- or Gauss-shaped likelihood. Theoretical uncertainties lack a
frequentist interpretation, and are described by flat likelihoods in
\sfitter, corresponding to a range of equally likely theory
predictions. An important difference between employing a flat
likelihood compared to a Gaussian one is that the uncorrelated profile
likelihood adds the uncertainties from the flat distributions
linearly, while Gaussian error bars are added in quadrature. The
profile likelihood combination of a flat and a Gaussian uncertainty
gives the well-known RFit prescription~\cite{Hocker:2001xe}.
Correlations among certain classes of systematic uncertainties are
also included.

From the technical point of view, the new aspect of the
\sfitter\ analysis presented in this paper is the translation of the
SMEFT likelihood into the parameter space of the UV model.  In the
fit, all observables are parameterized in the SMEFT using the operator
set provided in Tab.~\ref{tab:smeft_basis}, that is based on the HISZ
basis~\cite{Hagiwara:1993ck}. All SMEFT predictions are at LO in QCD
and scaled by the same corrections as the SM-rates used for the actual
experimental analysis. Terms obtained from squaring amplitudes with
one operator insertion, that are quadratic in the Wilson coefficients,
are retained.  The Wilson coefficients are then expressed in terms of
$\gt_i$ parameters of the UV model, Eq.\eqref{eq:lagrangian1}, using
the one-loop matching expressions onto the Warsaw basis provided in
Ref.~\cite{code} and the Warsaw-to-HISZ basis translation in
Appendix~\ref{app:basis}. In this way, the likelihood can be directly
sampled in the parameter space of the UV model.

In addition, we employ a new likelihood sampling
method~\cite{nextSFitter} compared to previous \sfitter\ analyses,
that ensures a much more efficient sampling close to the SM point,
where all Wilson coefficients vanish. By contrast, the previous
sampling method was optimized for the detection of potential secondary
maxima in the likelihood, by giving higher weight to the edges of the
parameter space.

\subsubsection*{Dataset}

The SMEFT analysis presented in this work builds directly on the
dataset employed in Ref.~\cite{Biekotter:2018rhp}, which includes
electroweak precision observables (EWPO) at LEP (14 measurements),
Higgs measurements (275) and di-boson measurements at the LHC
(43). The latter contain results from both Run~1 and
Run~2~\cite{Butter:2016cvz}.  In addition, we include differential
measurements from three resonance searches by ATLAS, that reach up to
invariant masses in the multi-TeV range and that we re-interpret
within the SMEFT framework.  One of these~\cite{Aaboud:2017cxo} was
already included in the analysis of Ref.~\cite{Biekotter:2018rhp}. The
other two~\cite{Aad:2020tps,Aad:2020ddw} are more recent and have been
added specifically for this work.  These measurements are not usually
included in the SMEFT analyses and are not covered by the simplified
template cross section framework~\cite{Brehmer:2019gmn}. Nevertheless,
it can be instructive to explore their sensitivity, particularly to
operators that induce momentum-enhanced corrections. Moreover, all the
resonance searches considered here target heavy vector triplets
decaying into $WH$ or $WW$ as a potential signal. Therefore they allow
to compare directly the constraining power of the SMEFT analysis to that
of the direct search.

\subsubsection*{Theory uncertainties}

In view of the upcoming LHC runs and their rapidly growing data sets,
the treatment of theory uncertainties in global analyses is becoming
critical. In our analysis, we include theory uncertainties associated
to parton distribution functions, to missing higher orders in the SM
or SMEFT predictions, and to the matching scale to the EFT. The latter
will be discussed in more detail in Sec.~\ref{sec:toy_scale}.

For the time being, we do not include uncertainties associated to
missing SMEFT operators due to the truncation of the SMEFT
Lagrangian~\cite{Trott:2021vqa} or to symmetry assumptions, such as
CP-conservation. Nevertheless, the impact of missing higher orders in
the EFT expansion becomes obviously manifest in the comparison between
constraints extracted from the SMEFT analysis and from direct
searches.

Concerning higher orders in the loop expansion, Higgs analyses in
\sfitter\ currently adopt the most accurate SM predictions available,
which are implemented so as to match the state-of-the art predictions
reported in the experimental analyses.  The corresponding $K$-factors
are then applied onto the tree-level SMEFT predictions as well, which
is tantamount to assuming that QCD corrections scale evenly for all
SMEFT operators and in the same way as in the SM.  Although this
assumption is, strictly speaking, not correct~\cite{Baglio:2020oqu},
for the rate measurements considered here we do not expect large
variations in the $K$-factors between different operators.  For some
kinematic distributions these effects can be larger. We therefore
assign conservative theory uncertainties in order to reduce the
numerical impact of these effects.  A proper SMEFT simulation of Higgs
and di-boson production up to NLO in QCD is postponed to a future
work.

\section{Toy fits and matching uncertainty}
\label{sec:toy}

In this section we discuss two aspects of the vector triplet model and
of its matching onto the SMEFT, that are preliminary to a correct
SMEFT global analysis. The first is the decoupling limit of the model,
and the second is the numerical impact of varying the scale at which
the 1-loop matching is performed. Both issues are analyzed via
simplified toy fits.

\subsection{Decoupling}
\label{sec:toy_decoup}

\begin{figure}[t]\flushleft
  \includegraphics[height=0.35\textwidth,trim=0 0 4cm 0,clip]{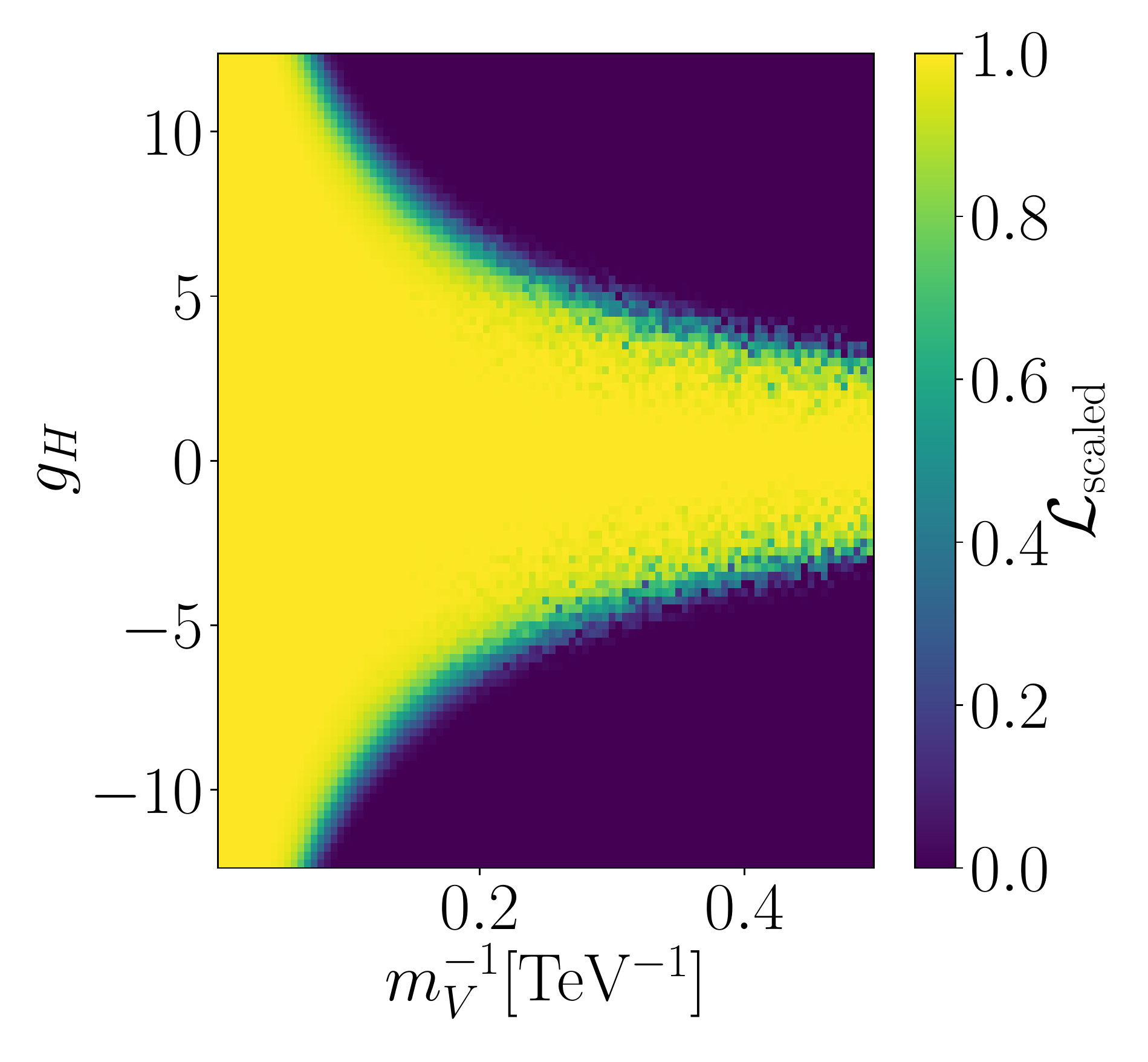}
  \includegraphics[height=0.35\textwidth,trim=0 0 4cm 0,clip]{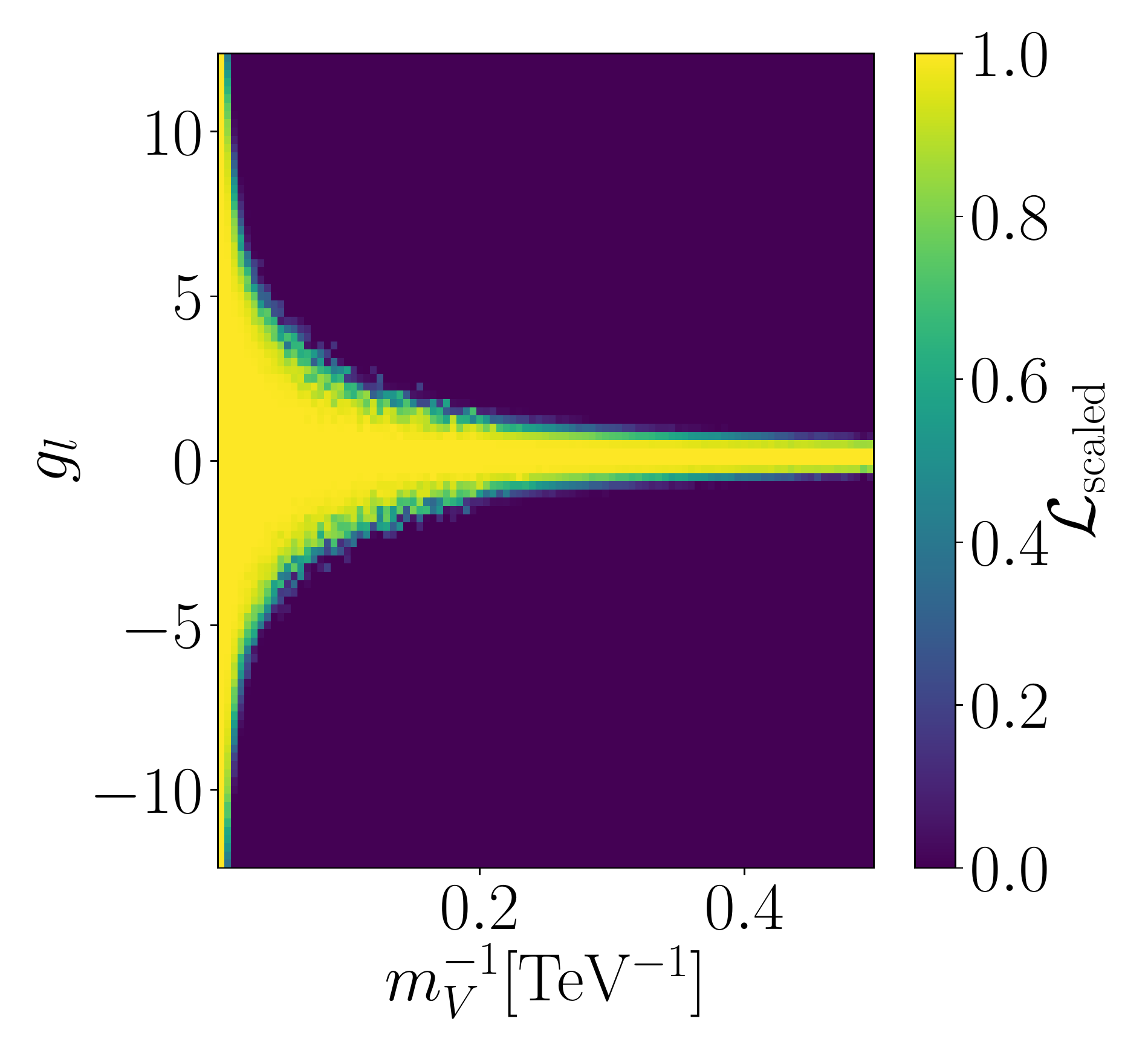}
  \includegraphics[height=0.35\textwidth,trim=0 0 4cm 0,clip]{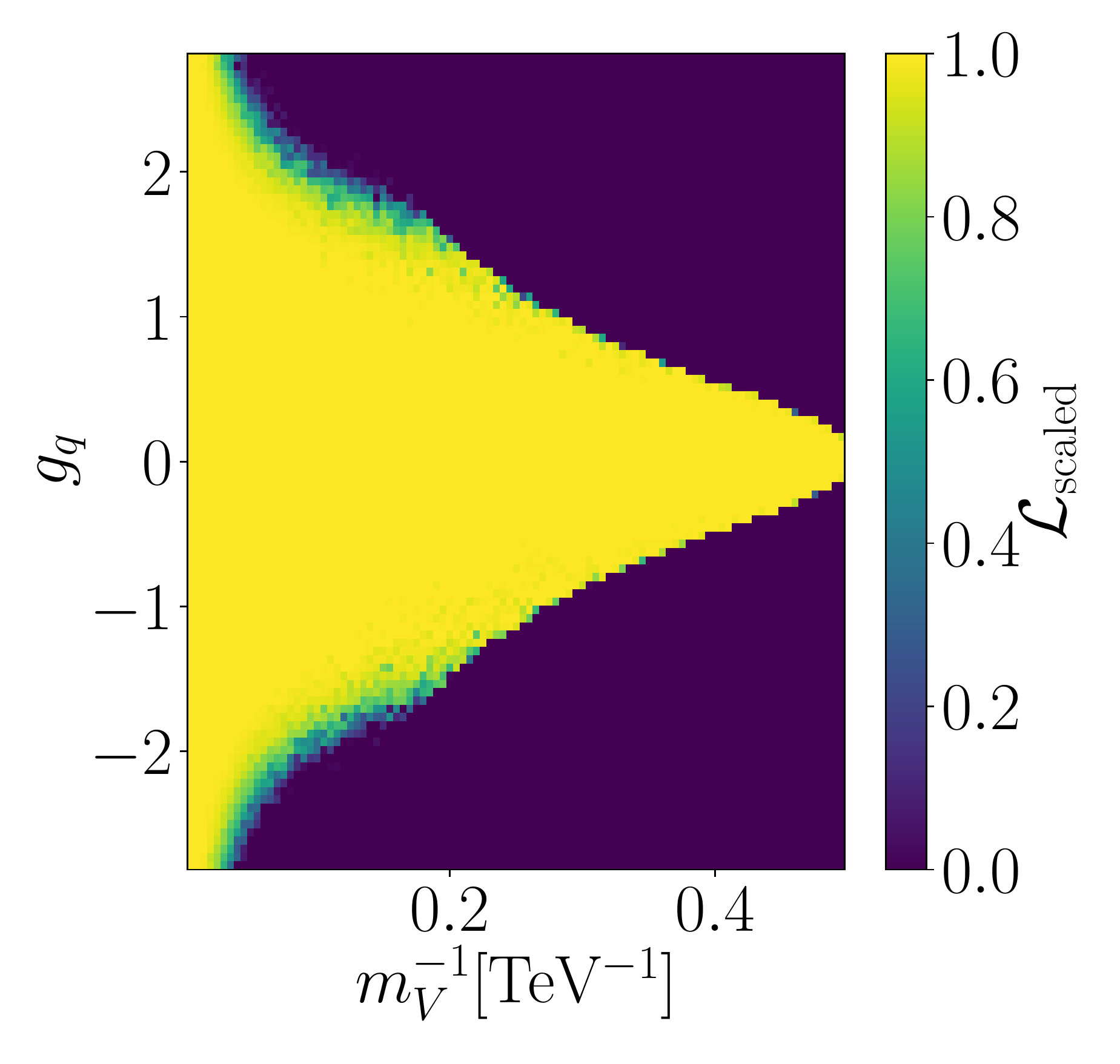} \\
  \includegraphics[height=0.35\textwidth,trim=0 0 4cm 0,clip]{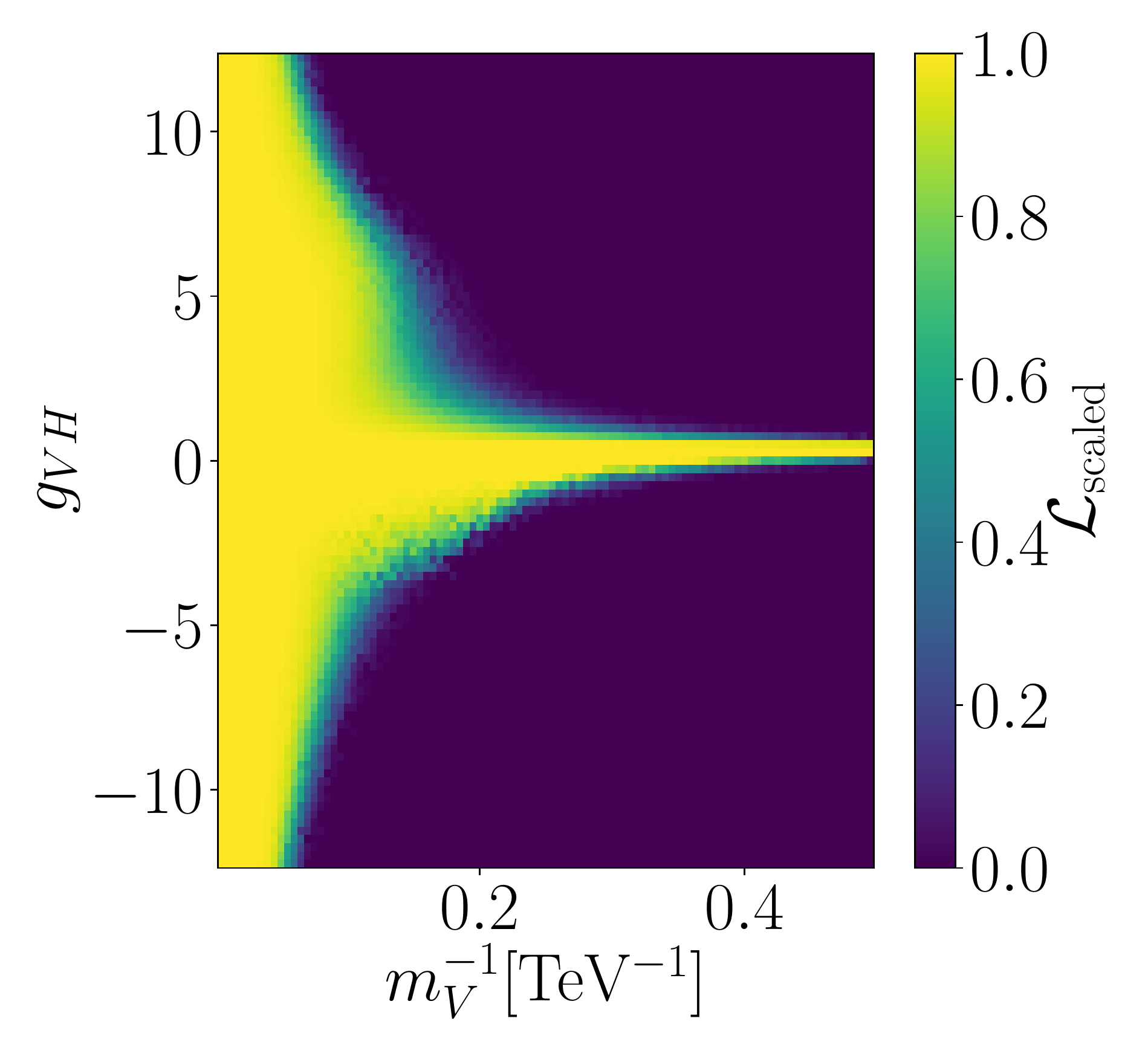}
  \includegraphics[height=0.35\textwidth,trim=0 0 4cm 0,clip]{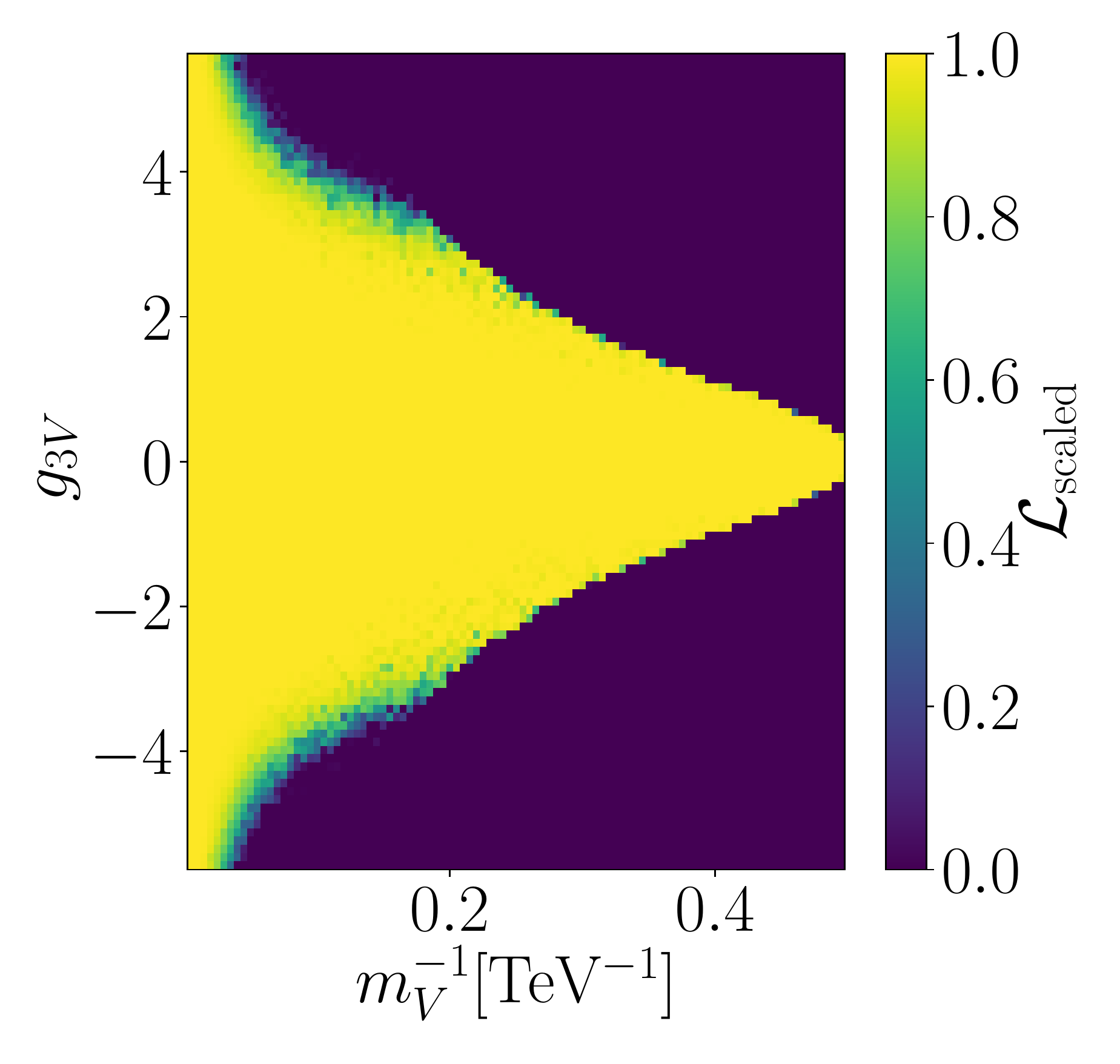}
  \includegraphics[height=0.35\textwidth]{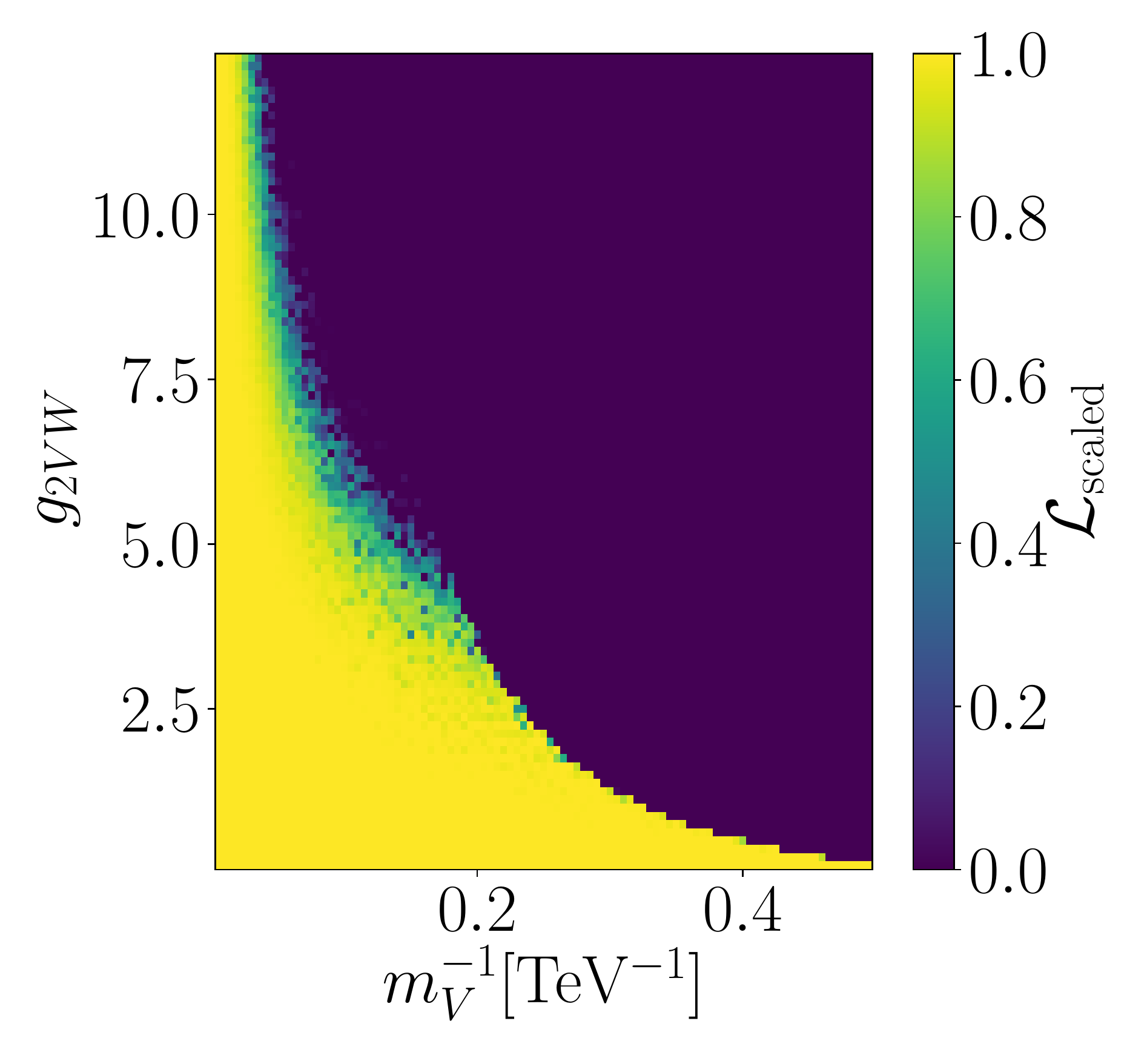}
  \caption{Decoupling pattern for the vector triplet model. Global fit
    with all measurements at their SM values and to the 4 free
    parameters $\tilde{m}_V, \gt_M, \gt_H, \gt_l$, and subsequently
    projected onto the 7 parameters of the unmixed Lagrangian
    Eq.\eqref{eq:lagrangian2}.}
  \label{fig:dec_sm}
\end{figure}

\begin{figure}[t]
  \includegraphics[height=0.35\textwidth ,trim=0 0 4cm 0,clip]{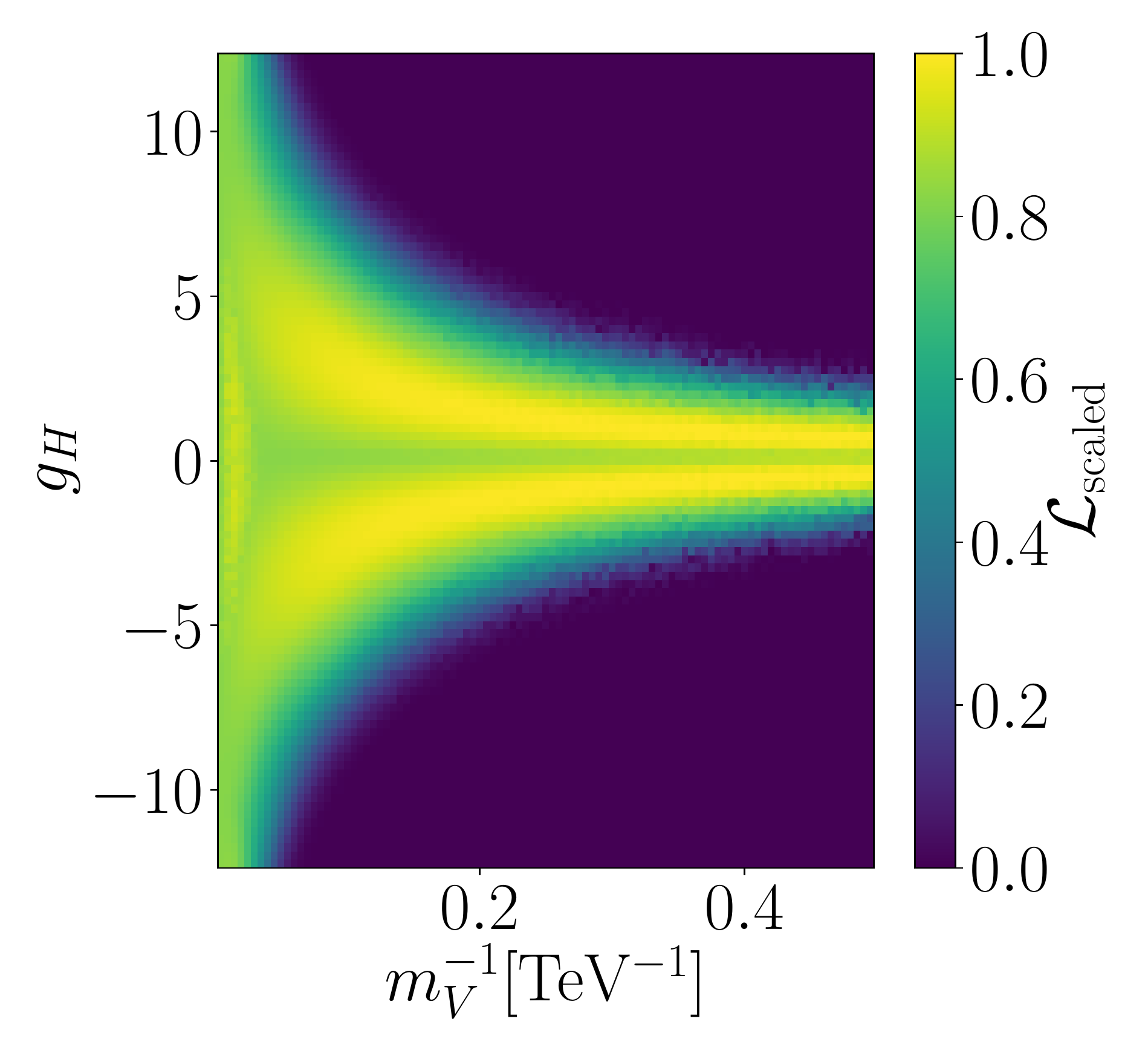}
  \includegraphics[height=0.35\textwidth,trim=0 0 4cm 0,clip]{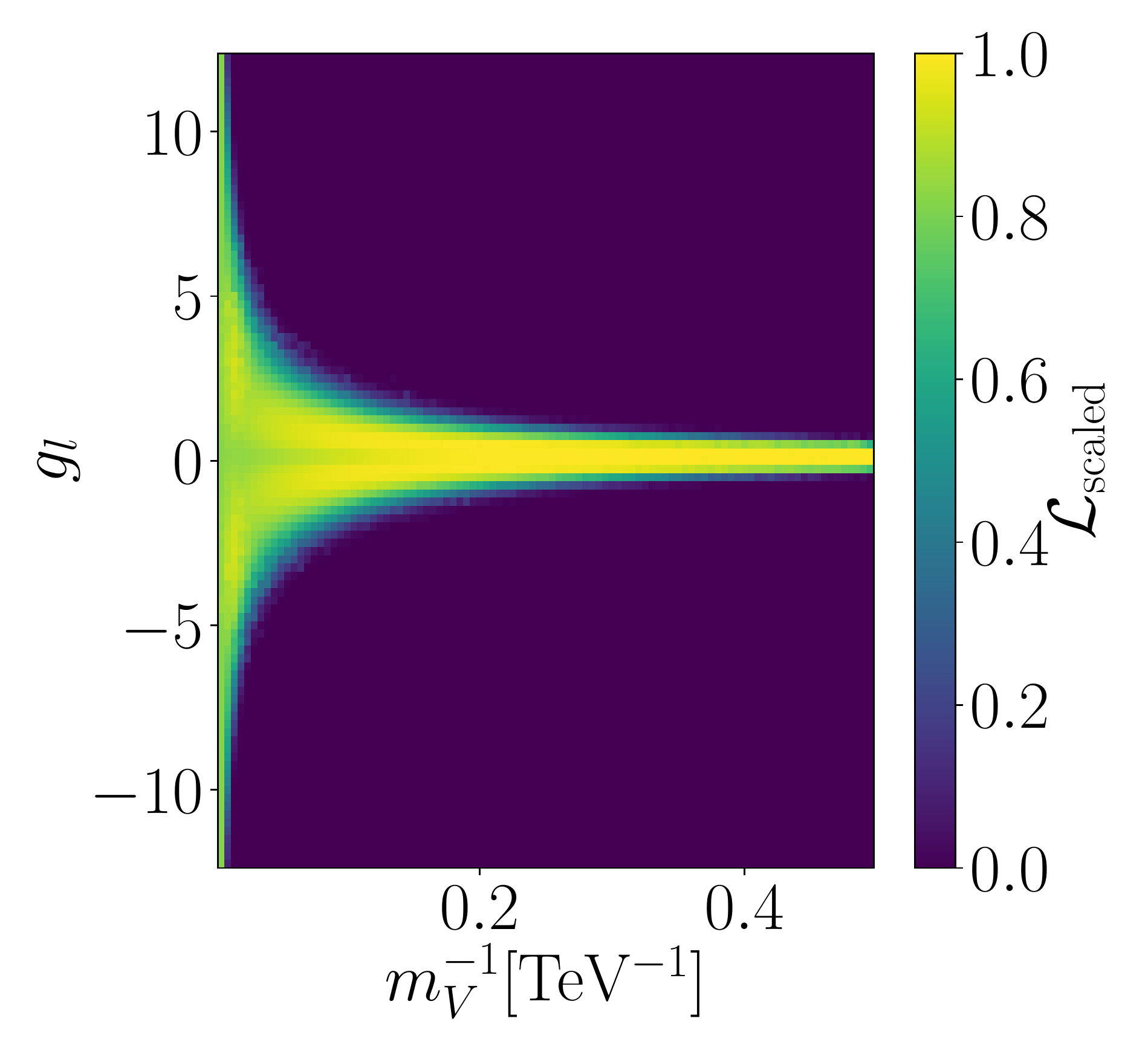}
  \includegraphics[height=0.35\textwidth]{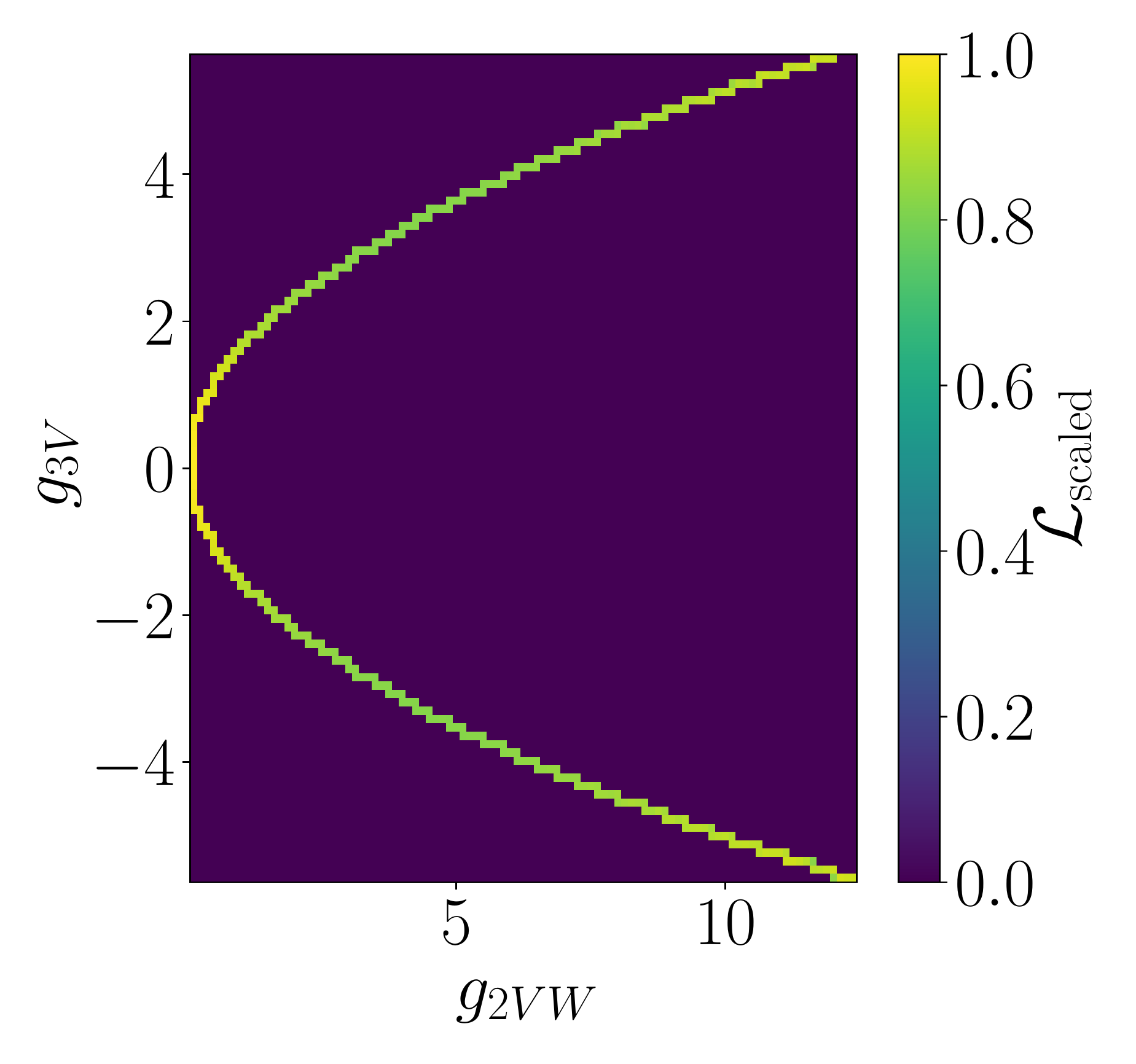}
  \caption{Results of the same global analysis as in
    Fig.~\ref{fig:dec_sm}, but with measurements set to their actual
    values.}
  \label{fig:dec_data}
\end{figure}

The decoupling limit of the vector triplet model considered in this
work is most easily identified starting from the Lagrangian of
Eq.\eqref{eq:lagrangian2}, where, as long as the EW symmetry is
unbroken, the heavy triplet and the SM gauge bosons do not mix.  In
this case, it is easy to see that the BSM states decouple for large
values of the physical mass, $m_V\to \infty$. This is directly
reflected in the matching formulas, which give $\lim_{m_V\to\infty}
C_i\equiv 0$ for all the dimension-6 Wilson coefficients.  At the
level of a global fit, the decoupling limit can be visualized by
setting the central values of all the measurements included to match
the corresponding SM predictions. Figure~\ref{fig:dec_sm} shows the
results obtained in this way from \sfitter: the likelihood is first
computed as a function of 4 free parameters in
the Lagrangian of Eq.\eqref{eq:lagrangian1}
\begin{align}
\{\tmv, \gt_M, \gt_H, \gt_l\}\,,
\label{eq:dec_tilde}
\end{align}
setting other parameters $\gt_q, \, \gt_{VH}$ to zero. We then project
them onto the 7 parameters for the rotated Lagrangian of
Eq.\eqref{eq:lagrangian2},
\begin{align}
  \{ m_V, g_H, g_l, g_q, g_{VH}, g_{3V}, g_{2VW}\} \,.
\label{eq:dec_notilde}
\end{align}
At this stage, we fix the matching scale to $Q=m_V=\unit[4]{TeV}$.
For each of the couplings we see that, as expected, the range of
allowed values increases as $m_V^{-1}\to 0$. It is worth noting that
the rate at which this happens varies between the $g$-parameters. This
is due to the fact that the matching expressions do not scale
homogeneously with $g_i^2/m_V^2$, but generally have a more complex
polynomial structure. The degeneracy between $g_i$ and $1/m_V$ in
these expressions is also broken by the $V-W$ interactions
proportional to the weak gauge coupling. The homogeneity of the yellow
regions indicates that there the likelihood is flat and no point is
preferred.  Setting all measurements to their actual measured values,
which generally depart from the SM predictions, has the effect of
introducing a substructure in the likelihood, thereby identifying a
more restricted preferred region. This is shown, for a subset of
panels, in Fig.~\ref{fig:dec_data}. Here, for instance, the best fit
point moves to finite $m_V$ and prefers non-vanishing values of
$g_H$. Note that, to good approximation, the entire region highlighted
in green is allowed at 68\%CL. The yellow points simply identify a
best-fit region and should not be interpreted as statistically
significant.
Finally, the reduced number of parameters in the Lagrangian
Eq.\eqref{eq:lagrangian1} as compared to the setup without kinetic
mixing induces strong correlations through $\gt_M$, as illustrated in
the $g_{2VW} - g_{3V}$ plane of Fig~\ref{fig:dec_data}.

As the
matching procedure highlighted in Sec.~\ref{sec:basics_match} requires
a separation between light and heavy degrees of freedom, defining the
decoupling limit in the notation of Eq.\eqref{eq:lagrangian1} requires
some more care, due to the explicit kinetic mixing between the heavy
triplet and the SM gauge fields.

From Eq.\eqref{eq:couplings}, we see that $m_V\to\infty$ can be
achieved for $\tmv\to\infty$ or for $|\gt_M|\to 1$.  However, the
condition $|\gt_M|=1$ does not lead to a well-defined decoupling
condition, because in this limit $\tilde V_\mu^A$ become auxiliary
fields, i.e. the theory loses three dynamical degrees of freedom. This is
not sufficient for a proper decoupling in the EFT sense because even
as an auxiliary field $\tV_\mu^A$ still induces mass-suppressed
vertices that enter correlation functions and we enter a strongly
interacting regime where our perturbative approach fails.

To see the impact of $\gt_M$ we resum insertions of gauge mixing into
the $\tW_\mu^A$ and $\tV_\mu^A$ propagators. The corrected propagators
of these fields become
\begin{align}
  \hat{D}^{\tV}_{\mu \nu} &=
  - \frac{i}{p^2-\tmv^2-\gt_M^2 p^2}\left(g_{\mu \nu} - (1-\gt_M^2) \frac{p_\mu p_\nu}{\tmv^2}\right) \notag \\
  \hat{D}^{\tW}_{\mu \nu} &=
  - \frac{i}{p^2}\left(g_{\mu \nu}-(1-\xi)\frac{p_\mu p_\nu}{p^2}\right)
  - \frac{i \gt_M^2}{p^2-\tmv^2-\gt_M^2 p^2}\left(g_{\mu \nu}-\frac{p_\mu p_\nu}{p^2}\right) \; .
\end{align} 
It is easy to see that for $|\gt_M|=1$ the resummed $\tV_\mu^A$
propagator loses its momentum dependence, which is indicative of the
field becoming auxiliary. For $|\gt_M|>1$, $\tV_\mu^A$ becomes
tachyonic while, for $|\gt_M|<1$, $\tV_\mu^A$ is a dynamical degree of
freedom. In this case its propagator has a physical pole at
$p^2=m_V^2$ as defined in Eq.\eqref{eq:couplings}, and it can be
expanded in $p^2/m_V^2 \ll 1$.  The resummed $\tW_\mu^A$ propagator
includes a term with a pole at $p^2=m_V^2$, contaminating $\tW_\mu^A$
with a contribution from $\tV_\mu^A$.  Therefore this field cannot be
directly identified with the SM weak bosons. However, in the
tree-level matching procedure, once the 1LPI effective action is
expanded in $p^2/m_V^2$, the component associated with the $\tV_\mu^A$
pole is shifted from the propagators to the interaction terms, which
are unambiguously fixed at this order by the matching condition of
Eq.\eqref{eq:DeltaGamma}.  At one loop, the fact that the EFT is the
low-energy limit of the UV model is manifest in the fact that only the
`hard' region of the momentum integral contributes to the functional
trace in the matching formula of
Eq.\eqref{eq:general-one-loop-matching}.  As a consequence, the first
term of the $\tW_\mu^A$ propagator cancels against the corresponding
EFT contributions, while the second term genuinely contributes to the
matching in the hard region. Equivalently, one can match in the
shifted basis directly identifying $W_\mu^A$ in the UV model with the
corresponding weak bosons in the SMEFT.

\begin{figure}[t]\flushleft
  \includegraphics[height=0.35\textwidth,trim=0 0 4cm 0,clip]{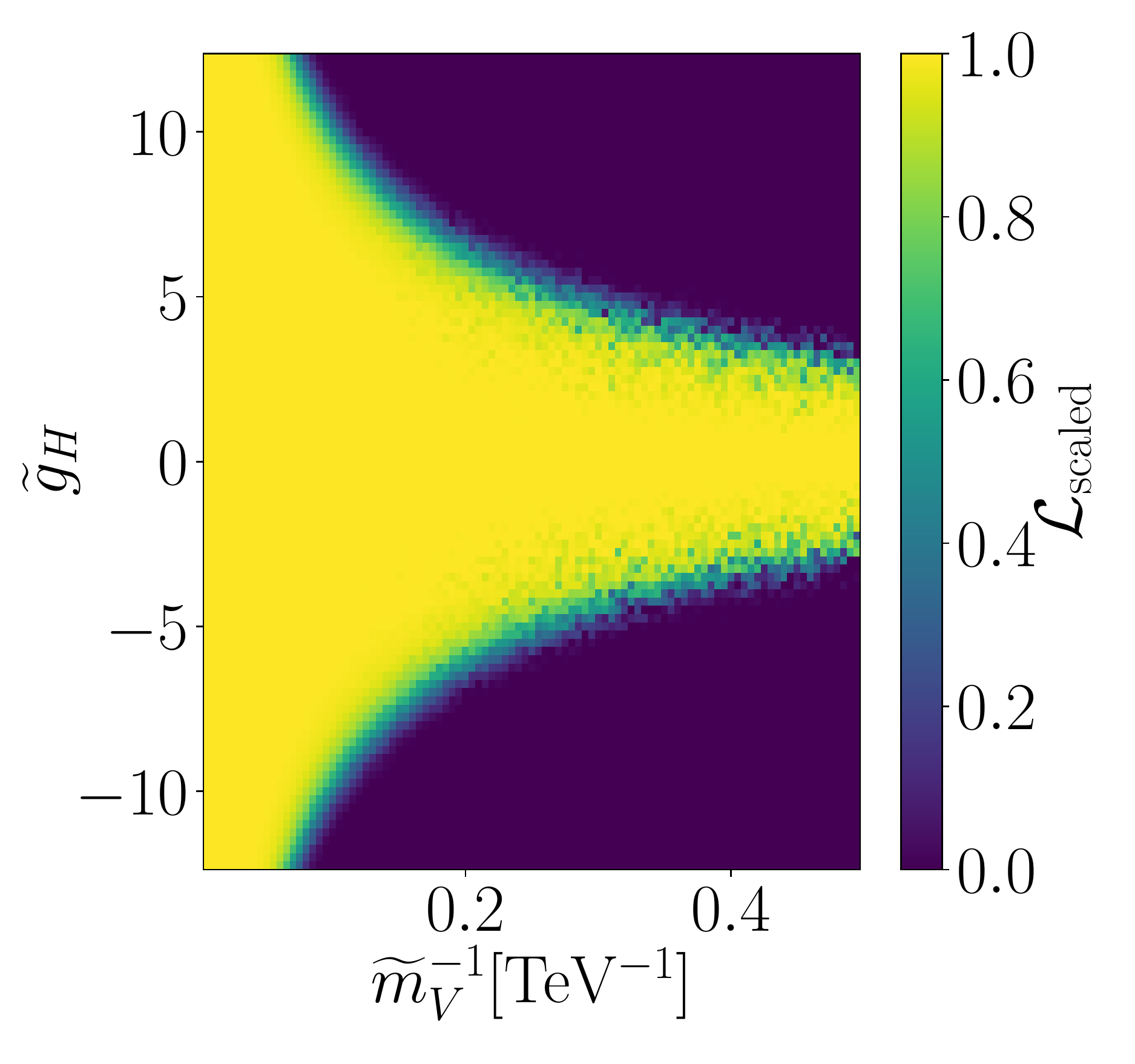}
  \includegraphics[height=0.35\textwidth,trim=0 0 4cm 0,clip]{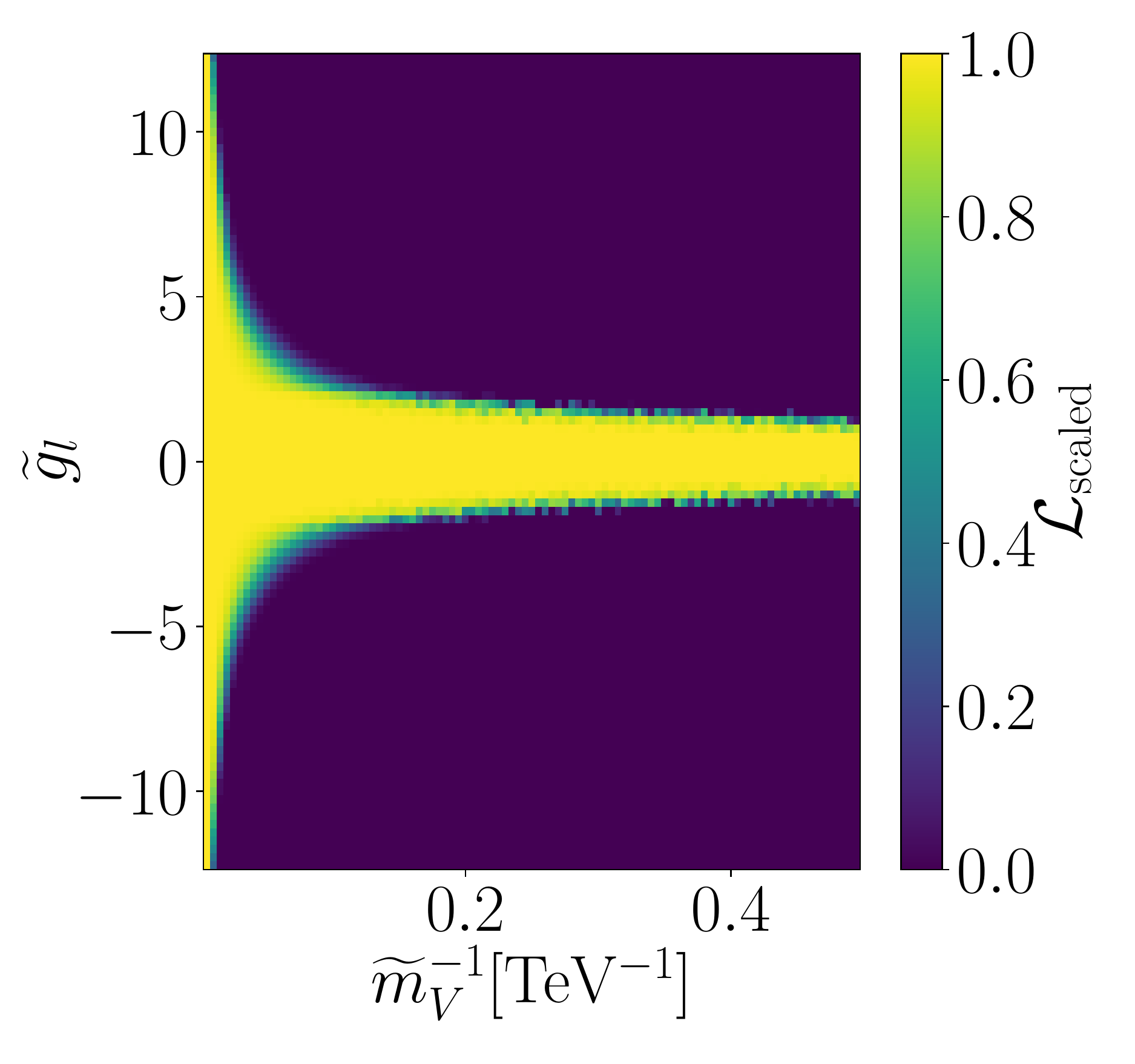}
  \includegraphics[height=0.35\textwidth,trim=0 0 4cm 0,clip]{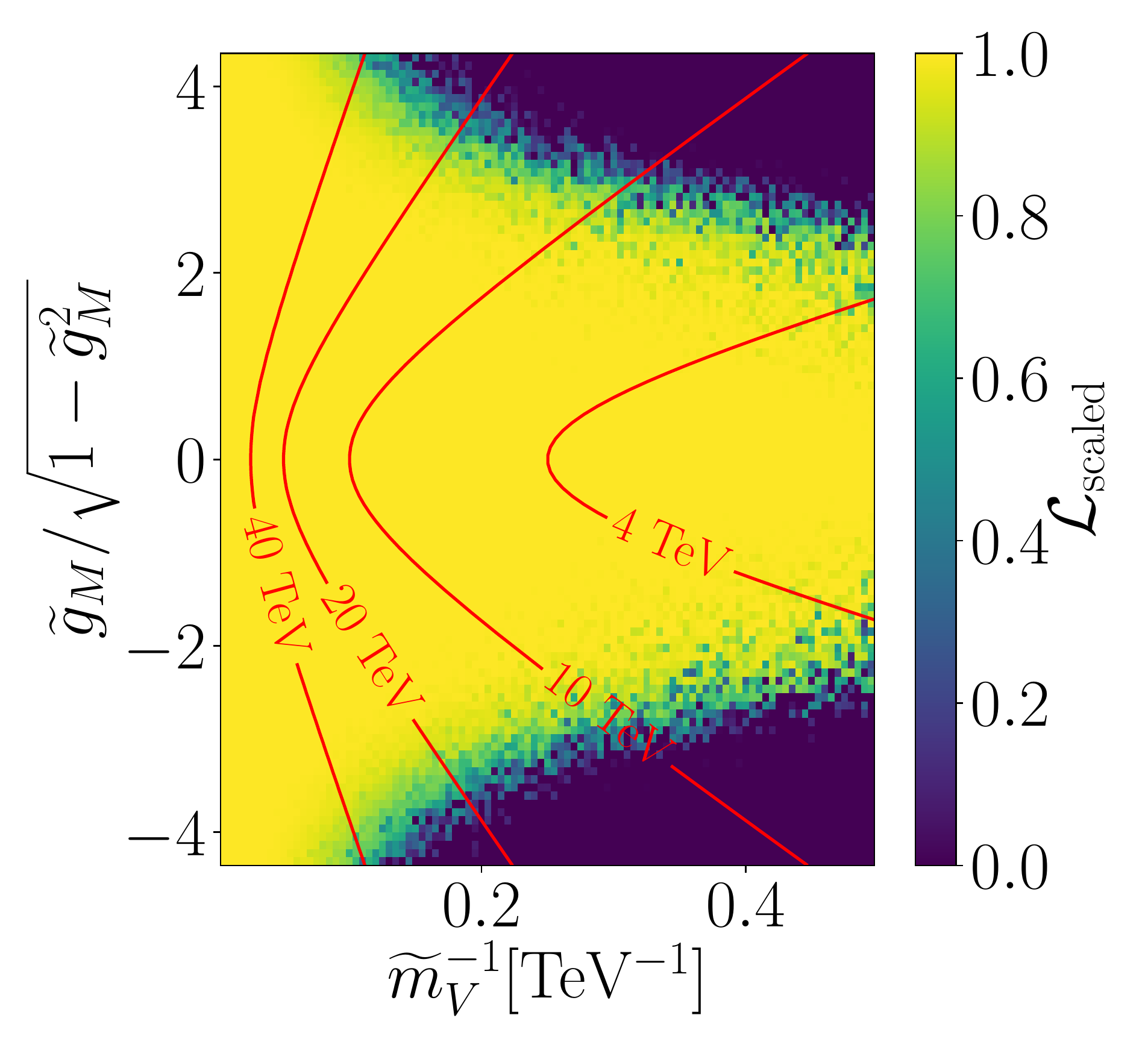}
  \\ 
  \includegraphics[height=0.35\textwidth,trim=0 0 4cm 0,clip]{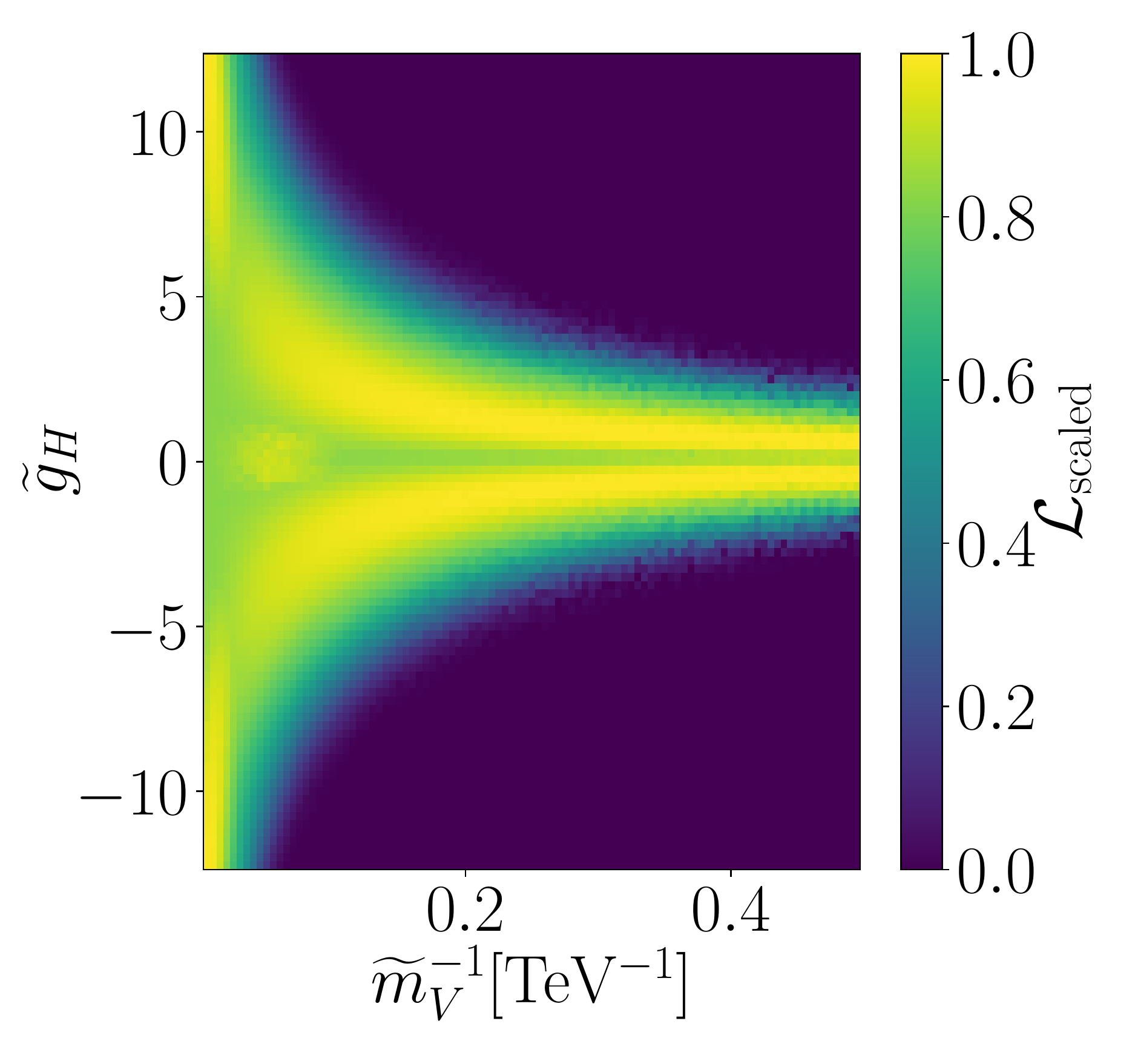}
  \includegraphics[height=0.35\textwidth,trim=0 0 4cm 0,clip]{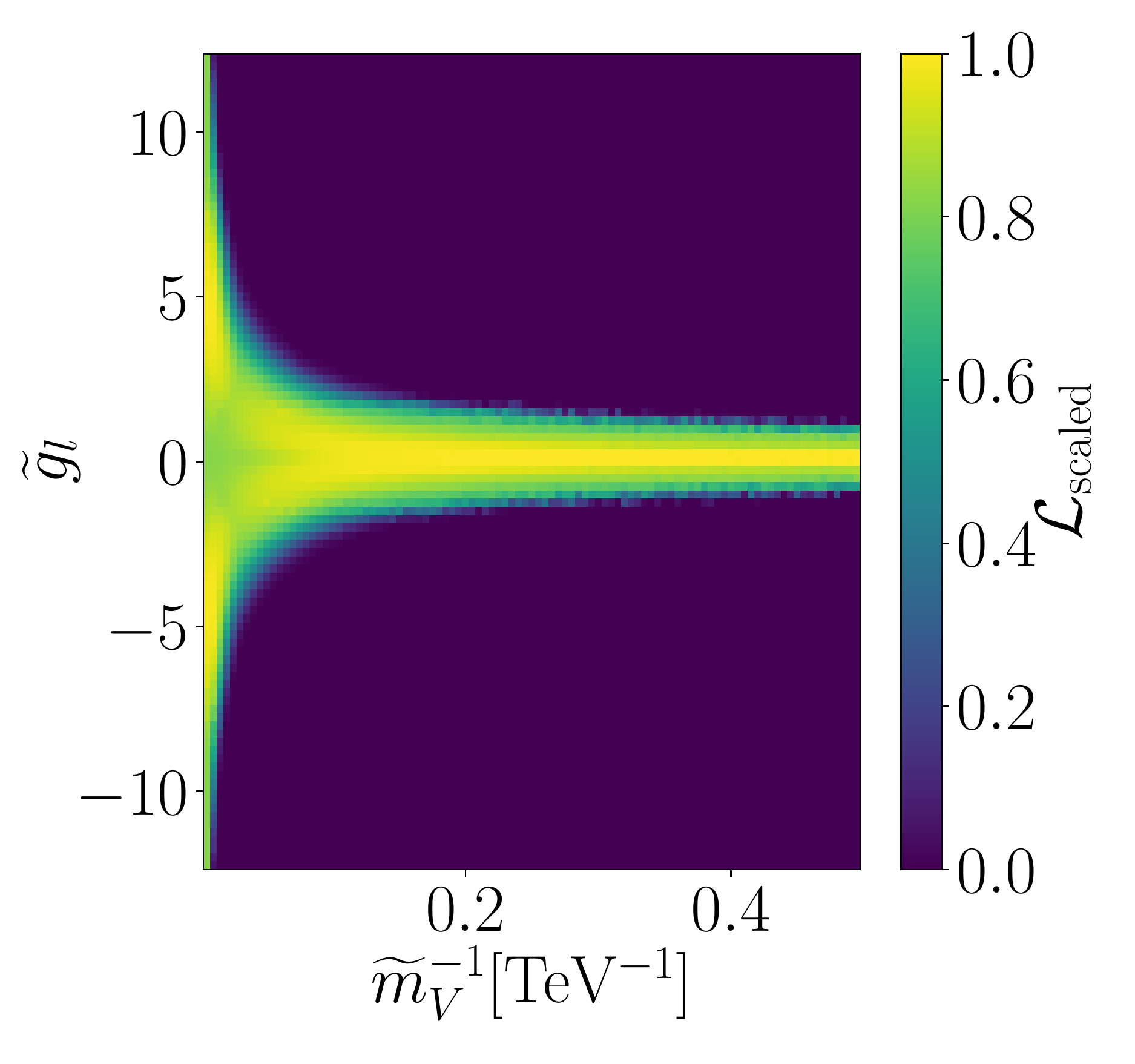}
  \includegraphics[height=0.35\textwidth]{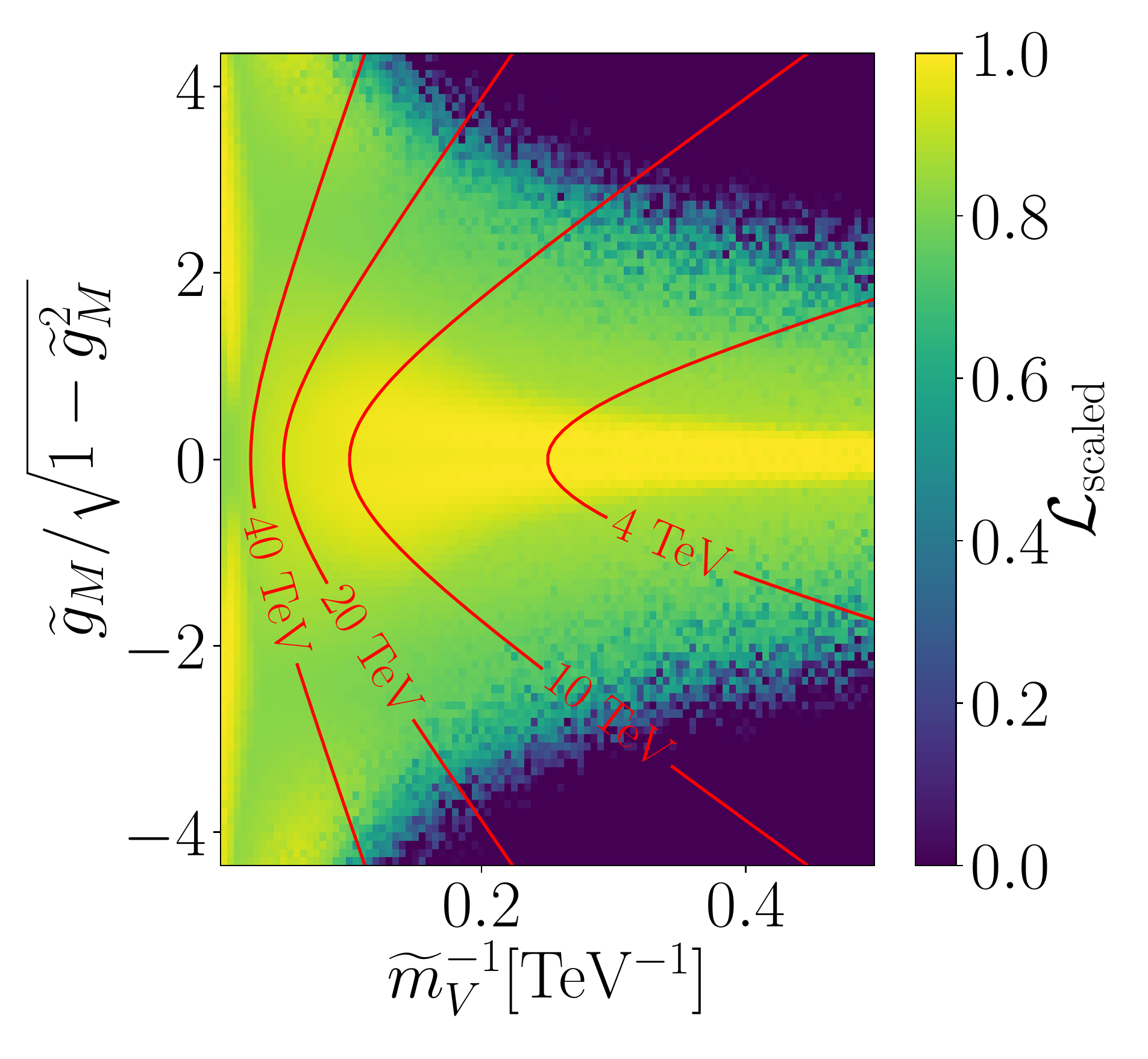}
  \caption{Results of the same global analyses as in
    Figs.~\ref{fig:dec_sm} (upper) and~\ref{fig:dec_data} (lower),
    projected on $\tilde{m}_V, \gt_H, \gt_l$ and the combination
    $\gt_M/\sqrt{1-\gt_M^2}$.}
  \label{fig:dec_tilde}
\end{figure}

In the top (bottom) panels of Fig.~\ref{fig:dec_tilde} we again show
the results of a global analysis where all measurements are set to
their SM prediction (to their actual values), this time projected onto
a subset of the $\gt$-parameters and onto the combination
$\gt_M/\sqrt{1 - \gt_M^2}$ that drives most $\gt-g$ relations, see
Eq.\eqref{eq:couplings}.  For reference, the right panels also show
lines of constant $m_V$, such that the decoupling limit $m_V\to\infty$
flows orthogonally to the lines.  Consistent with the results in the
unmixed basis (Fig.~\ref{fig:dec_sm}), the expected likelihood is
mostly flat in the entire preferred region, while the observed one
exhibits a substructure that identifies a best-fit region where
$\gt_H\neq 0$ and both $m_V$ and $\tmv$ are finite. The reason can be
identified in a few EWPO measurements that exhibit small ($<1\sigma$)
deviations from the SM expectation: $A_l(SLD)$ and $m_W$.

For $|\gt_M|\to1$ the theory becomes strongly interacting and some
perturbative unitarity considerations are therefore pertinent.
Requiring the couplings of the unmixed UV theory to remain
perturbative, the most stringent constraints on $\gt_M$ stem from
$g_{2VW}$
\begin{align}
  g_{2VW} \approx \frac{g_2 \gt_M^2}{1-\gt_M^2} < 4\pi
  \qquad \Leftrightarrow \qquad
 | \gt_M| < 0.975 \; .
\end{align}
Therefore for all our fits we require $| \gt_M| < 0.975$.

\subsection{Matching scale}
\label{sec:toy_scale}

In perturbative predictions of LHC observables, at least two
unphysical scales are known to reflect a theory uncertainty, the
factorization scale and the renormalization scale. Both arise
from a separation of an observable into different regimes with
different perturbative expansions, and the scale dependence would
vanish if we would include all orders in all predictions. For a
calculation at finite perturbative order we instead use the scale
variation as one measure of a theory uncertainty and treat it as an
unphysical nuisance parameter in theory
predictions~\cite{Lafaye:2009vr,Brivio:2019ius}.

One unphysical scale is the renormalization scale, which in the
context of dimensional regularization appears as a free parameter.  In
more physical terms, the renormalization scale is the energy scale
associated with those observables that we select for defining the
numerical parameters of the theory, the renormalization conditions.
Whenever scale choices are arbitrary, we often identify them with each
other and a typical energy scale of the scattering process to avoid
large logarithms.  Clearly, this does not work if renormalization
conditions involve widely distinct energy scales, such as in the
relation of UV-model parameters to the low-energy observables of the
SM.

The renormalization group equation apparently solves this problem.  It
relates observables at different scales, properly resumming logarithms
and absorbing them into running parameters.  However, it works only in
the absence of mass thresholds.  This strongly suggests to match
a UV model with a heavy mass $M$ to a low-energy EFT even if the
algebraic simplifications of the latter are not essential for a
specific calculation.

In a one-loop matching calculation that uses dimensional
regularization, the matching scale enters as an additional parameter.
However, in contrast to the original renormalization scale this
parameter is not entirely arbitrary. If we want to avoid large
logarithms, its reasonable range is bounded from above and below.  In
line with the generic discussion of one-loop matching above, we
illustrate this property in the following section.  We consider
examples of increasing complexity, starting from the QCD coupling,
turning next to the SM extended by a scalar singlet and finally
returning to the vector triplet model of Sec.~\ref{sec:basics_model}.

\subsubsection*{Running strong coupling}

We can illustrate the appearance of the matching scale using the
simple example of the running strong coupling. It provides the key
ingredients to understanding the EFT matching scale: the separation of
low-energy and UV regimes and contributions beyond tree level. In
general, the relation between the bare coupling and the renormalized
coupling in the $\msbar$ scheme is
\begin{align}
  \alpha_s^\text{bare}
  &= \alpha_s(p^2) \left[ 1 - \alpha_s b_0 \left( \frac{1}{\bar{\epsilon}} + \log \frac{\mu_R^2}{p^2} \right) \right] 
\quad \text{with} \quad 
&  b_0^{(n_f)} = \frac{1}{4\pi} \left( \frac{11}{3} N_c - \frac{2}{3} n_f \right) \; .
\label{eq:alphas1}
\end{align}
Here, $p^2$ is the energy scale of the scattering, $\mu_R^2$ is
introduced by dimensional regularization, and $1/\bar{\epsilon}=
1/\epsilon -\gamma_E+\log 4\pi$.  We identify our UV-regime as momenta
above the top mass, with six propagating quark flavors, and the
low-energy regime as described by five propagating quark flavors.  The
running of $\alpha_s$ in the two regimes is described by the beta
function with five or six flavors, respectively. The UV-divergences in
the low-energy and full UV-theories arise from five or six propagating
flavors, so the renormalization prescription Eq.\eqref{eq:alphas1} is
different in the two regimes.

The low-energy and UV-regimes are separated by a matching scale $\qm$,
which we choose to be of the order of the top mass to avoid large
logarithms or inconsistent symmetry structures. Matching conditions
guarantee that the two predictions for any observable are the same at
least at this scale.  Instead of looking at a full set of amplitudes
or correlation functions, we limit ourselves to the quasi-observable
$\alpha_s$.  Following Eq.\eqref{eq:alphas1}, the definitions of
$\alpha_s(p^2)$ in relation to the bare parameter are different, but
they have to agree when evaluated at the matching scale. This defines
a threshold correction
\begin{align}
  1 - \frac{\alpha_s b_0^{(6)}}{4\pi} \left( \frac{1}{\bar \epsilon} + \log \frac{\mu_R^2}{p^2} \right)
  \Bigg|_{\qm^2}
  = 1 - \frac{\alpha_s b_0^{(5)}}{4\pi} \left( \frac{1}{\bar \epsilon} + \log \frac{\mu_R^2}{p^2} \right) \Bigg|_{\qm^2} 
    + \frac{\alpha_s}{6 \pi}  \log \frac{\mu_R^2}{\qm^2} \; .
\end{align}
The relation of the threshold correction to loop effects is reflected
in the logarithmic form $\log \mu_R^2/\qm^2$. Together with the
five-flavor $\msbar$ counter term it defines $\alpha_s$ in the
low-energy regime as
\begin{align}
  \alpha_s^\text{bare}
  &= \alpha_s(p^2) \left[ 1 - \frac{\alpha_s b_0^{(5)}}{4\pi} \left( \frac{1}{\bar \epsilon} + \log \frac{\mu_R^2}{p^2} \right)
    + \frac{\alpha_s}{6 \pi}  \log \frac{\mu_R^2}{\qm^2}\right] \; .
\label{eq:ct_low}
\end{align}
This definition includes three scales for a given scattering process,
the physical scale $p^2$, the renormalization scale $\mu_R^2$, and the
matching scale $\qm^2$. In simple problems, the renormalization scale
and the physical scale can be identified to avoid potentially large
logarithms. The matching scale is usually set to the mass of the
decoupled particle, $\qm = m_t^2$, leading to a threshold correction
that is non-zero in general.

From our toy example we can immediately see the role of the threshold
correction at the matching scale and the renormalization group
running. If we start from the UV, all parameters of the theory evolve
based on the full particle spectrum. In the low-energy theory part of
the spectrum decouples also from the running, which can even break the
underlying symmetries~\cite{Kilian:2004uj}, and we will follow a
completely different renormalization group flow.
The matching corrections adjust for this effect. They move us to the
same flow line in the EFT, independent of the choice of matching scale
and with all the caveats of maintaining perturbative control,
accounting for changes of the spectrum, changing symmetries, etc.

\subsubsection*{Singlet extension}

When we interpret a SMEFT calculation for an LHC process as a
low-energy approximation to a UV-prediction, we again break the phase
space of the scattering process into two parts.  We first illustrate
SMEFT matching using the singlet-extended SM~\cite{Dawson:2021jcl,Haisch:2020ahr},
\begin{align}
  \lag \supset \frac{1}{2}\left(\partial_\mu S\right)\left(\partial^\mu S\right)
  -\frac{1}{2} M^2 S^2
  -A |\phi|^2 S
  -\frac{\kappa}{2}|\phi|^2 S^2
  -\frac{\mu}{3!}S^3
  -\frac{\lambda_S}{4!}S^4 \; . 
\end{align}
The singlet mass is given by $M_S^2 = M^2+\mathcal{O}(v^2)$; we
integrate it out under the condition $M_S \sim M \gg v$, ensuring a
consistent expansion in $v/M$~\cite{Cohen:2020xca}.  As a
simplification, we also assume $A$ to be of the order of $M$.  The
leading term in $v/M$ is defined by $v=0$ and can be obtained by matching in the unbroken phase. 
In the broken phase the Higgs VEV enters via the masses
of the SM-particles which properly belong to the EFT Lagrangian, below
the matching scale.  Matching in the broken phase would allow us to
include partial higher-order corrections in the EFT
expansion~\cite{Freitas:2016iwx}.  Since the mass scales in question
are not widely separated, it depends on the detailed numerics which
setup yields a more reliable approximation.  The SMEFT Lagrangian
reads
\begin{align}
\lag_\text{SMEFT} =  \lag_\text{SM} + \sum_i f_i(p/\mu_R) \, \mathcal{O}_i \; ,
\label{eq:SMEFT_LAG}
\end{align}
where the Wilson coefficients are scale dependent.  Specifically, we
want to define these coefficients such that the SMEFT reproduces all
low-energy observables of the UV-theory up to
$\mathcal{O}(v^3/M_S^3)$. As matching condition we use
Eq.\eqref{eq:matching-master}. In the functional approach we compute
this once and for all using functional traces. To illustrate some
features related to the matching scale, we compute some contributions
to the Wilson coefficient $f_{\phi,2}$ of the operator $\ope_{\phi,2}
= \partial_\mu (\phi^\dagger \phi) \partial^\mu (\phi^\dagger \phi)/2$
diagrammatically. As discussed in Appendix~\ref{app:basis}, it is related to
$Q_{\phi\Box} = |\phi|^2\Box |\phi|^2$ as $c_{\phi\Box} \approx
-f_{\phi,2}/2$, modulo fermionic operators. The operator contributes
to the correlation function with two external fields $\phi$ and two
external fields $\phi^\dagger$ and depends on $p^2$, so we fix it by
requiring
\begin{align}
     \partial_{p^2} \Gamma_\text{SMEFT} (\phi^\dagger,\phi^\dagger,\phi,\phi) \Bigg|_{p^2=0} = \partial_{p^2} \Gamma_\text{L,UV} (\phi^\dagger,\phi^\dagger,\phi,\phi) \Bigg|_{p^2=0} \; , 
\label{eq:singlet_match1}
\end{align}
order by order in the coupling. With some abuse of notation we also
denote specific correlation functions by $\Gamma$, arguments
indicating the external fields. Since both sides of the equation
involve running parameters, the matching has to be imposed at a given
scale,
\begin{center}
\begin{tikzpicture}[scale=0.5]
\draw[dashed,thick] (12,6) -- (13,5);
\draw[dashed,thick] (12,4) -- (13,5);
\draw[thick] (13,5) -- (15,5);
\draw[dashed,thick] (15,5) -- (16,6);
\draw[dashed,thick] (15,5) -- (16,4);
\node[align=left, below] at (14,5)%
{$S$};
\node[align=left, above] at (10.5,4.4)%
{\Large $\partial_{p^2}$};
\node[align=left, above] at (11.4,4.0)%
{\Huge(};
\node[align=right, above] at (19.5,4.5)%
{ $+$ $t$-channel + SM};
\node[align=right, above] at (22.8,4.0)%
{ \Huge)};
\node[align=right, above] at (9.3,4.6)%
{=};
\draw[](5,5) circle (0.3);
\draw[dashed,thick] (4,6) -- (6,4);
\draw[dashed,thick] (6,4) -- (4,6);
\draw[dashed,thick] (4,4) -- (6,6);
\draw[dashed,thick] (6,6) -- (4,4);
\node[align=left, above] at (2.6,4.4)%
{\Large $\partial_{p^2}$};
\node[align=left, above] at (3.5,4.0)%
{\Huge (};
\node[align=right, above] at (7.3,4.5)%
{+ SM};
\node[align=right, above] at (8.5,4.0)%
{\Huge )};
\node[align=right, above] at (25.3,4.5)%
{at $p^2=0$ \; .};
\end{tikzpicture} \\[2mm]
\end{center}
The SM-contributions contain the same diagrams on both sides, with
appropriately adjusted parameters through the matching conditions, so
their contributions cancel. Only diagrams with at least one heavy
propagator actually contribute to the matching, so
Eq.\eqref{eq:singlet_match1} becomes
\begin{align}
    \partial_{p^2} \left( 8p^2 f^{(0)}_{\phi,2} \right) \Bigg\vert_{p^2=0}
  = \partial_{p^2} \frac{2 A^2}{4p^2-M^2} \Bigg\vert_{p^2=0} 
  \qquad \Rightarrow \qquad
   f^{(0)}_{\phi,2} = \frac{A^2}{M^4} \; .
  \label{eq:singlet_match2}
\end{align}
At tree level, the scale dependence only appears implicitly for $A$
and for $f^{(0)}_{\phi,2}$.
\begin{figure}
    \centering
\begin{tikzpicture}
\draw[thick, dashed] (3.9,1.0) -- (4.3,0.5);
\draw[thick, dashed] (3.9,0.0) -- (4.3,0.5);
\draw[thick, dashed] (5.7,1.0) -- (5.3,0.5);
\draw[thick, dashed] (5.7,0.0) -- (5.3,0.5);
\draw[thick] (4.8,0.5) circle [radius=0.5];;
\node[align=right, above] at (4.8,1)%
{$S$};
\node[align=right, below] at (4.8,0)%
{$S$};
\draw[thick, dashed] (7.0,1.0) -- (7.6,0.5);
\draw[thick, dashed] (7.0,0.0) -- (7.6,0.5);
\draw[thick, dashed] (8.4,0.5) -- (9.0,1.0);
\draw[thick, dashed] (8.4,0.5) -- (9.0,0.0);
\draw[thick] (7.6,0.5) -- (8.4,0.5);
\node[align=right, below] at (8.0,0.5)%
{$S$};
\draw[thick] (8.0,0.9) circle [radius=0.4];;
\node[align=right, above] at (8.0,1.3)%
{$S$};
\end{tikzpicture} 
\caption{Feynman diagrams contributing to $f^{(1)}_{\phi,2}$. Left: Diagram yielding a $\kappa^2$-contribution. Right: Diagram yielding a $A^2\lambda_S/M^2$-contribution. The dashed line corresponds to the Higgs field, whereas the solid line corresponds to the singlet.}
\label{fig:singlet_matching_diagrams}
\end{figure}
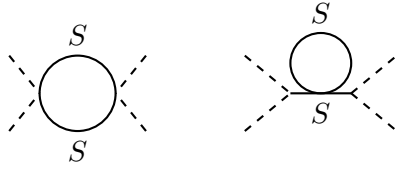

Next, we compute the $\kappa^2$-contribution to $f^{(1)}_{\phi,2}$ at
one loop. This contribution is induced by the diagram on the left in Figure \ref{fig:singlet_matching_diagrams}, where the external particles are as specified in Eq.\eqref{eq:singlet_match1}.
We again set all external scales to $p^2$ and find for the diagram
\begin{align}
  \kappa^2 \mu_R^{4-d}  \int \frac{d^d q}{(2\pi)^d} \frac{1}{((2p+q)^2-M^2)(q^2-M^2)}
  = \kappa^2 \frac{i}{16 \pi^2}  B_0(4 p^2,M,M) \notag \\
\text{with} \quad   
B_0(4p^2,M,M) = \frac{1}{\bar{\epsilon}} - \log \frac{M^2}{\mu_R^2} + \frac{2p^2}{3M^2} + \mathcal{O} \left( \frac{p^4}{M^4} \right) \; .
\end{align}
In the full expression the renormalization
scale appears, but taking the derivative in the matching condition for
this contribution to $f_{\phi,2}$ removes it,
\begin{align}
    \partial_{p^2}B_0(4 p^2,M,M) \Bigg|_{p^2 = 0} = \frac{2}{3 M^2} 
  \qquad \Rightarrow \qquad
  f^{(1)}_{\phi,2} \supset \frac{1}{16 \pi^2}\frac{\kappa^2}{12 M^2} \; .
\end{align}
Just as at tree level, the matching scale does not appear explicitly.

Finally, we compute the $A^2\lambda_S/M^2$-contribution to
$f^{(1)}_{\phi,2}$ to illustrate the appearance of matching scale
logarithms. This contribution arises from the diagram on the right in Figure \ref{fig:singlet_matching_diagrams}.
The diagram is not 1PI, but is 1LPI and therefore has to be included in the
matching.
With all external scales again set to $p^2$ this diagram gives
\begin{align}
  - \frac{\lambda_S A^2}{(4p^2-M^2)^2} \mu_R^{4-d} \int \frac{\rd ^d q}{(2\pi)^d}\frac{1}{q^2-M^2}
  = -\frac{\lambda_S A^2}{16 \pi^2} \frac{M^2}{(4p^2-M^2)^2}
  \left(\frac{1}{\bar{\epsilon}} + 1 -\log \frac{M^2}{\mu_R^2}\right) \; .
\end{align}
Taking the derivative with respect to $p^2$ and evaluating it at
$p^2=0$ we find the one-loop matching condition
\begin{align}
    f^{(1)}_{\phi,2} \supset - \frac{1}{16\pi^2} \frac{\lambda_S A^2}{M^4} \left(-1+\log \frac{M^2}{Q^2}\right) \; ,
\end{align}
where the Wilson coefficient explicitly depends on the matching
scale. This scale dependence is expected since the corresponding correlation function 
is divergent.
As mentioned before, in models with one new mass scale, we can
of course avoid these logarithms by identifying $\qm = M$.

\subsubsection*{Vector triplet}

\begin{figure}[t]
  \centering
  \includegraphics[scale=0.6]{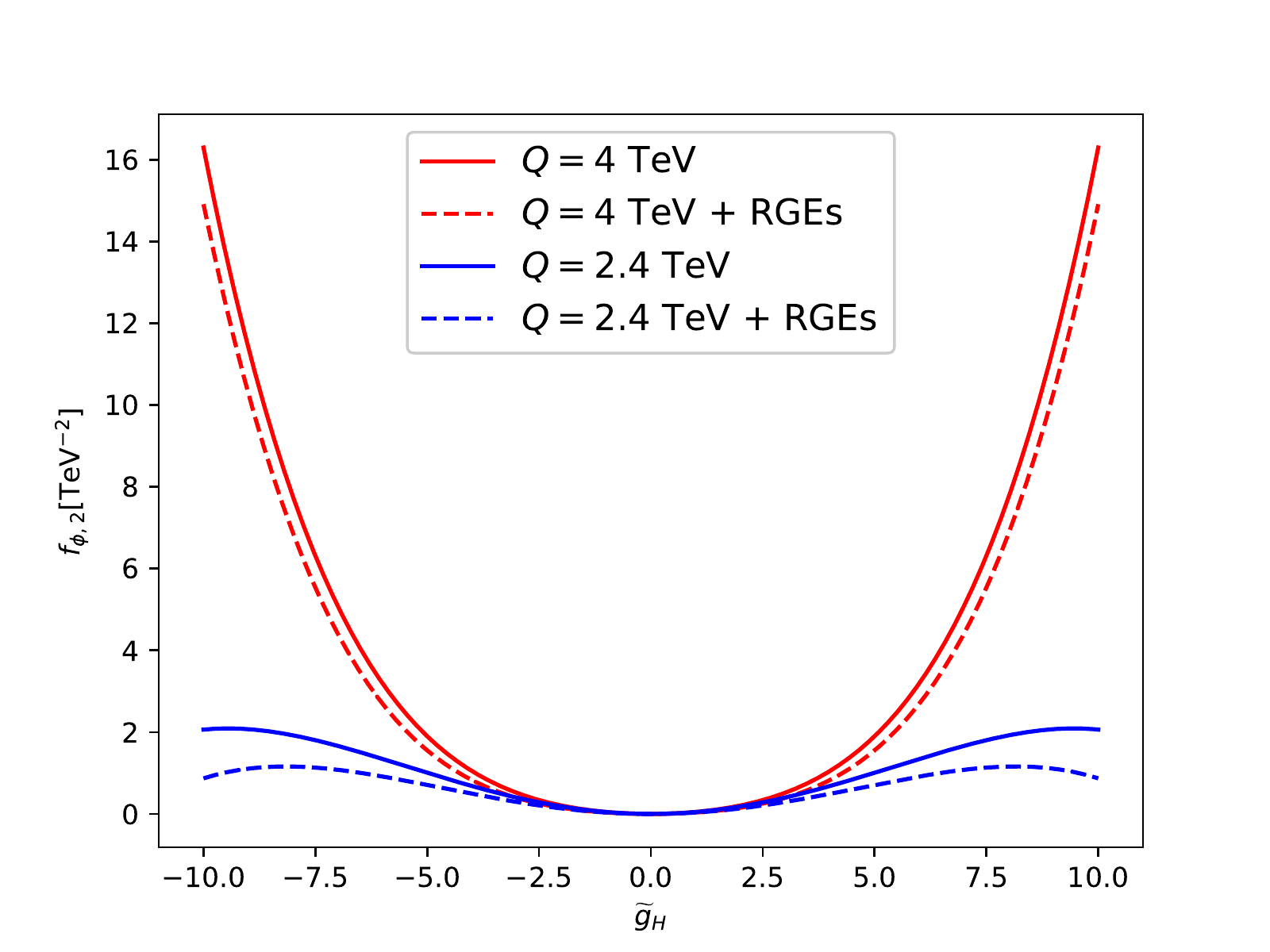}
  \caption{Wilson coefficient $f_{\phi2}$ as a function of $\gt_H$ at
    different values of the matching scale $\qm$ for fixed $m_V=4$~TeV
    and all other UV couplings set to zero. The dashed lines include approximate RG running.}
  \label{fig:qm_dep}
\end{figure}

Moving to the triplet model defined by the Lagrangian of
Eq.\eqref{eq:lagrangian1}, we will not attempt to show analytic
results and instead illustrate the matching scale dependence for one
finite coupling $\gt_H$ and a mass term $\tilde{m}_V$ numerically.  In
this simplified setup, $m_V=\tmv$. Among the various Wilson
coefficients, it is instructive to consider $f_{\phi,2}$, as its
dependence on the matching scale exhibits interesting
features. Including both tree and loop contributions, the matching
expression has the form
\begin{align}
  \!\!\frac{f_{\phi, 2}}{\Lambda^2} \simeq
  \frac{1}{m_V^2}\left[
  g_2^4 \left( c_0+c_1 \log \frac{m_V}{\qm} \right)
  + \gt_H^2 \left( c_2 + c_3 \log \frac{m_V}{\qm} \right)
  + \gt_H^4 \left( c_4 + c_5 \log \frac{m_V}{\qm} \right)\right]  , 
\label{eq:fphi2_1}
\end{align} 
where $c_0 = c_1/2$ emerges from 1-loop diagrams inducing the operator
structure $(D_\mu W^{\mu\nu})^2$, which maps to $\ope_{\phi,2}$ via
the equations of motion. Of the additional constants, the
$\gt_H^2$-coefficient is dominated by the tree-level contribution to
$c_2$, while the $\gt_H^4$-coefficient is completely determined by the
one-loop matching. Numerically, we find
\begin{alignat}{7}
  c_0 &= \frac{c_1}{2} = \frac{3}{128\pi^2} = 0.0024\,, \notag \\
  c_2 &= 0.75\,, \qqquad 
  c_3  = 0.0069\,,  \qqquad
  c_4  = 0.019\,, \qqquad 
  c_5  = - 0.045  \; .
\end{alignat}
In Fig.~\ref{fig:qm_dep} we show the numerical dependence of $f_{\phi,
  2}$ on $\gt_H$ for different choices of $\qm$. For $\qm = m_V =
4$~TeV the Wilson coefficient has a simple power dependence on $\gt_H$
driven by $c_4$.  For $\qm\approx 0.66 \, m_V=\unit[2.6]{TeV}$ the
$\gt_H^4$-term cancels exactly. For $\qm$ below this threshold, the
coefficient in front of $\gt_H^4$ becomes negative, which flips the
sign of $f_{\phi,2}$ at $\gt_H\gg 1$ and allows a solution of
$f_{\phi,2}=0$ for $\gt_H\neq 0$. For $\qm\lesssim\unit[2.4]{TeV}$ the
solution is within the range $|\gt_H|<4\pi$ and leads to visible
effects in our global analysis.

\begin{figure}[t]
  \includegraphics[height=0.35\textwidth,trim=0 0 4cm 0,clip]{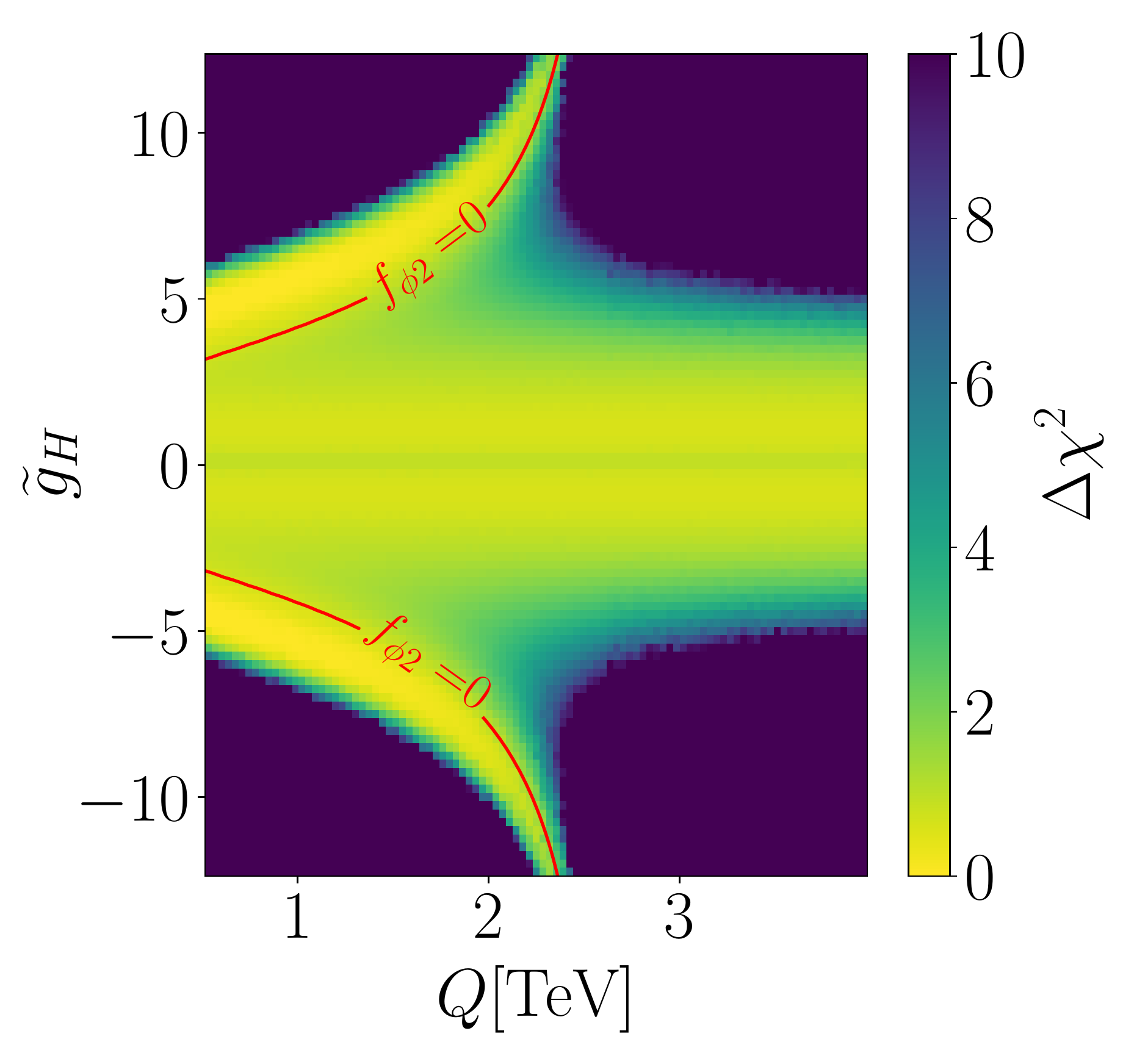}
  \includegraphics[height=0.35\textwidth,trim=0 0 4cm 0,clip]{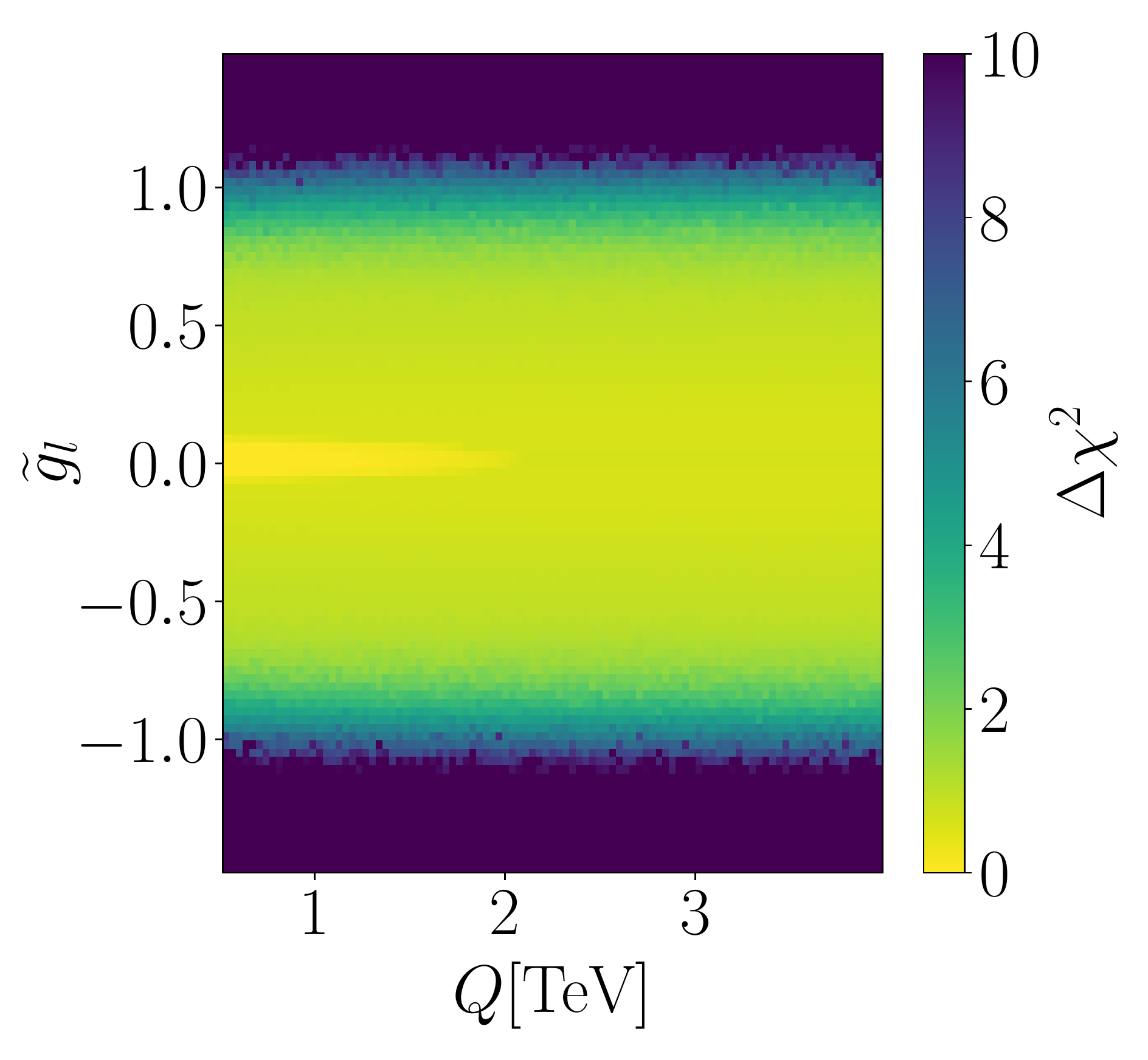}
  \includegraphics[height=0.35\textwidth]{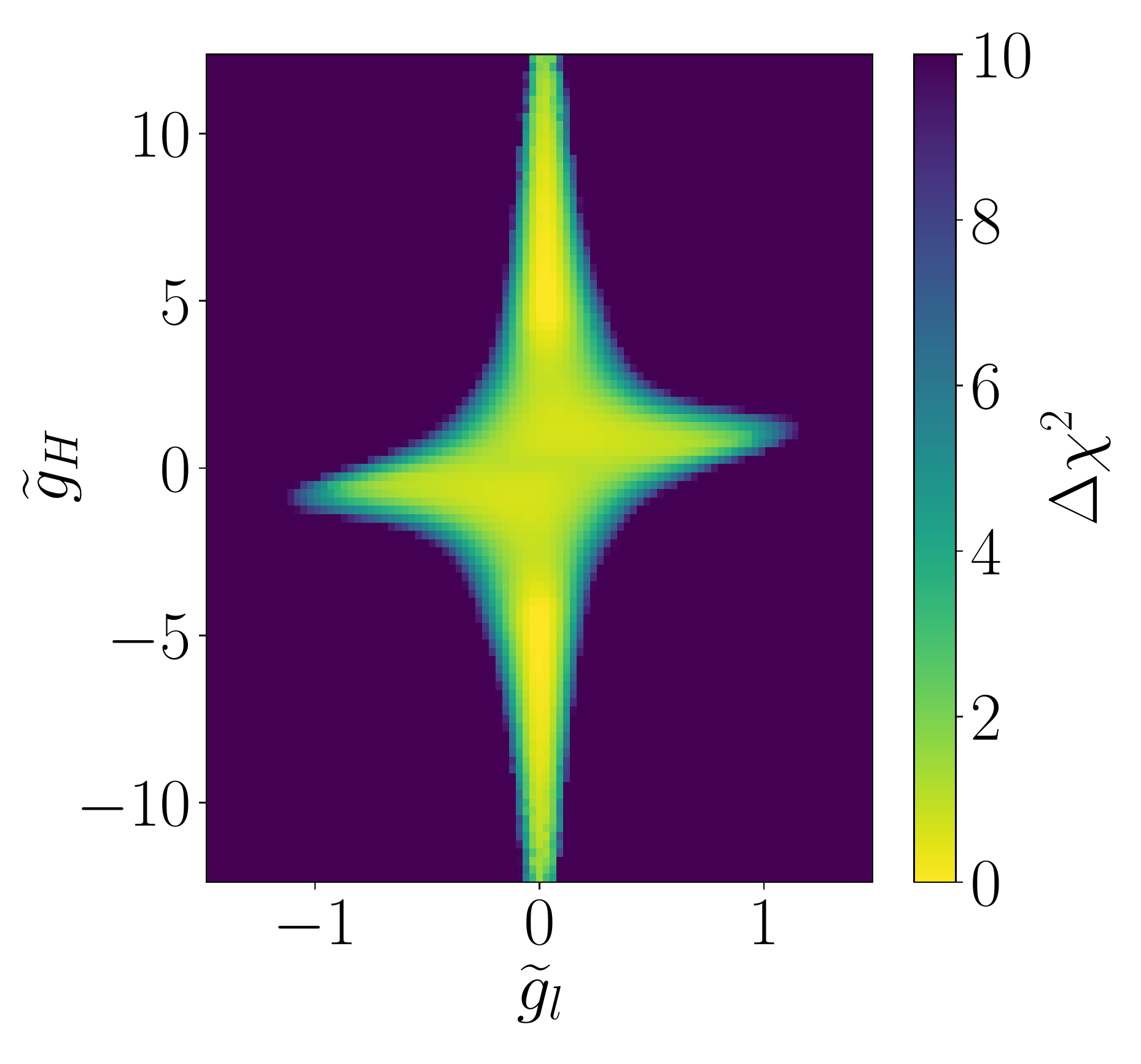}
  \caption{The impact of the variation of the matching scale $Q$ at a
    mass of $m_V=\unit[4]{TeV}$ for a reduced model with free $\gt_M,
    \gt_H, \gt_l$, expressed in the unmixed Lagrangian
    Eq.\eqref{eq:lagrangian2} with actual measurements.}
  \label{fig:matching}
\end{figure}

Figure~\ref{fig:matching} shows the results of the same global
analysis as in Sec.~\ref{sec:toy_decoup}, where now we fix $m_V =
\unit[4]{TeV}$. The free parameters are
\begin{align}
  \{\gt_H, \gt_l, \gt_M, \qm\} \; ,
\end{align}
where the matching scale is varied in the range
$\qm=\unit[500]{GeV}~...~\unit[4]{TeV}$.  The left panel shows a
central allowed region for $|\gt_H|\lesssim 4$ that is independent of
$\qm$.  In addition, a beautiful \textsl{fleur-de-lis shape} arises in
$\gt_H$ vs $\qm$ for $\qm < 2.4$~TeV. It roughly follows the curves
along which $f_{\phi,2}=0$ marked in red.  The Wilson coefficients
$f_t, f_b, f_\tau$ have a similar behavior and vanish approximately in
the same region, because they are induced by the same or similar loop
contributions. As these are the operators that dominate the constraint
on $\gt_H$, the fleur-de-lis feature persists in the full global fit,
see Sec.~\ref{sec:global}.  When we profile over $\qm$ as a nuisance
parameter, this correlation broadens the 1-dimensional and
2-dimensional profile likelihood in $\gt_H$ by roughly a factor 2.  As
shown in the second and third panels of Fig.~\ref{fig:matching}, the
broadening affects significantly only the constraints in the $\gt_H$
direction, while those on $\gt_l$ are essentially unchanged compared
to when $\qm=m_V$. Although not shown, this is also verified for
$\gt_M$.

We emphasize that the tree-loop cancellations that drive this effect
are only very slightly affected by the renormalization group evolution
of $f_{\phi, 2}$, as illustrated by the dashed lines including
approximate RGE contributions in Fig.~\ref{fig:qm_dep}. They really
correspond to a choice of the unphysical matching scale, which cannot
be compensated by the well-defined change of renormalization scale of
the low-energy SMEFT description. Adding higher orders in the loop
expansion to the matching decreases the sensitivity to the matching
scale. Similar effects, but with a much smaller numerical impact have
been observed in Ref.~\cite{Dawson:2021jcl}.

\section{SMEFT global analysis}
\label{sec:global}

In this section we discuss the results of the SMEFT global analysis,
mapped to the parameter space of the heavy vector triplet model
defined in Section~\ref{sec:basics_model} using 1-loop matching
relations.  We derive constraints on the UV-parameters $\{\gt_H,
\gt_q, \gt_l, \gt_M, \gt_{VH} \}$ defined by the Lagrangian in
Eq.\eqref{eq:lagrangian1} for fixed values of the heavy vector triplet
mass. We consider two benchmark values: $m_V = \unit[4]{TeV}$, to be
compared with direct resonance searches by the ATLAS Collaboration,
and $m_V = \unit[8]{TeV}$ for a consistent SMEFT analysis safely below
any on-shell pole.

\subsection{Resonance searches at high invariant masses}
\label{sec:global_meas}

As mentioned in Sec.~\ref{sec:sfitter}, in addition to more standard
Higgs measurements, the global analysis includes constraints from
searches for exotic particles in the $WH$ and $WW$ channels by the
ATLAS Collaboration. In particular, two of these
analyses~\cite{Aad:2020tps,Aad:2020ddw} have been newly implemented in
\sfitter.

\begin{figure}[t]\flushleft
\hspace*{3cm}
\includegraphics[height=5.3cm,trim=0 0 4cm 0, clip]{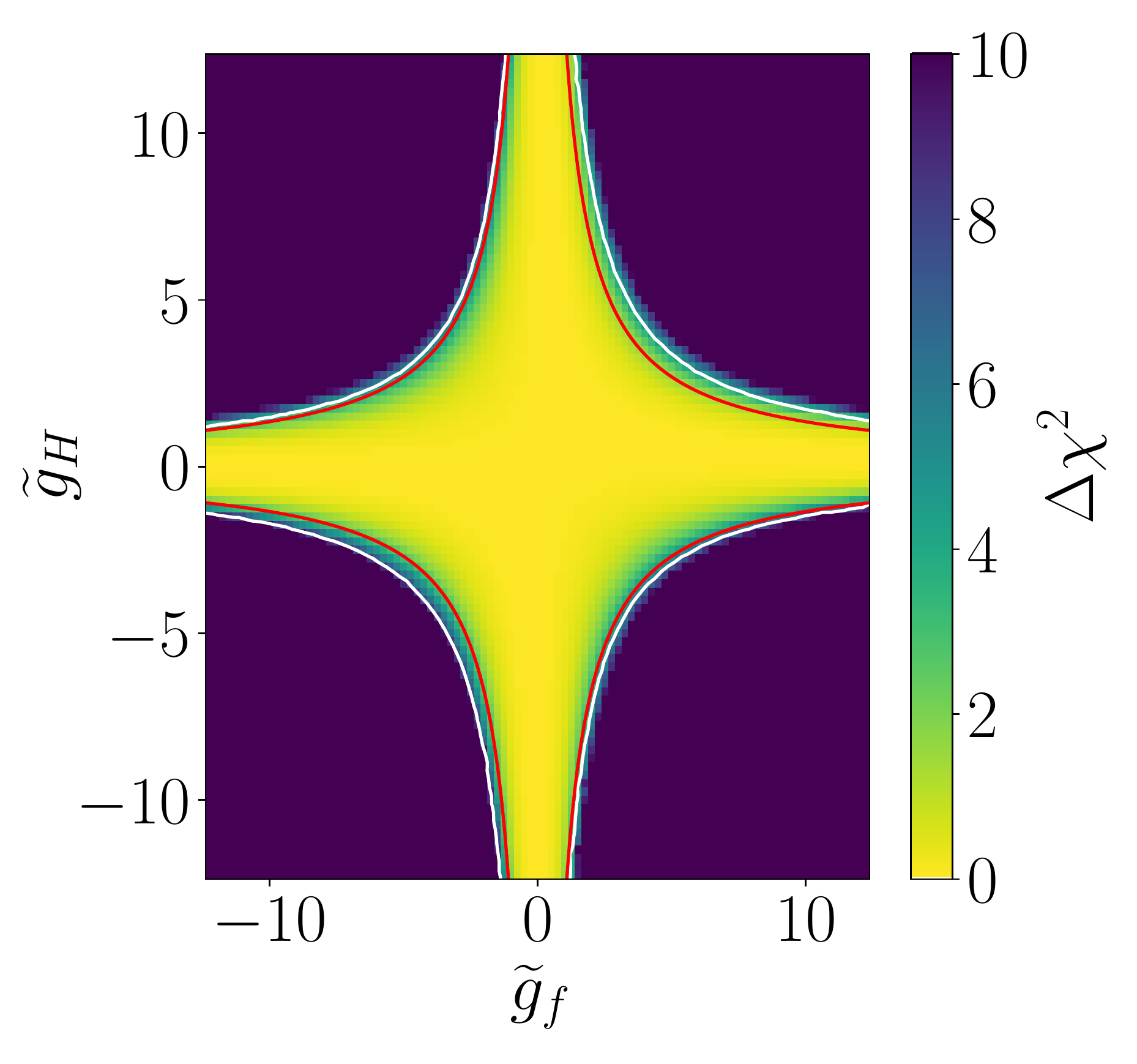}
\includegraphics[height=5.3cm, trim=0 0 4cm 0, clip]{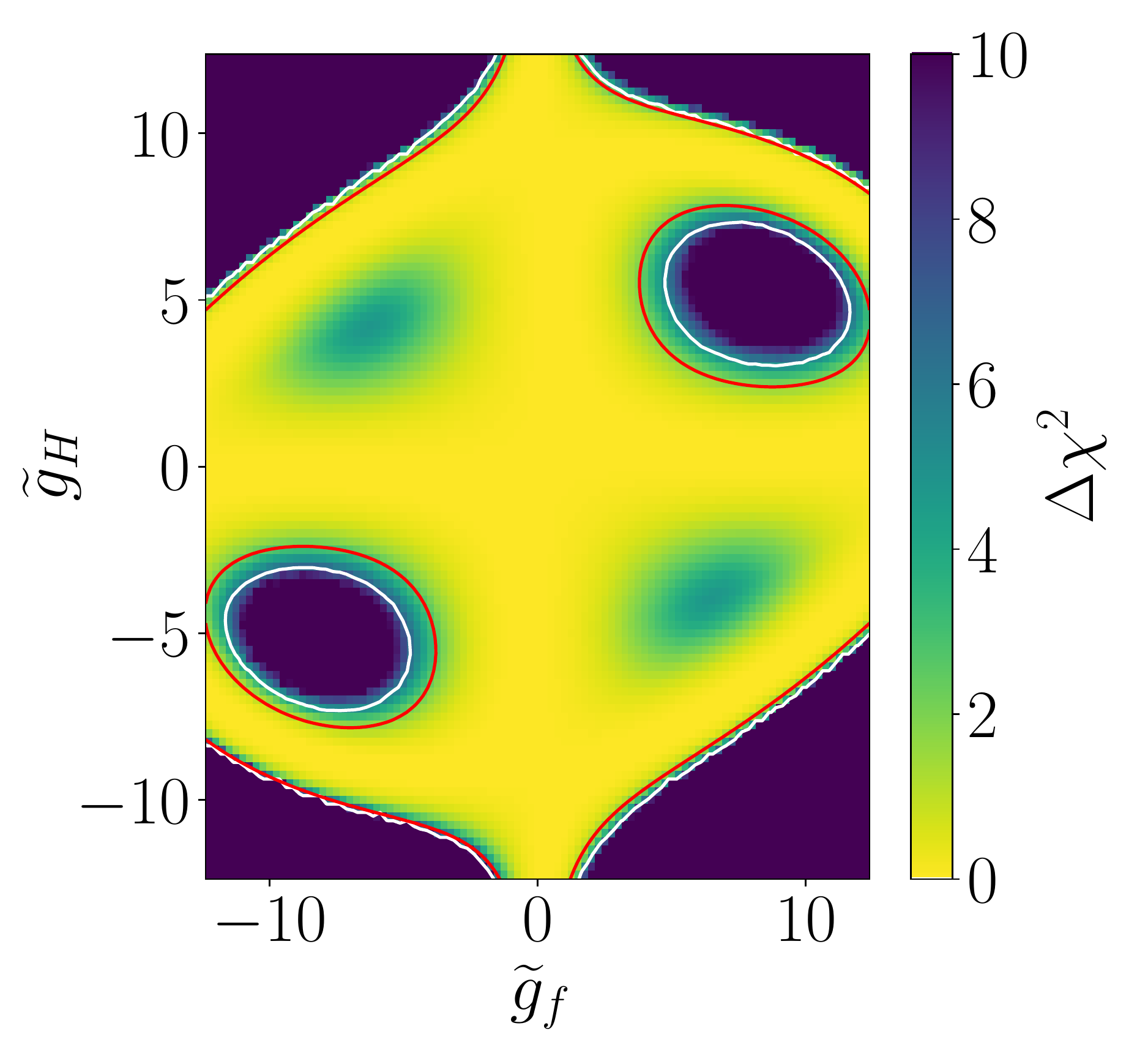}\\
\hspace*{3cm}
\includegraphics[height=5.3cm,trim= 0 0 4cm 0, clip]{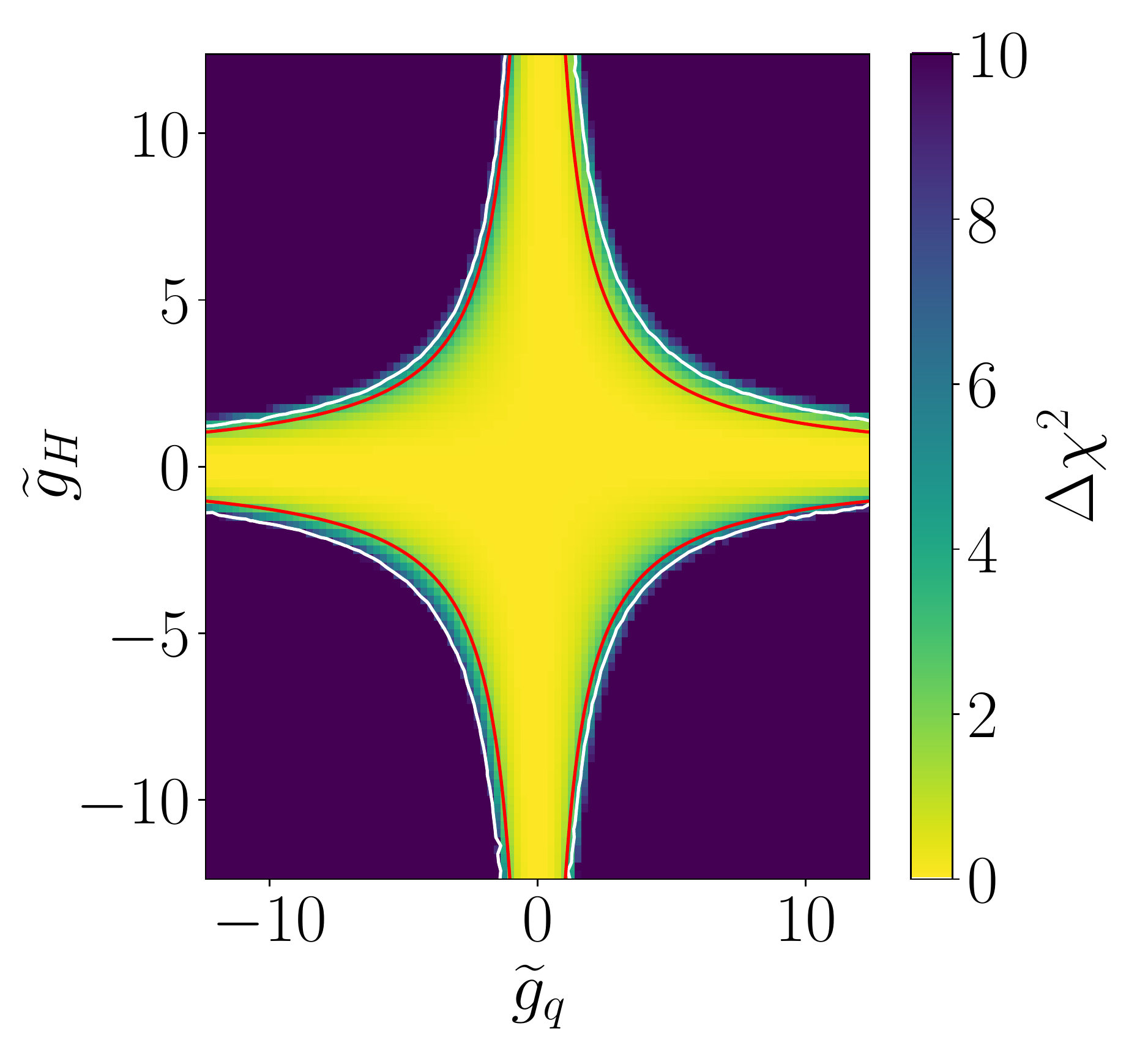}
\includegraphics[height=5.3cm]{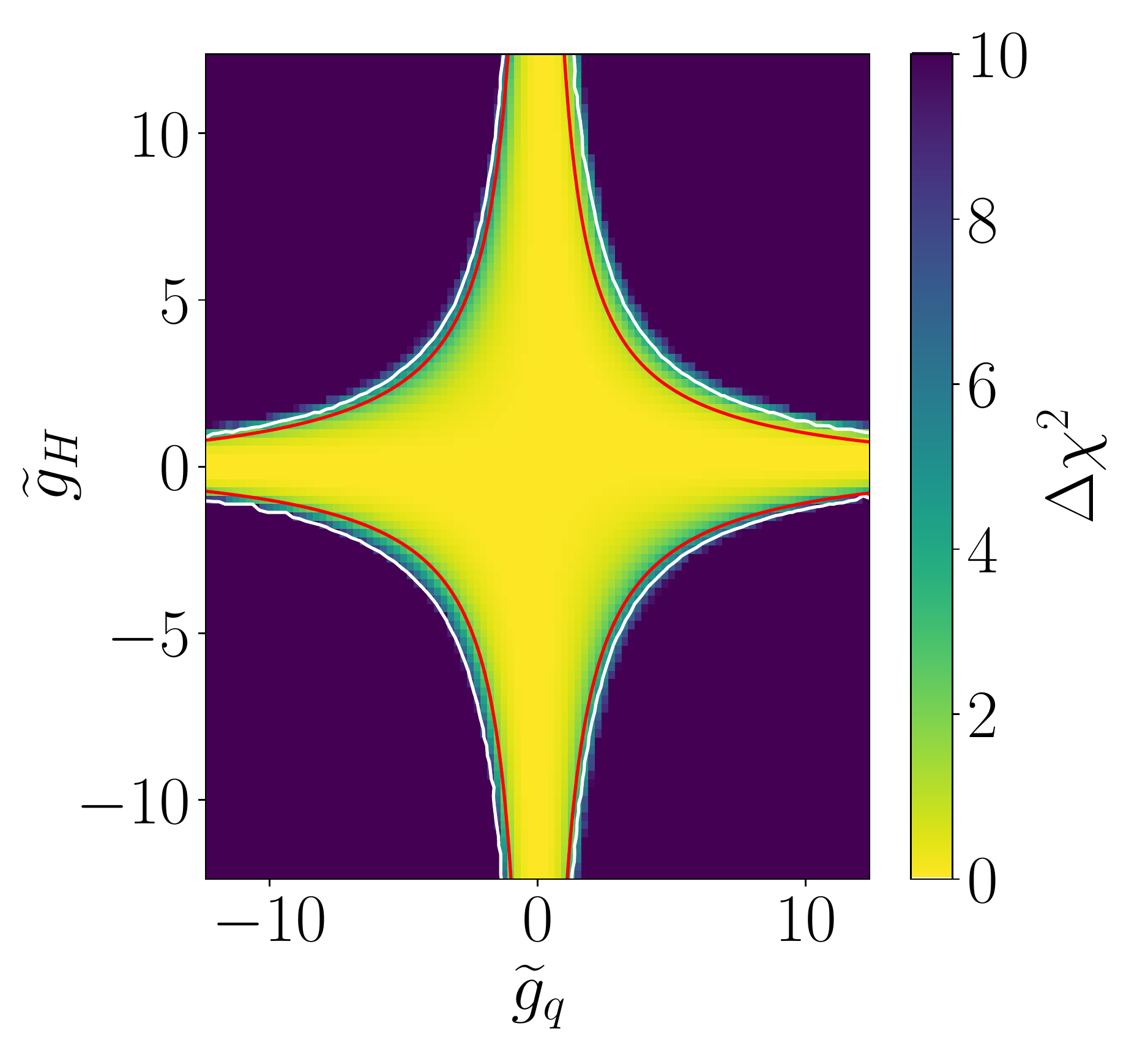}
\caption{2D fits of the $WH$ resonance search of
  Ref.~\cite{Aad:2020tps} only. We fix $m_V=\unit[4]{TeV}$ and $\gt_M=
  \gt_{VH} =0$. Left: tree-level matching. Right: Loop-level
  matching. Top: with $\gt_{l} = \gt_{q} = \gt_{f}$. Bottom: with
  $\gt_{l} = 0$. In the top (bottom) row, red contours indicate
  $f_{W}=\pm4$ ($\fpWt=\pm0.8$) with $\Lambda=\unit[1]{TeV}$ and white
  contours indicate $\Delta\chi^2=5.991$.}
\label{fig:VH_fit}
\end{figure}

\subsubsection*{WH search}

We consider the $m_{WH}$ invariant mass distribution measured
in Ref.~\cite{Aad:2020tps} in the $WH$ 1-tag category, and we compare
it to a $WH$ signal including dimension-6 corrections.  This kinematic
distribution extends up to $m_{WH}=\unit[5]{TeV}$
and the strongest constraints on BSM effects stem from the region around
$m_{WH}=\unit[2-2.5]{TeV}$, where the measurement exhibits large
under-fluctuations. A detailed description of the implementation of this analysis will be provided in a future work~\cite{nextSFitter}.

For equal values of the Wilson coefficients, the largest correction to
the $m_{WH}$ spectrum is induced by the operator $\O_{\phi
  Q}^{(3)}$~\cite{Zhang:2016zsp,Corbett:2017qgl,Baglio:2017bfe,Baglio:2018bkm,Alves:2018nof,Dawson:2018jlg,Brehmer:2019gmn},
that contributes via corrections to the $qqV$ vertex and via a 4-point
$qqVH$ interaction. The latter exhibits an enhancement at large
partonic energies due to the missing $s$-channel propagator and is
therefore dominant in the high-invariant-mass regime. Further
significant corrections, albeit less momentum-enhanced, are induced by
$\O_W$. All other SMEFT operators in the HISZ basis do not contribute
significantly to $WH$ production in the high-energy regime.

Figure~\ref{fig:VH_fit} shows the results from a 2D-analysis of the
$m_{WH}$ distribution alone, fixing the matching scale $\qm=
m_V=\unit[4]{TeV}$ and considering only two $\gt$-couplings at a
time. The top row in Fig.~\ref{fig:VH_fit} shows $\gt_f\equiv
\gt_q=\gt_l$ vs $\gt_H$, which matches the benchmark considered in the
ATLAS analysis~\cite{Aad:2020tps}. In this limit, the matching
contribution to $\fpWt$ cancels exactly, both at tree and loop levels.
As a consequence, the constraints are driven by $f_{W}$, whose
matching expressions reduce to
\begin{align}
 \frac{f_W}{\Lambda^2} &= 4.76\,\frac{\gt_H \gt_l}{m_V^2}
 &&\text{(tree)}
 \notag \\
 \frac{f_W}{\Lambda^2}&\simeq \gt_l\gt_H
 \frac{
  4.71\,
 + 0.019\, \gt_l \gt_H
 - 0.023\, \gt_l^2 
 - 0.057\, \gt_H^2
 }{m_V^2}
 && \text{(tree+loop).}
 \label{eq:fw}
\end{align}
The red contours in the plots indicate
$f_W/\Lambda^2=\unit[\pm4]{TeV^{-2}}$, which is representative of the
$2\sigma$ boundaries$f_W/\Lambda^2\in [-3.6, 4.4]~\unit{TeV^{-2}}$
found in a 1D fit to the SMEFT parameters.  In a slight abuse of
language, here and in the following the $\Delta\chi^2\leq1\, (2.3)$
and $\Delta\chi^2\leq 3.841\, (5.991)$ regions in 1D (2D) fits are
sometimes referred to as $1\sigma$ and $2\sigma$ intervals,
respectively.  The fact that these lines coincide to a very good
approximation with the $2\sigma$ contours (indicated in white) in
Fig.~\ref{fig:VH_fit} shows that the constraint on $f_W$ is indeed the
leading one.  The bottom row shows $\gt_q$ vs $\gt_H$ for
$\gt_l=0$. In this case the cancellation in $\fpWt$ is spoiled and the
constraints are dominated by this Wilson coefficient. Numerically, the
matching expression is
\begin{align}
\frac{\fpWt}{\Lambda^2} &= \frac{\gt_H(\gt_l-\gt_q)}{m_V^2}
 &&\text{(tree)}
 \notag \\
 \frac{\fpWt}{\Lambda^2} &\simeq  0.99\,  \frac{\gt_H(\gt_l-\gt_q)}{m_V^2}
 &&\text{(tree+loop),}
 \label{eq:fphiq3}
\end{align}
and the bottom panels in Fig~\ref{fig:VH_fit} show contours for
$\fpWt/\Lambda^2=\unit[\pm 0.8]{TeV^{-2}}$, which is representative of
the $2\sigma$ interval $\fpWt/\Lambda^2\in [-0.90,
  0.76]~\unit{TeV^{-2}}$ obtained in a 1D fit.

Finally, comparing the left and right panels in Fig.~\ref{fig:VH_fit},
it is worth noting that the impact of loop contributions to the
matching is negligible in the case $\gt_l=0$, but significant for
$\gt_l=\gt_q$. This is a direct consequence of the form of the
matching expression in the particular model considered. Loop terms
only induce a very minor overall rescaling in the expression of
$\fpWt$, Eq.\eqref{eq:fphiq3}, but they introduce a series of new
terms in the expression of $f_W$, Eq.\eqref{eq:fw}. Although
numerically subdominant, the latter have a strong impact on the
likelihood structure.

\subsubsection*{WW search}

We consider the $m_{WW}$ distribution measured in
Ref.~\cite{Aad:2020ddw} in the $WW$ 1-lepton category and ggF/DY
merged, high-purity signal region, that targets neutral resonances
decaying to $W^\pm W^\mp$ pairs and covers invariant masses up to
$m_{WW}=\unit[4]{TeV}$. We compare the measured distribution to a
$W^\pm W^\mp$ production signal including SMEFT corrections. Again, we
postpone a detailed discussion of the implementation to a later paper.

The $W^\pm W^\mp$ production process exhibits a greater complexity in
the SMEFT compared to $W^\pm H$ in the high-energy limit. We find
that, fixing all Wilson coefficients to the same numerical value, the
largest corrections are induced by the operators $\O_{\phi u},
\O_{\phi d}, \O_{\phi Q}^{(1)}, \O_{\phi Q}^{(3)}$ at quadratic level,
that exhibit a large enhancement $\propto m_{WW}^2$. The origin of
this behavior can be identified as a $qq\phi\phi$ contact interaction
between two quarks and two Goldstone bosons induced by these
operators, that dominates at high energies due to the equivalence
theorem~\cite{Falkowski:2016cxu}.  Effects induced by $\O_W, \O_B$,
and $\O_{WWW}$ have a weaker momentum-enhancement and are roughly two
orders of magnitude smaller. Nevertheless, they were retained in the
fit, as they are relevant for the global analysis in terms of both
SMEFT and UV model parameters.  In the former case, this measurement
contributes significantly to improving the constraints on $f_W$, by
roughly a factor two~\cite{nextSFitter}.  In the latter, it is
important to stress that the matching expressions for a given UV model
generally do \emph{not} give homogeneous values for the Wilson
coefficients. Therefore a suppression of two orders of magnitude in
the SMEFT predictions can be easily compensated in the matching, and
the corresponding contributions to the signal may lead to significant
constraints on the UV model parameters.  In fact, for the $WW$
analysis implemented here we find that the constraints projected on
the $\gt_q - \gt_H$ and $\gt_f - \gt_H$ planes are entirely dominated
by the contributions of $f_W$ and $\fpWt$, the same two operators that
lead in the $WH$ case.

\begin{figure}[t]
\flushleft
\hspace*{3cm}
\includegraphics[height=5.3cm,trim=0 0 4cm 0,clip]{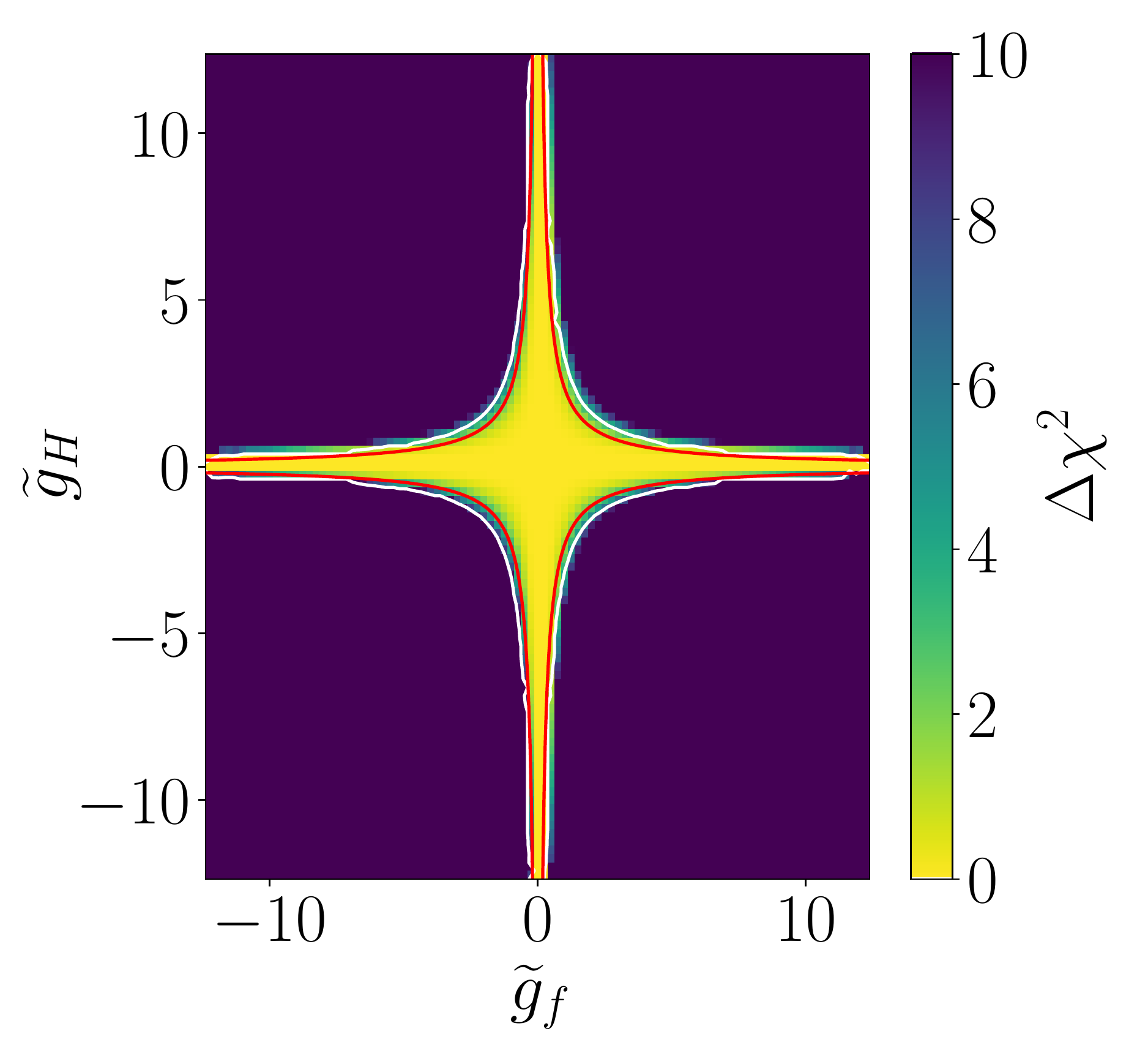}
\includegraphics[height=5.3cm,trim=0 0 4cm 0,clip]{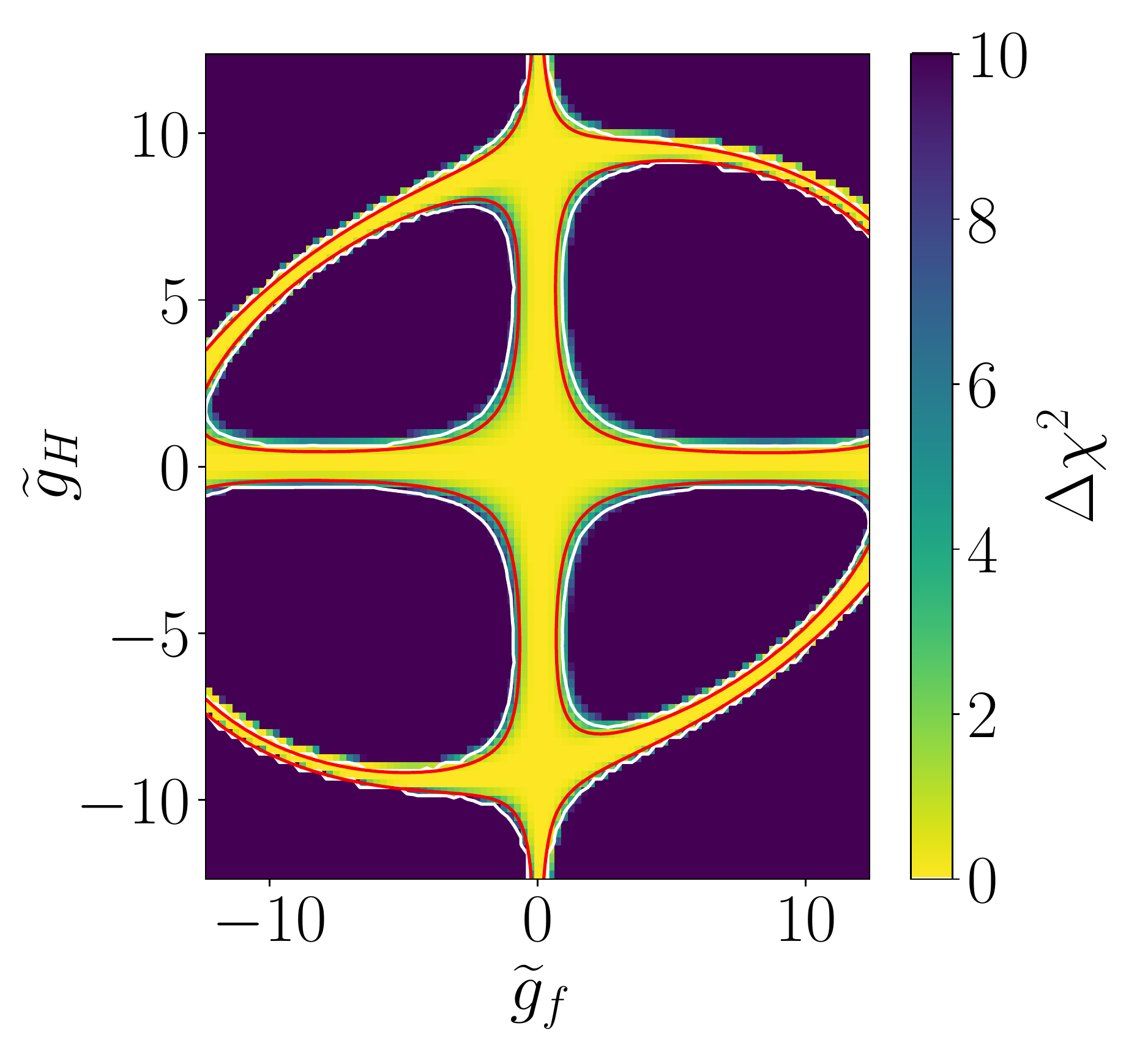}\\
\hspace*{3cm}
\includegraphics[height=5.3cm,trim=0 0 4cm 0,clip]{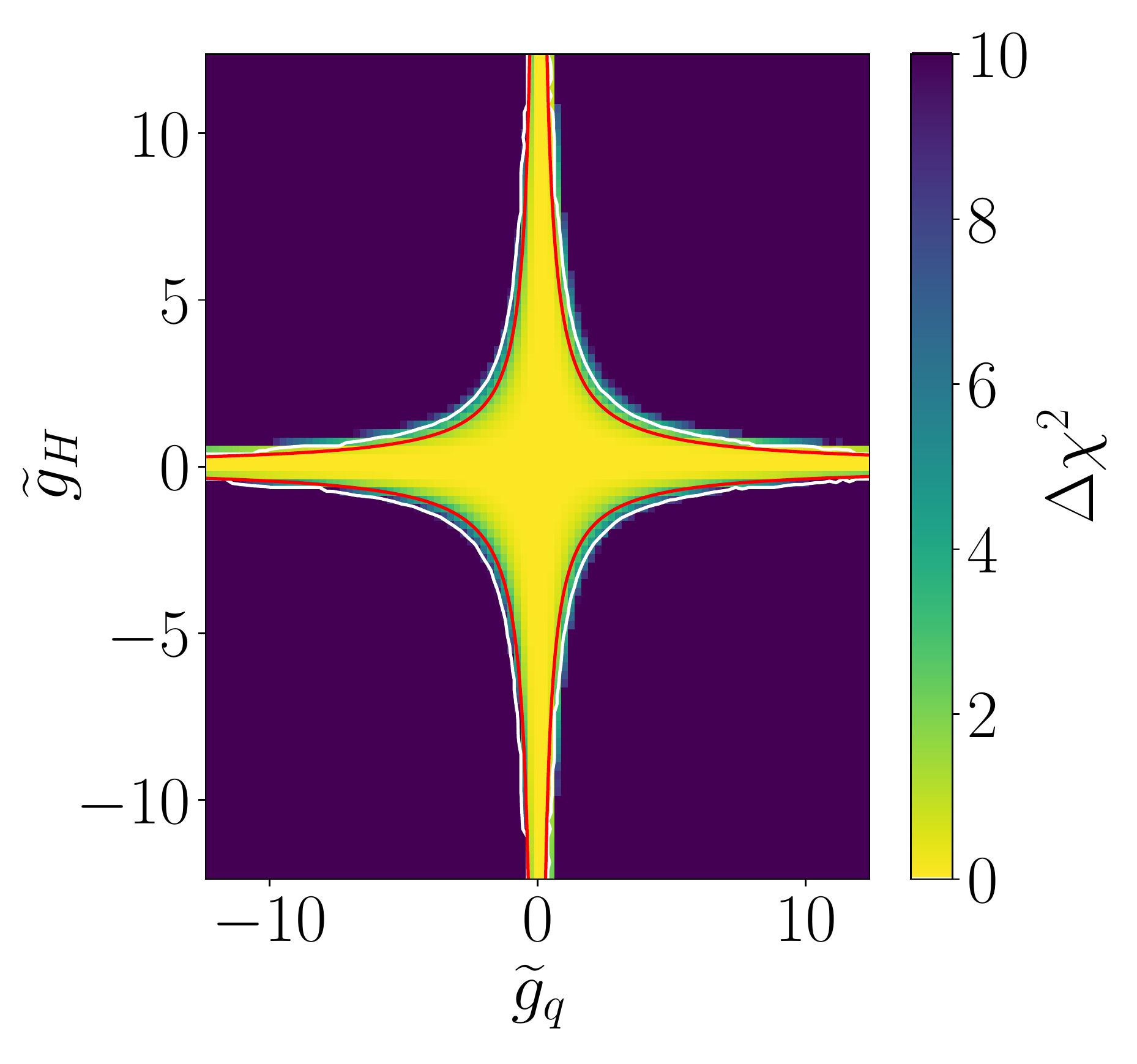}
\includegraphics[height=5.3cm]{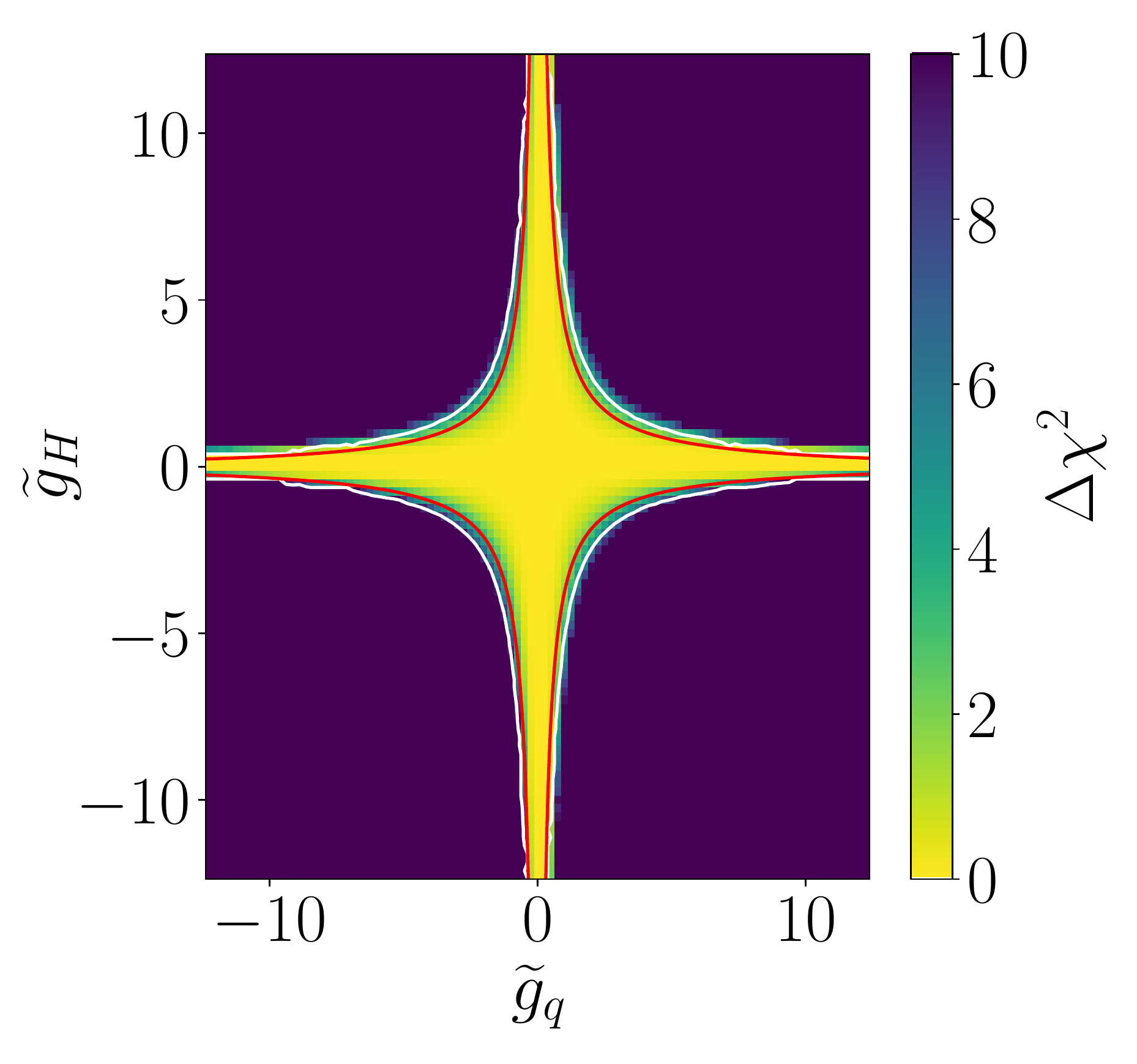}
\caption{2D fits of the $WW$ resonance search of
  Ref.~\cite{Aad:2020ddw} only. We fix $m_V=\unit[4]{TeV}$ and
  $\gt_M= \gt_{VH} =0$. Left: tree-level matching. Right: Loop-level matching. Top:
  with $\gt_{l} = \gt_{q} = \gt_{f}$. Bottom: with $\gt_{l} =
  0$. In the top (bottom) row, red contours
  indicate $f_{W}=\pm0.7$ ($\fpWt=0.2$ or $\fpWt=-0.3$) with
  $\Lambda=\unit[1]{TeV}$ and white contours indicate $\Delta\chi^2=5.991$.
}\label{fig:VV_fit}
\end{figure}

Figure~\ref{fig:VV_fit} shows the results from a 2D-analysis of the
$m_{WW}$ distribution alone, fixing $Q= m_V=\unit[4]{TeV}$ and
considering the same benchmarks as in Fig.~\ref{fig:VH_fit}.  The red
curves in Fig.~\ref{fig:VV_fit} are again given by Eq.\eqref{eq:fw}
and~\eqref{eq:fphiq3}, but for different values of $f_W$ and $\fpWt$,
namely $f_{W}/\Lambda^2=\unit[\pm0.7]{TeV^{-2}}$ and
$\fpWt/\Lambda^2=\unit[-0.27, +0.23]{TeV^{-2}}$. Again, these values
correspond to the $2\sigma$-boundaries identified in 1D fits.

This analysis yields stronger bounds compared to $WH$ because in this
particular case the constraints are dominated by the tail of the
distribution, in the region around $m_{WW}= 2.5-4$ TeV, which exhibits
under-fluctuations.  Again, the effect of introducing loop
contributions to the matching expressions is only visible in the
scenario dominated by $f_W$, for the same reasons as described above.

\subsection{Global analysis results}
\label{sec:model_uv}

Figure~\ref{fig:4tev_5param_fit} shows the results of our global
analysis, including the full data set described in
Sec.~\ref{sec:sfitter} as well as the resonance searches discussed in
Sec.~\ref{sec:global_meas}, for a fixed value of the heavy vector
triplet mass $m_V=\unit[4]{TeV}$. The analysis is performed varying
$\gt_M$ and $\gt_{VH}$ within the physical region $\gt_M = -1~...~1$,
$\gt_{VH}>0$ and all other coupling parameters in the perturbative range
$\gt =  -4\pi~...~4\pi$.

\subsubsection*{Fixed matching scale}

\begin{figure}[t]
 \includegraphics[width=.24\textwidth]{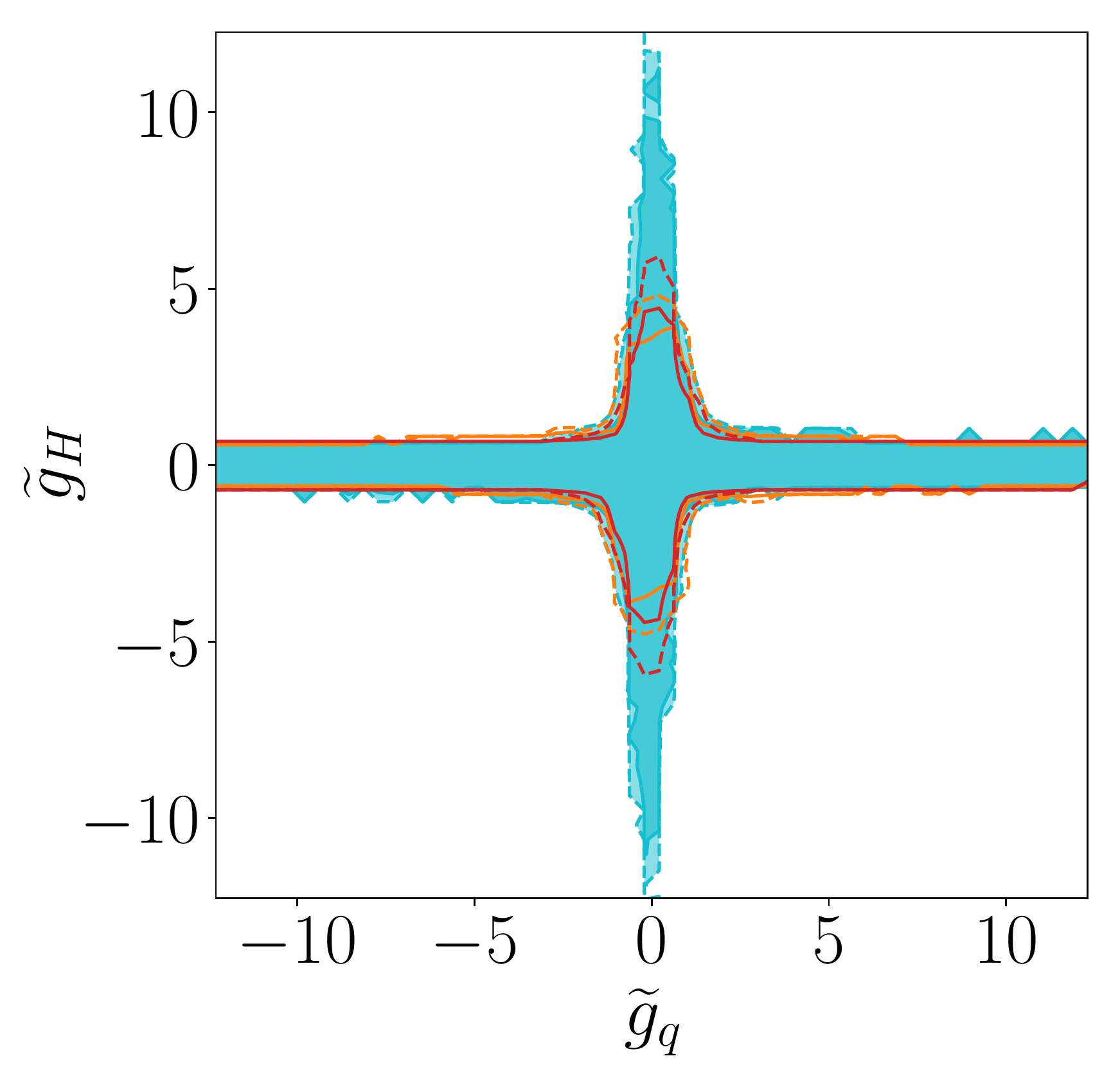}
 \includegraphics[width=.24\textwidth]{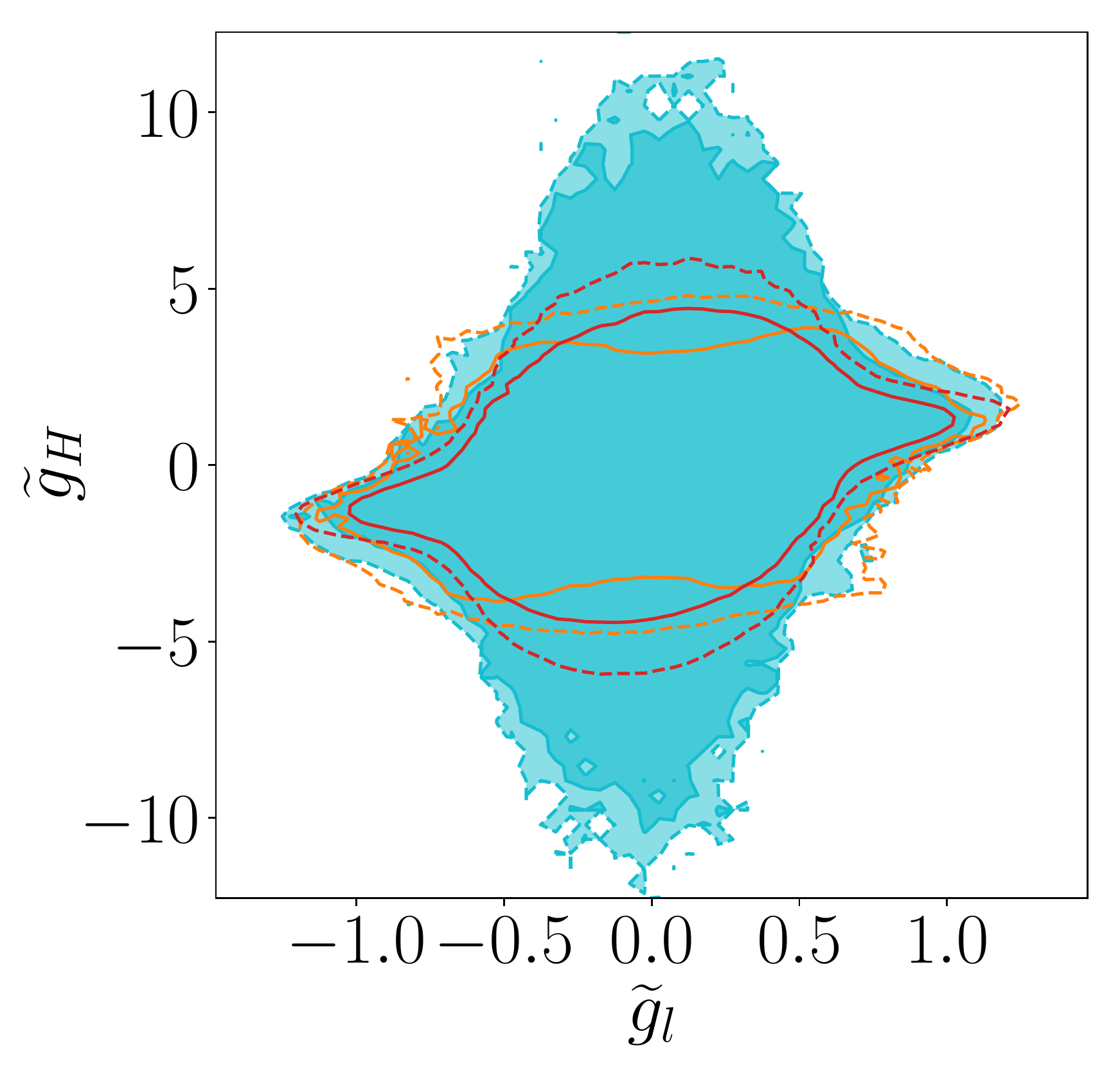}
 \includegraphics[width=.24\textwidth]{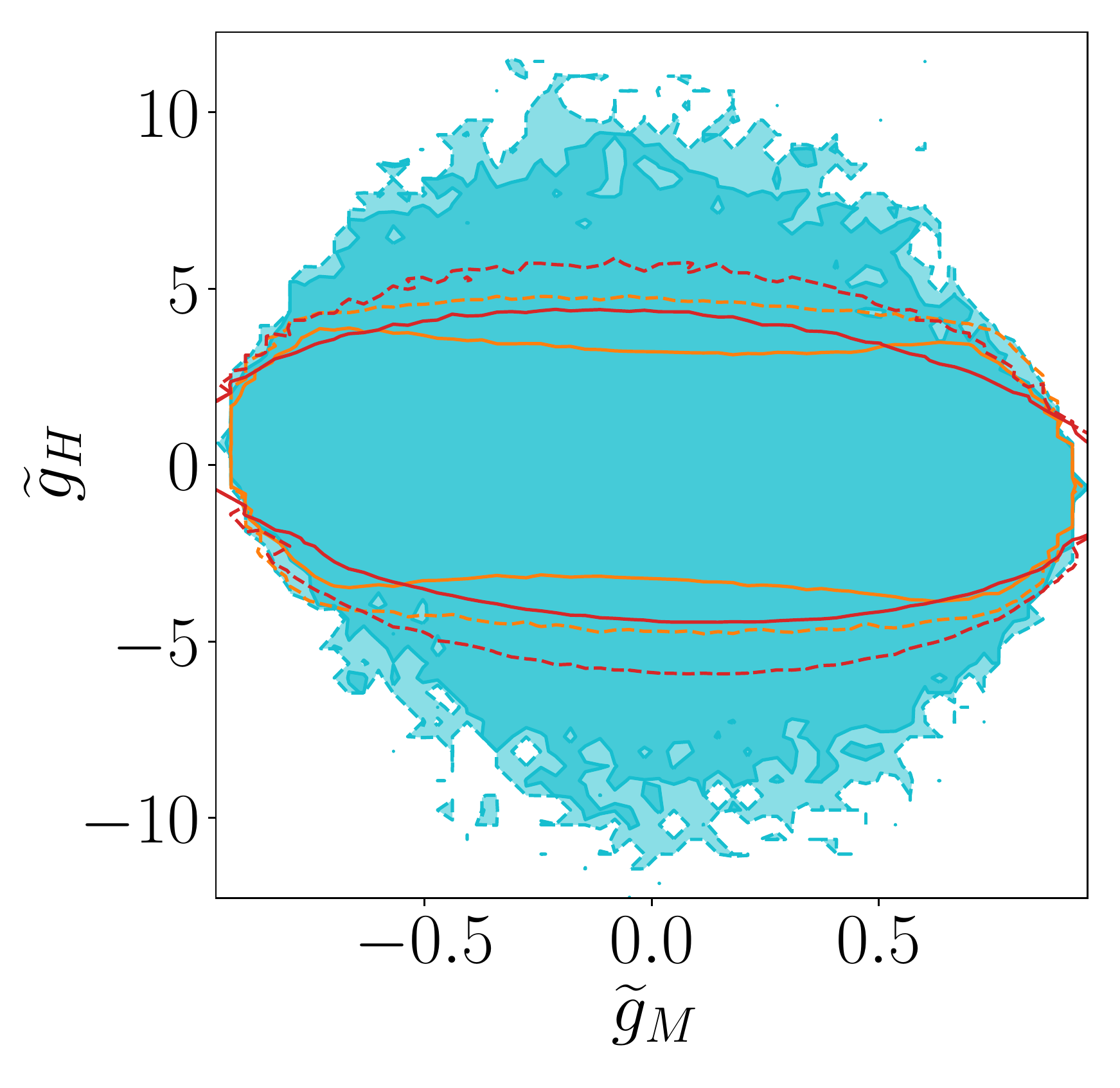}
 \includegraphics[width=.24\textwidth]{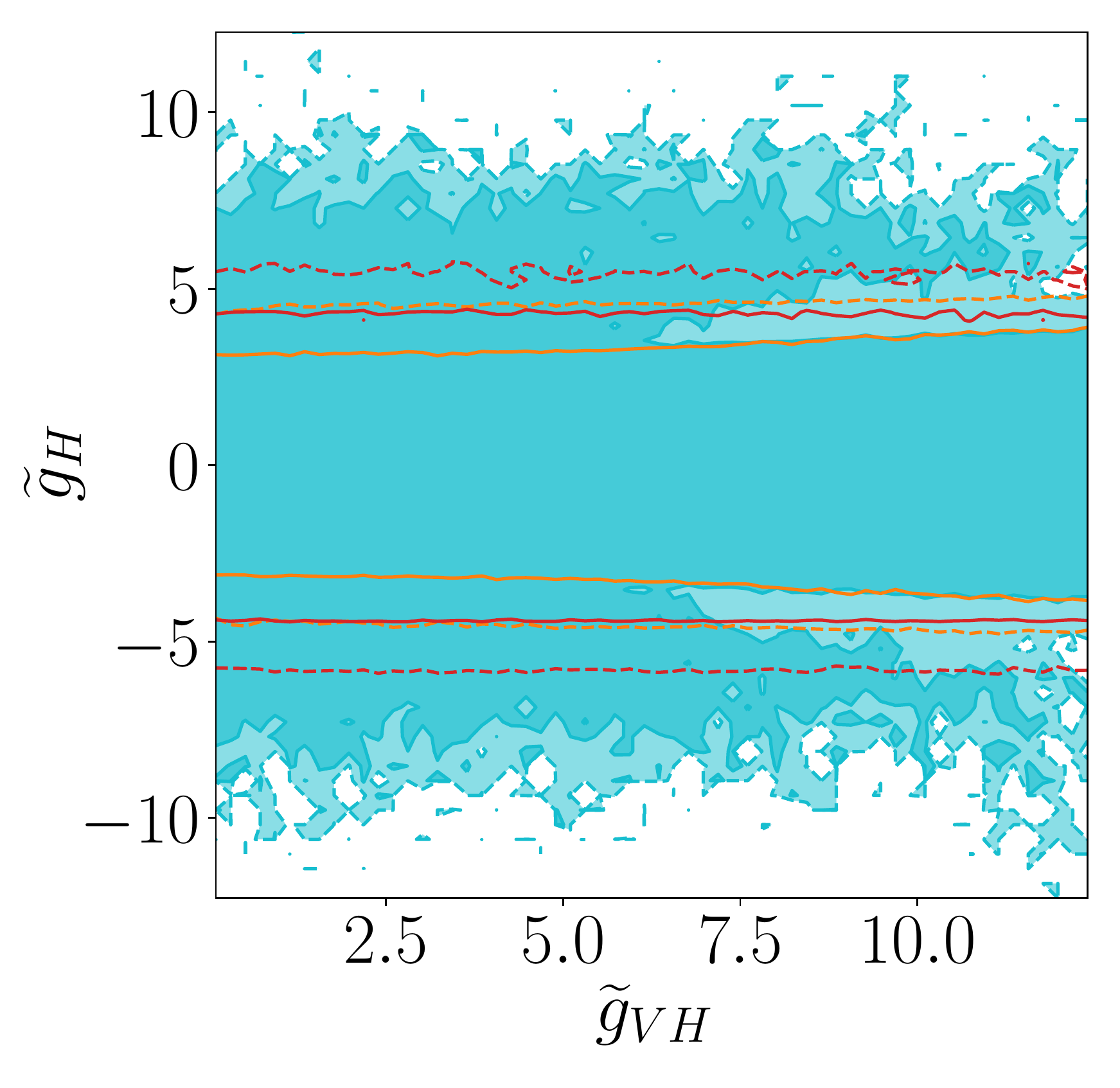}\\
 \hspace*{.24\textwidth}
 \includegraphics[width=.24\textwidth]{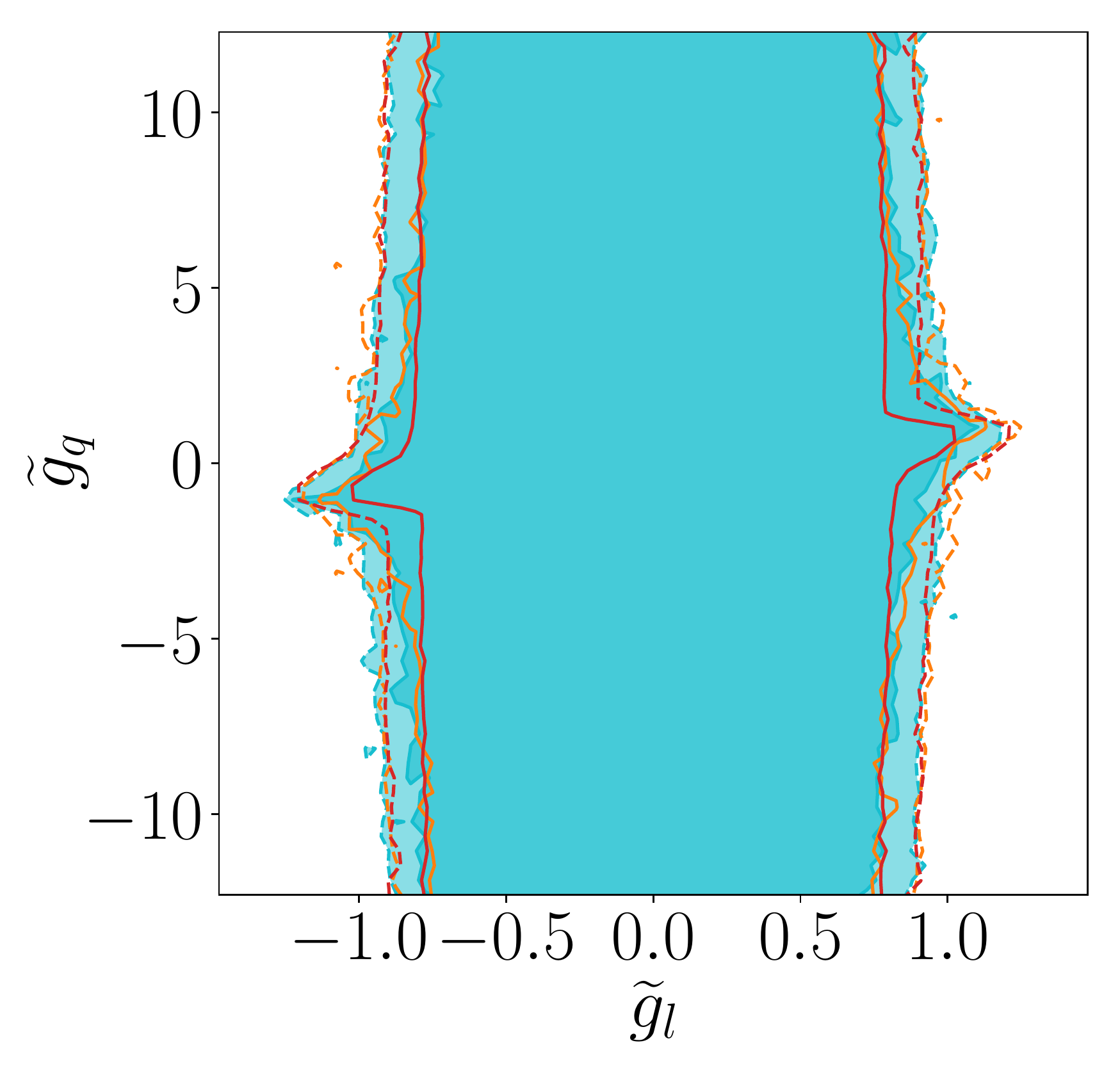}
 \includegraphics[width=.24\textwidth]{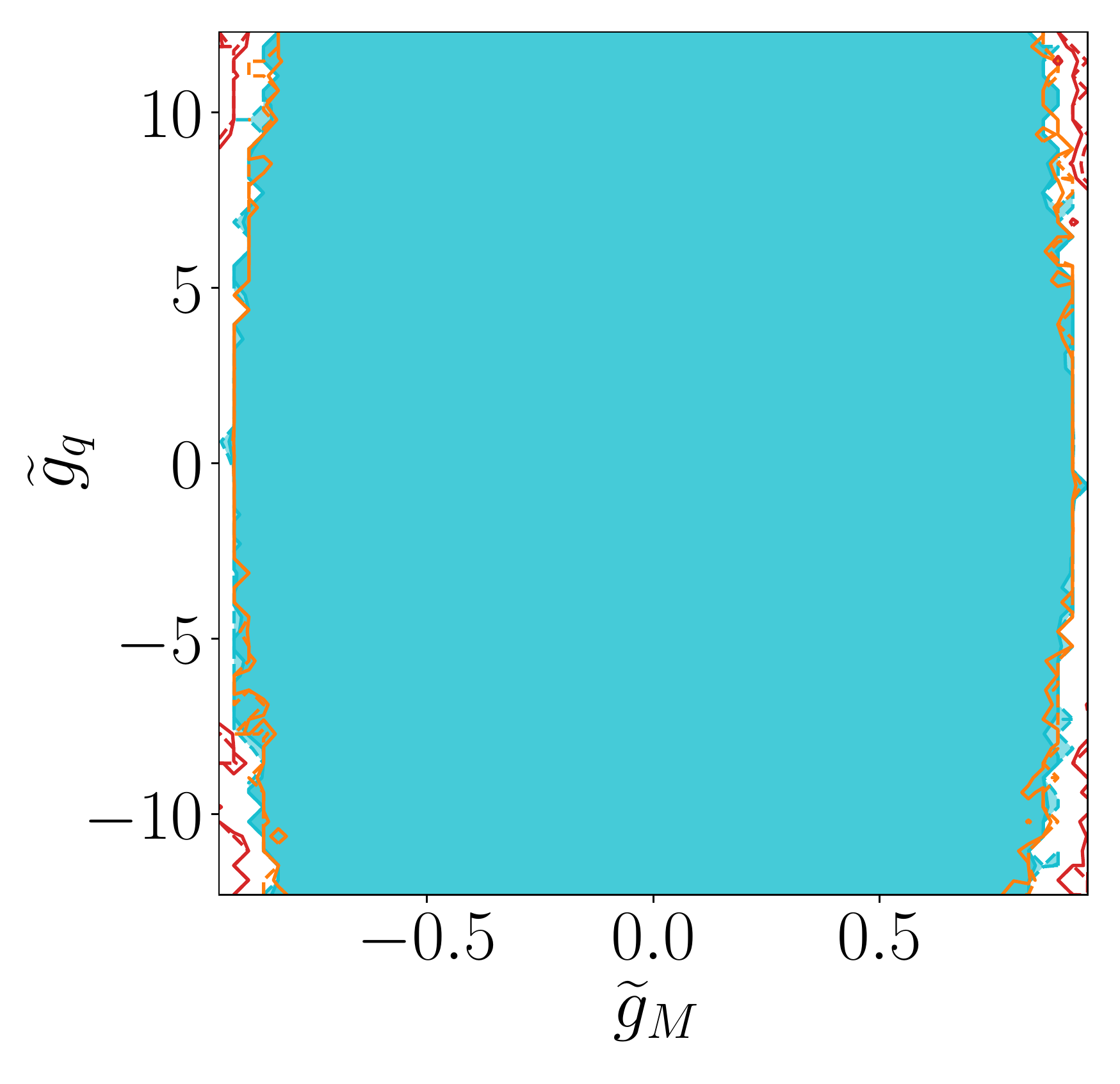}
 \includegraphics[width=.24\textwidth]{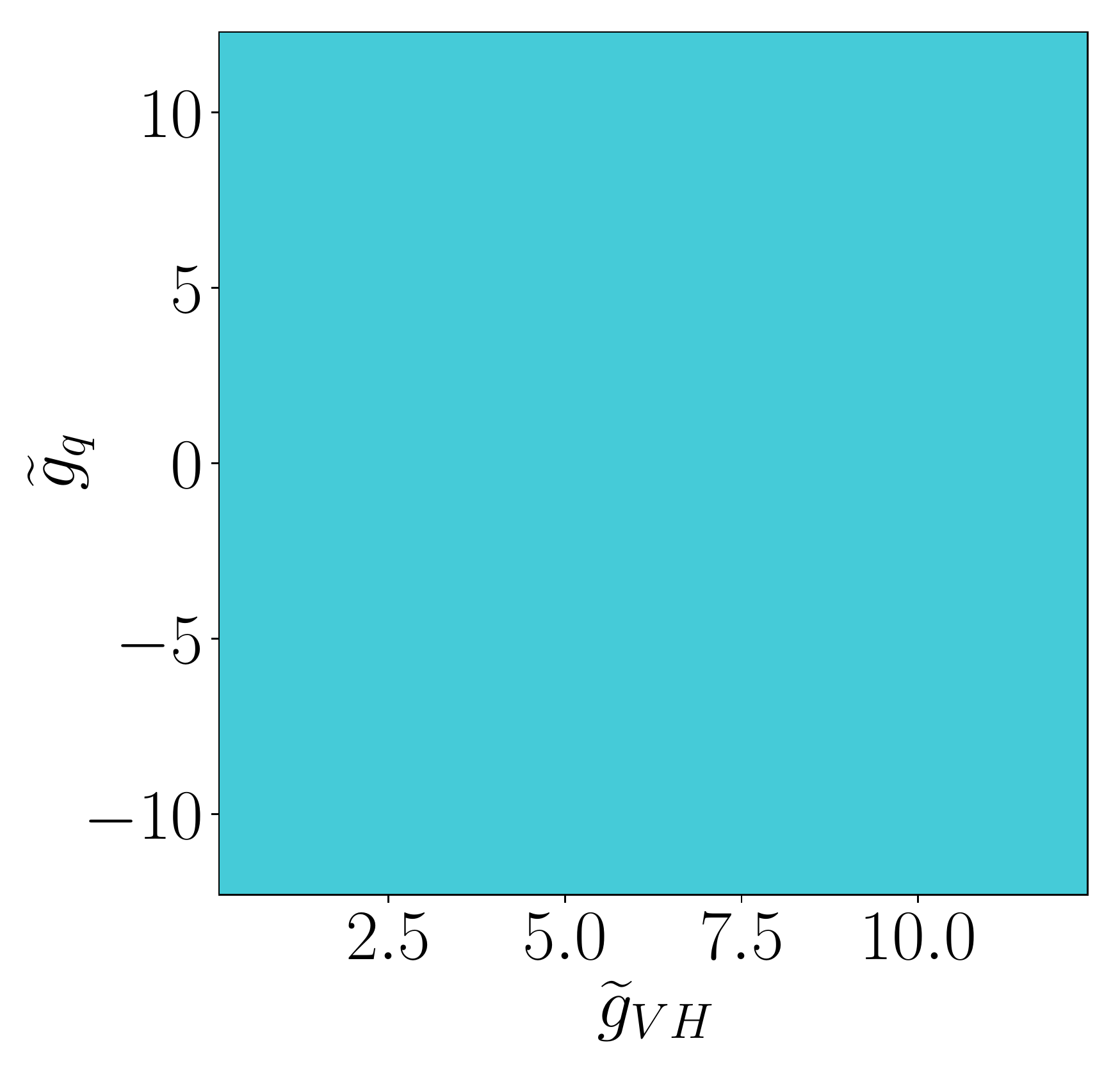}\\
 \hspace*{.48\textwidth}
 \includegraphics[width=.24\textwidth]{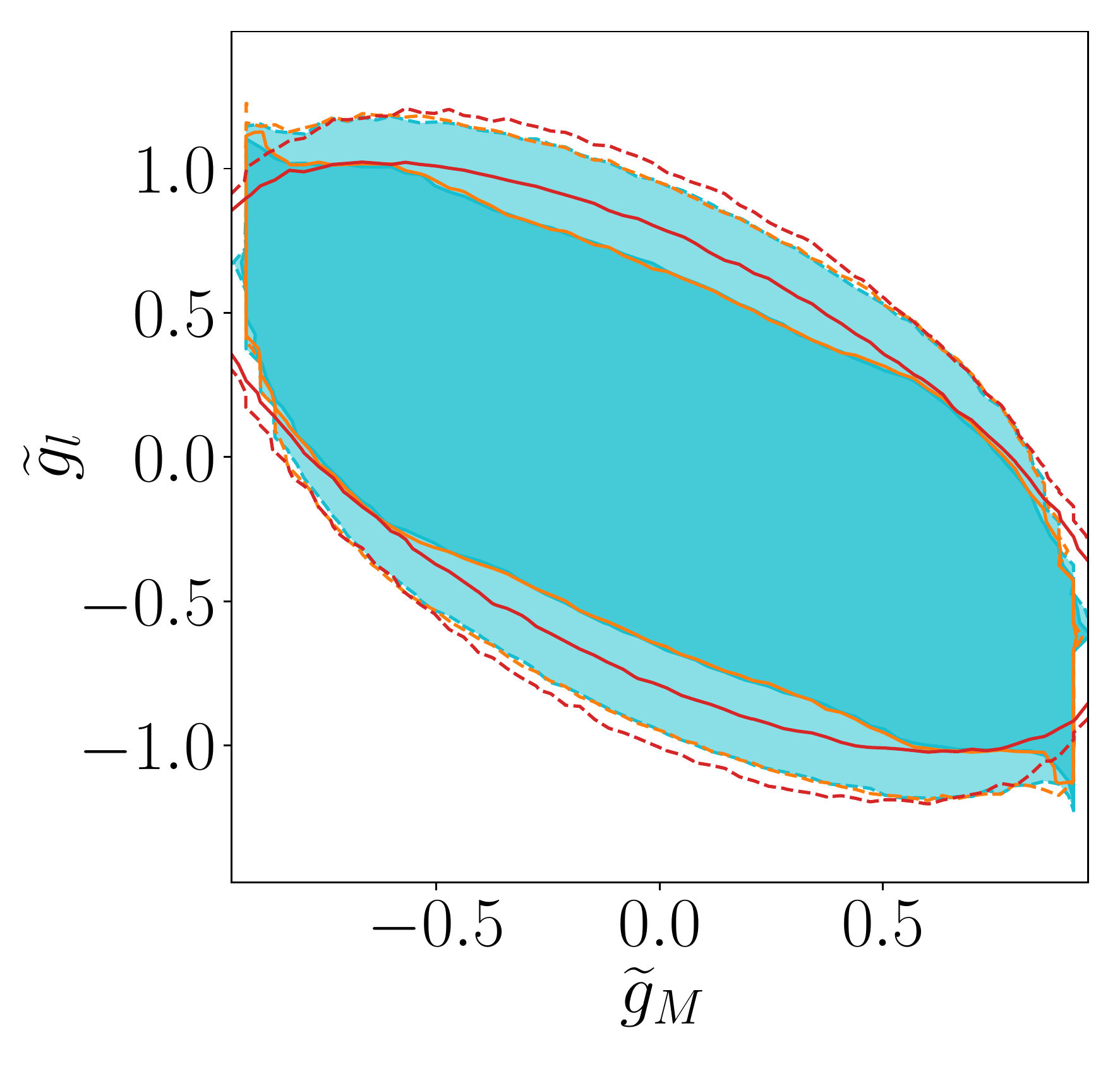}
 \includegraphics[width=.24\textwidth]{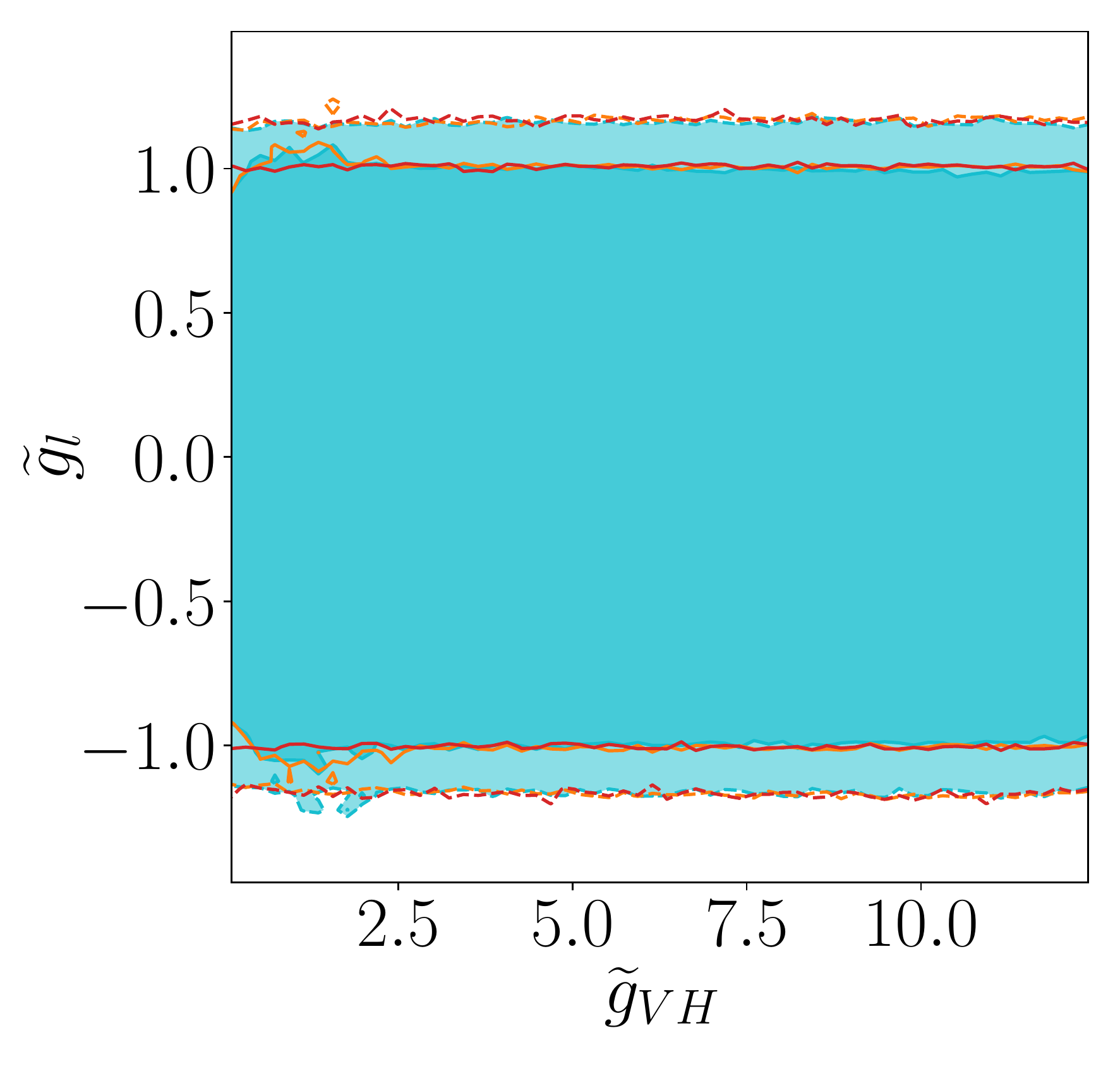}\\
 \hspace*{.72\textwidth}
 \includegraphics[width=.24\textwidth]{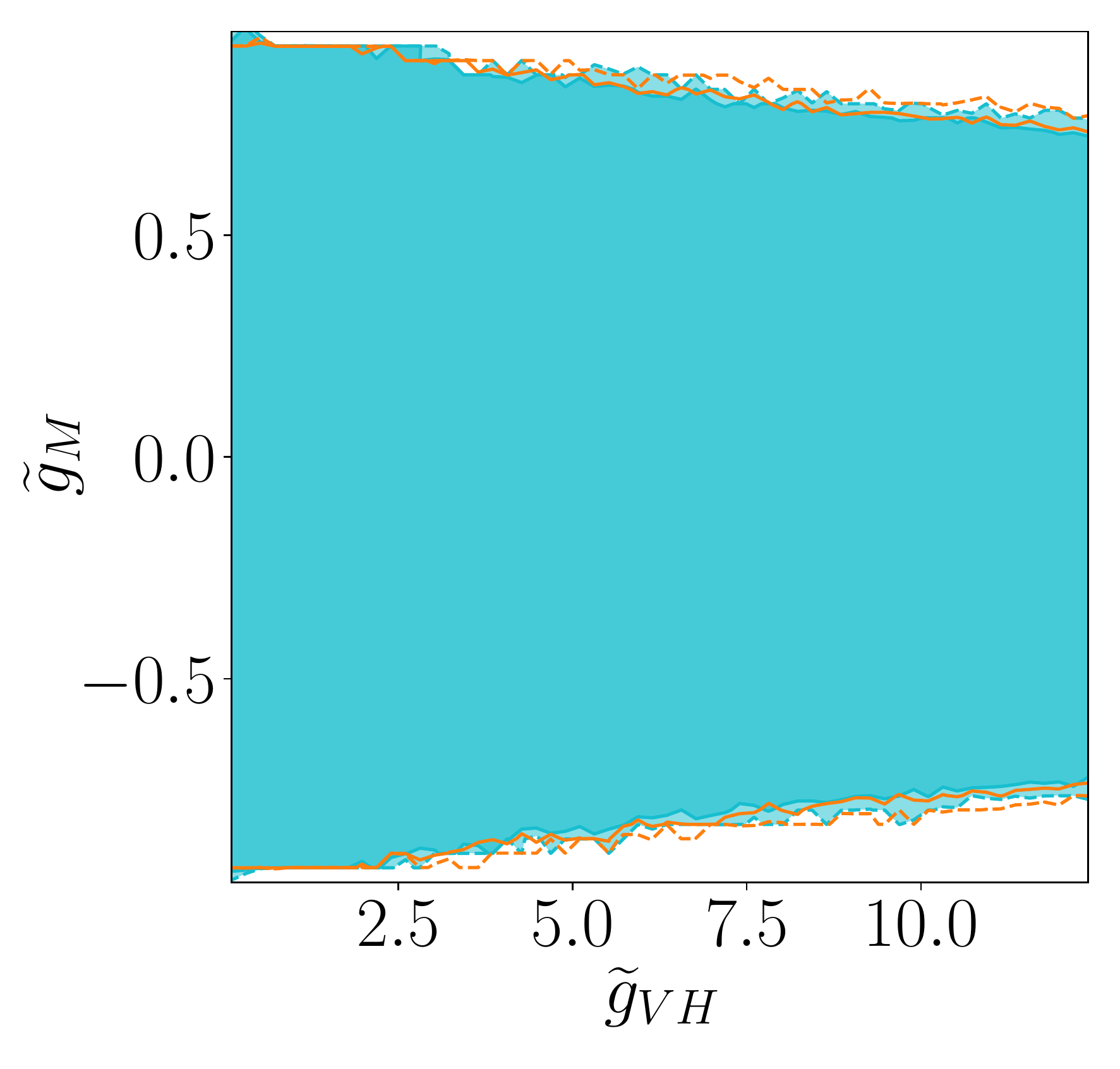}
 \caption{5-parameter global fit of the full data set to the model
   parameters from Eq.\eqref{eq:lagrangian1}. Profiled
   $\Delta\chi^2=2.3$ $(\Delta\chi^2=5.991)$ contours are shown as
   solid (dashed) lines.  Red (orange) curves indicate the results
   obtained with tree (1-loop) matching onto the SMEFT and a fixed
   matching scale $Q=m_V$. The light blue region shows the results
   from 1-loop matching, profiled over $\qm =
   \unit[500]{GeV}~...~m_V$.}
 \label{fig:4tev_5param_fit}
\end{figure}

For a fixed matching scale $Q=m_V$ (red and orange lines in
Fig.~\ref{fig:4tev_5param_fit}), we find that the SMEFT fit constrains
significantly $\gt_l$ and $\gt_H$, while $\gt_M$, $\gt_q$, and
$\gt_{VH}$ are essentially unconstrained.  The striking difference
between the constraints on the vector triplet couplings to leptons and
to quarks is largely due to the fact that the SMEFT fit is dominated
by EWPO constraints extracted at LEP, on which the leptonic
interactions have a much stronger impact.  We have verified that,
indeed, removing EWPO constraints from the fit relaxes significantly
the constraint on $\gt_l$.

\begin{figure}[t]
  \centering
  \includegraphics[height=5.3cm, trim = 0 0 4cm 0, clip]{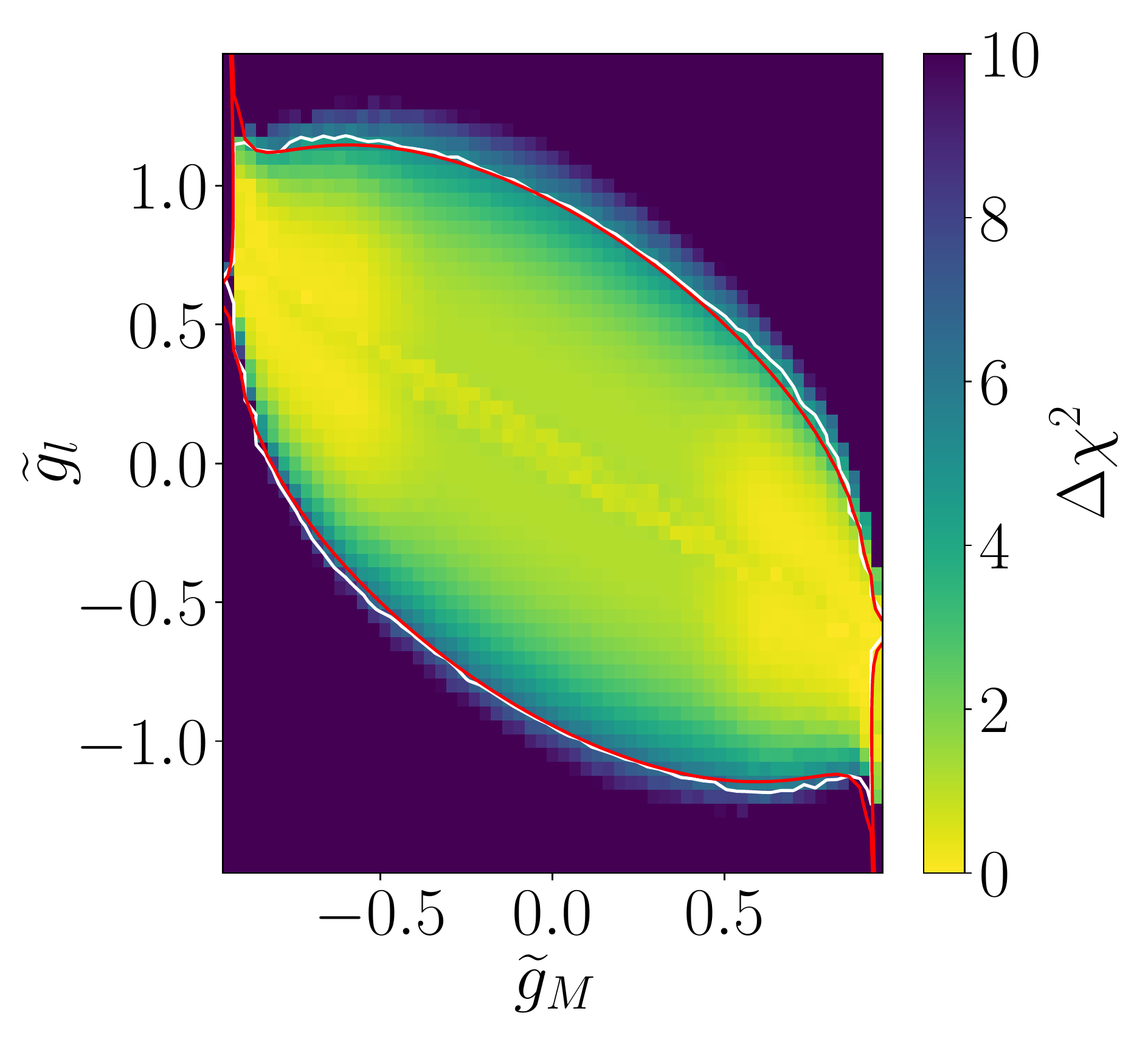}
  \hspace{0.5cm}
  \includegraphics[
  height=5.3cm]{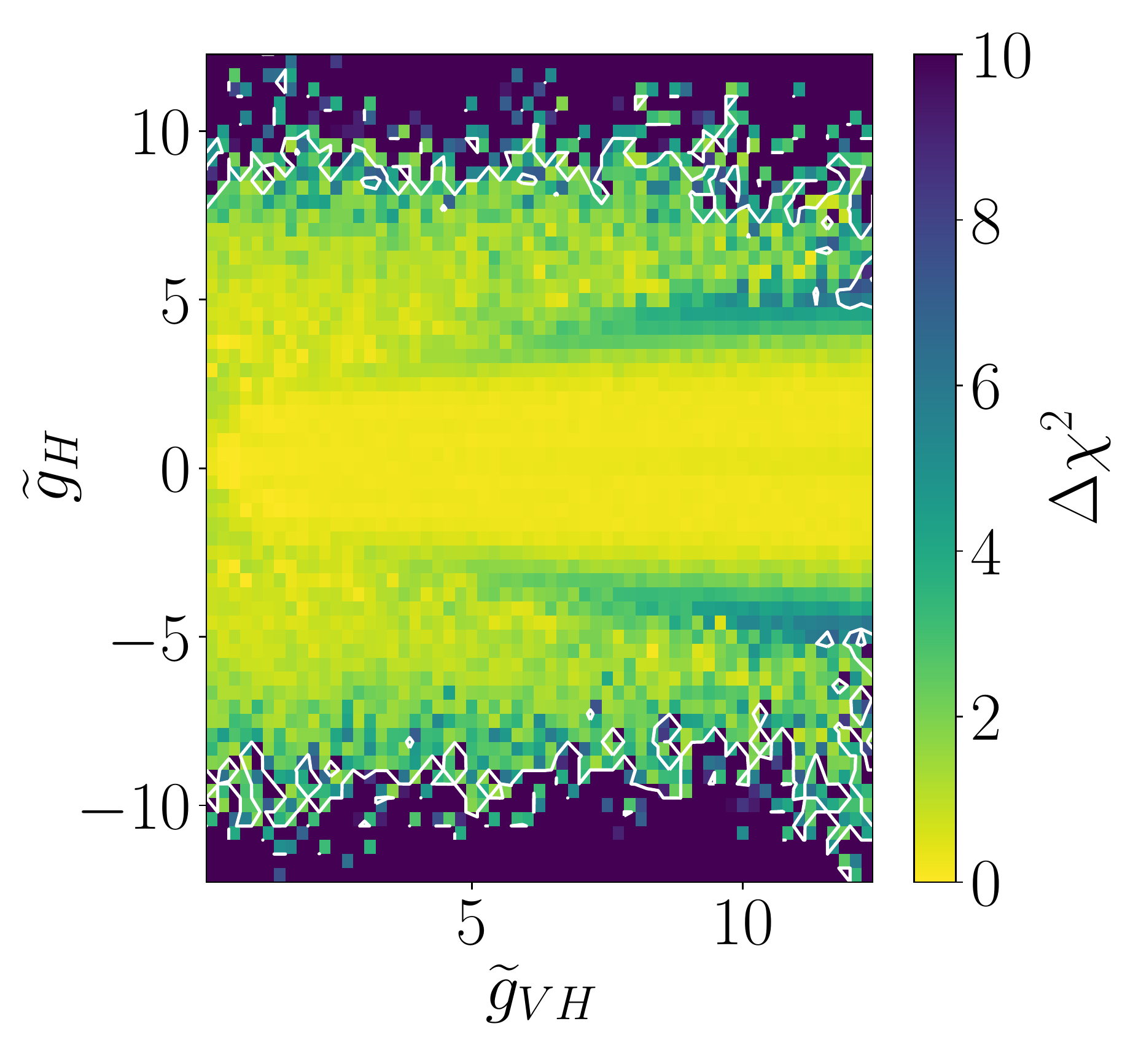}
  \caption{Heat map of the profiled $\Delta\chi^2$ distribution from
    the same fit as in Fig.~\ref{fig:4tev_5param_fit}, with 1-loop
    matching and profiling over the matching scale. The red contours
    indicate $f_{LLLL}/\Lambda^2=-0.014, +0.017\ $TeV$^{-2}$ and the
    white contours indicate $\Delta\chi^2=5.991$.}
  \label{fig:4tev_5param_fit_likelihood}
\end{figure}

The 2D projections show that $\gt_l$ is also anti-correlated to
$\gt_M$. The reason is that, at tree-level, $\gt_l$ enters the
matching expressions only in the combination $\gt_l + g_2 \gt_M$,
where $g_2$ is the $SU(2)$ coupling constant.  Specifically, we find
that the constraints in the $\gt_M - \gt_l$ plane are dominated by
the constraint on $f_{LLLL}$, whose tree-level matching expression is
quadratic in the relevant combination
\begin{align}
 \frac{f_{LLLL}}{\Lambda^2} &= - \frac{(\gt_l + g_2 \gt_M)^2}{4\tilde m_V^2}\; .
\end{align}
Therefore, for most values of $\gt_M$ and $\gt_l$, the constraints are
driven by the limit for negative values of this Wilson coefficient. At
1-loop, the matching expression is more complex and allows for
positive values of $f_{LLLL}$ in a region close to $|\gt_M|\simeq 1$
and $ |\gt_l|\simeq 1$.  The right panel in
Fig.~\ref{fig:4tev_5param_fit_likelihood} shows that the $2\sigma$
boundary from the 5D likelihood (in white) matches very well the
contours for $f_{LLLL}/\Lambda^2=\unit[-0.014, +0.017]{TeV^{-2}}$ (in
red), corresponding to the $2\sigma$ interval derived from a 2D fit of
$f_{LLLL}$ and $f_{BW}$.  Here a 2D fit is necessary owing to the
strong correlation between $f_{LLLL}$ and $f_{BW}$.  A 1D fit would
lead to an over-estimation of the constraints.

There are no major differences between tree and loop level matching
when keeping the matching scale fixed $Q=m_V$.  Only slight
differences can be observed in the limits on $\gt_M$ and $\gt_H$. The
effect on $\gt_H$ is completely washed out once the matching scale is
allowed to vary, as we discuss below.  Although less visible due to
the different scales, an analogous anti-correlation is present in the
$\gt_M - \gt_H$ plane, as $\gt_H$ also enters tree-level matching
expressions exclusively in the combination $\gt_H + g_2
\gt_M$. Because $\gt_H$ enters many Wilson coefficients, both at tree
and loop level, in this case it is not possible to identify one
particular SMEFT parameter, or combination thereof, that drives the
global bounds.

The constraint on $\gt_q$, on the other hand, is driven by that on
$\fpWt$, whose matching expression is given in
Eq.\eqref{eq:fphiq3}. This is consistent with the fact that $\gt_q$
only shows a non-trivial interplay with $\gt_H$. The cross-like shape
emerging in the $(\gt_q, \gt_H)$ panel results from the superposition
of the hyperbola-like shape expected from the $f_{\phi Q}^3$ matching
expression, and of additional constraints on $\gt_H$ that introduce
extra suppressions away from the two axes.  Finally, $\gt_{VH}$ does
not contribute to any dimension-6 operator at tree-level, so, in this
limit, the likelihood is exactly flat in the corresponding
direction. At 1-loop $\gt_{VH}$ gives contributions to $f_W, f_{WW},
f_{\phi2}, f_{t,b,\tau}$ and $\fpWt$. Among these, the dominant
constraint stems from $f_{\phi2}$, leading to the orange contours in
the $\gt_{VH} - \gt_M$ and $\gt_{VH} - \gt_H$ planes.

\subsubsection*{Variable matching scale}

Varying the matching scale as $\qm =
\unit[500]{GeV}~...~m_V=\unit[4]{TeV}$, as shown as light blue region
in Fig.~\ref{fig:4tev_5param_fit}, affects the constraints on $\gt_H$,
while for the other parameters the dependence is negligible. This is
what we expect from the toy results in Sec.~\ref{sec:toy_scale} and
Fig.~\ref{fig:matching}, and we have verified that extending the range
to $\qm \gtrsim m_V$ does not add any significant feature to the
results.  As for the 5-parameter fit, the main consequence of variable
$\qm$ is that, for $\qm\lesssim\unit[2.4]{TeV}$, the matching
expressions of $f_{\phi2}$ and $f_{t,b,\tau}$ acquire a new
zero. Because these operators are the dominant source of constraints
on $\gt_H$, this results in a broader allowed region for this
parameter, which is largest close to the $Q\simeq\unit[2.4]{TeV}$
threshold. This effect washes out the correlation between $\gt_H$ and
$\gt_M$ mentioned above.

At $Q\simeq\unit[2.4]{TeV}$, the most constraining Wilson coefficient
is $f_{\phi2}$, which is responsible for the outermost
region of the $2\sigma$ contours for $\gt_H$ in
Fig.~\ref{fig:4tev_5param_fit}.  The inner structure of the
likelihood, including the $1\sigma$ contour, cannot be explained in
terms of a single Wilson coefficient. It is the result of a
non-trivial interplay between several effects, including 
$\gt_H$ entering a large number of Wilson coefficients and
profiling over the matching scale.\bigskip

It is also interesting to look at the finer structure of the profiled
likelihood. In Fig.~\ref{fig:4tev_5param_fit_likelihood} we show
$\Delta\chi^2$ for the same 2D projections as before. We can see that
the best-fit points are focused in regions where $|\gt_M|>0.5$. This
effect emerges in the 5-parameter fit with 1-loop matching,
irrespective of whether the matching scale is fixed or varied. It
is the same effect as observed for the 3-parameter fit varying the
heavy vector mass in Fig.~\ref{fig:dec_tilde}, and it is due to the
EWPO preferring a best-fit point away from the SM. In particular, we
have checked that the observed substructures are entirely
dominated by less than $1\sigma$ deviations in $A_l(SLD)$ and
$m_W$. In addition, the measurements of $\sigma_h^0,
\,R^0_l,\,A_{FB}^{0,l}, \,A_c$ reinforce the deviation through
correlations. If future measurements with reduced uncertainties
confirmed the present deviations from the SM, this would lead to
exclusion limits with intricate patterns.

\subsubsection*{Impact of high energy measurements}

\begin{figure}[t]
  \centering
  \includegraphics[width=.32\textwidth]{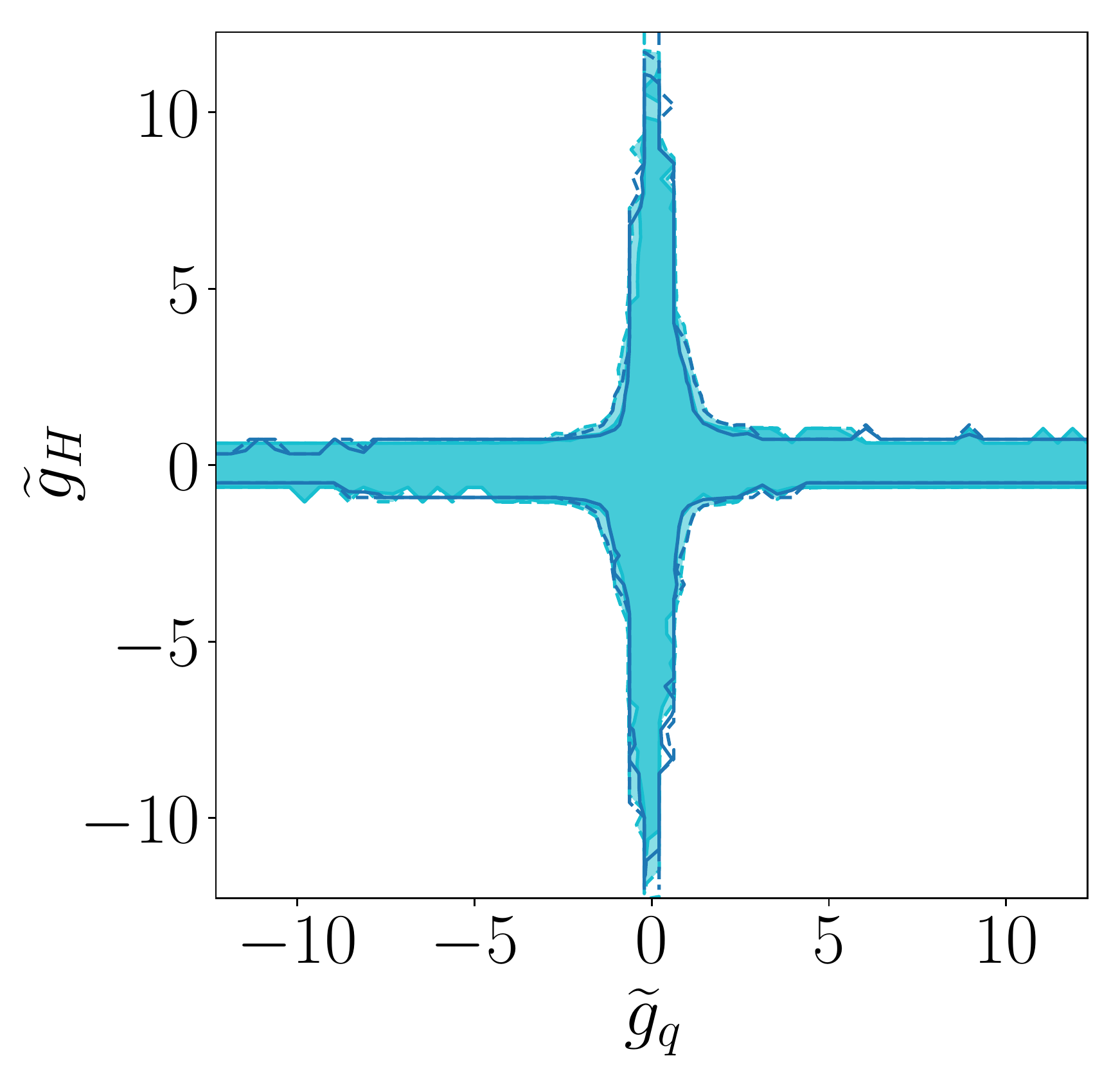}
  \includegraphics[width=.32\textwidth]{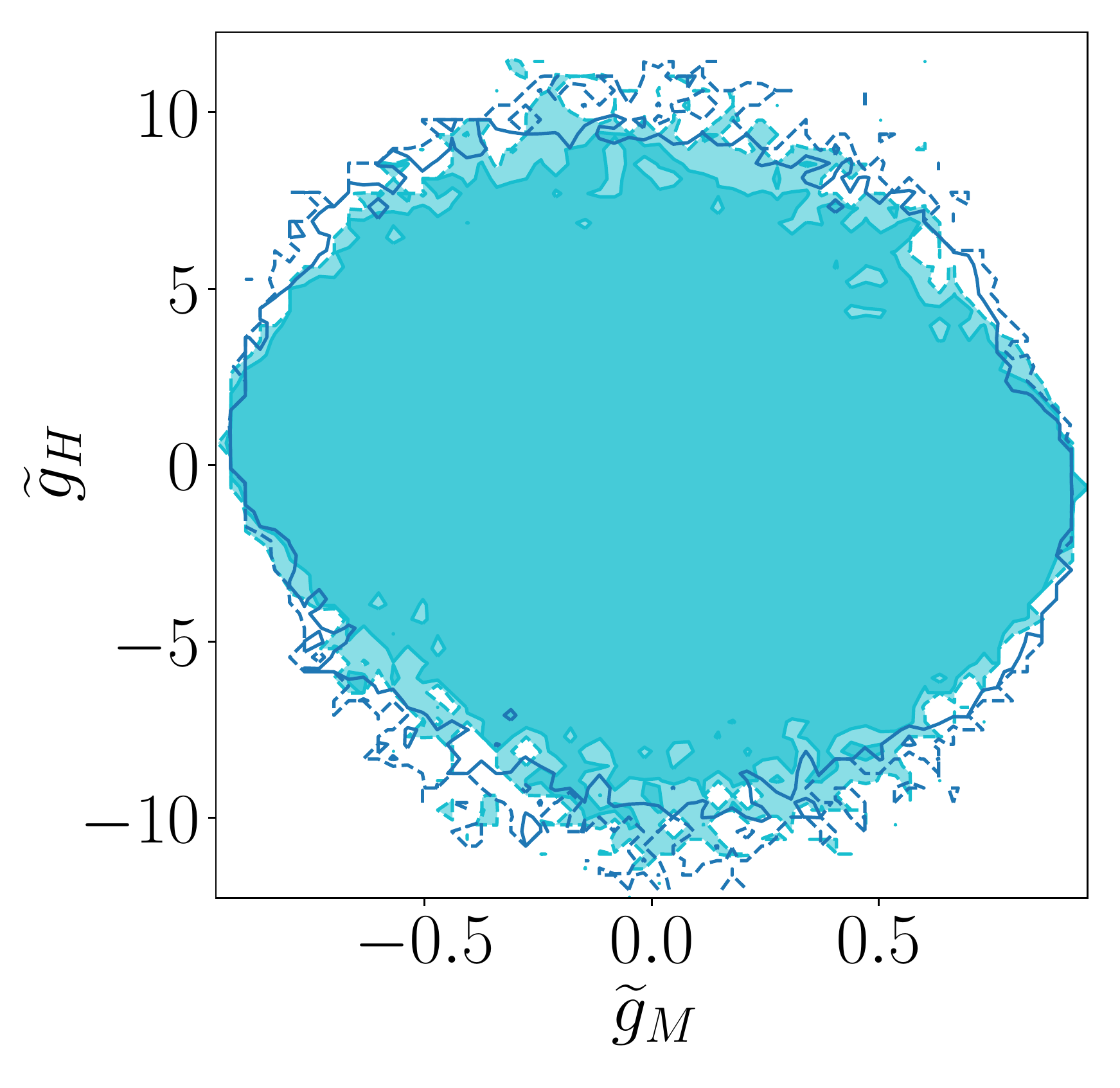}
  \includegraphics[width=.32\textwidth]{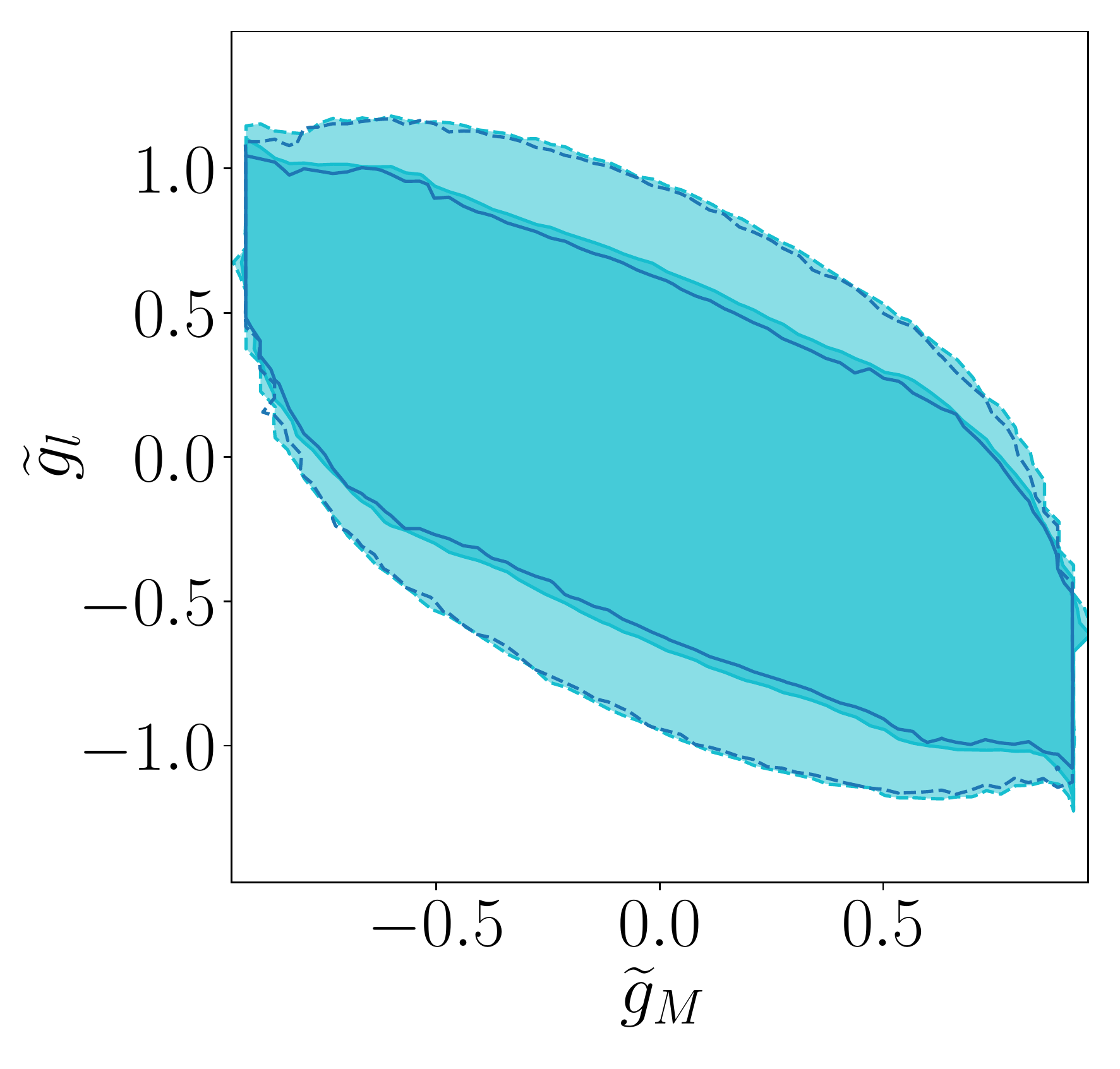}
  \caption{Impact of the high-energy kinematic
    distributions~\cite{Aaboud:2017cxo,Aad:2020tps,Aad:2020ddw} on the
    global 5-parameter SMEFT fit. The solid regions include the full
    data set (same as Fig.~\ref{fig:4tev_5param_fit}), while the dark
    blue lines exclude the high-energy kinematic distributions. Solid
    (dashed) lines mark the $\Delta\chi^2=2.3$ ($\Delta\chi^2=5.991$)
    contours.}
  \label{fig:high_vs_low}
\end{figure}

It is well known~\cite{Biekotter:2018rhp} that kinematic
distributions probing high invariant masses have significant impact on
global fits to the SMEFT parameters. In our analysis, we confirm this
behavior for the two analyses described in Sec.~\ref{sec:global_meas},
which are found to constrain significantly $f_W$, $f_{\phi d}$ and
$f_{WWW}$. Unfortunately, once the SMEFT is mapped onto the heavy
vector triplet model, the constraining power of these measurements is
diminished. This is shown in Fig.~\ref{fig:high_vs_low}, where
the results of Fig.~\ref{fig:4tev_5param_fit} are compared to
those from a 5-parameter fit where the three analyses of
Refs.~\cite{Aaboud:2017cxo,Aad:2020tps,Aad:2020ddw} are removed (dark
blue line).  The lack of visible impact of the high-energy kinematic
distributions is very much due to the specific model and the
corresponding numerical behaviour of the matching formulae.  As
discussed above, the main constraints on the vector triplet parameter
space are dominantly associated to those on $f_{LLLL}, f_{\phi2}$ and
$f_{\phi Q}^{3}$, which are only marginally improved by these
searches.

\subsubsection*{SMEFT vs direct searches}

\begin{figure}[t]
  \centering
  \includegraphics[width=.41\textwidth]{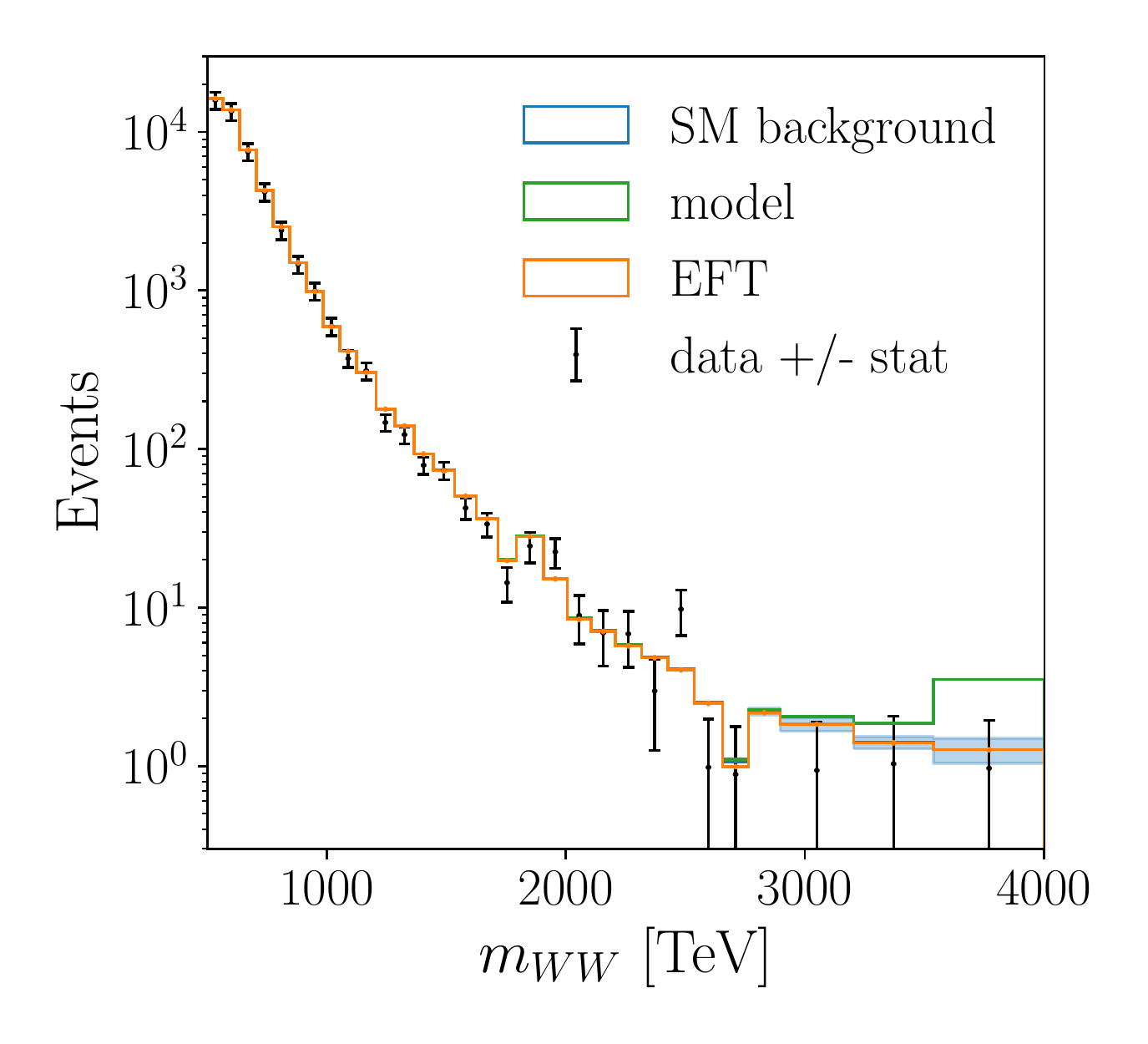}
  \includegraphics[width=.58\textwidth]{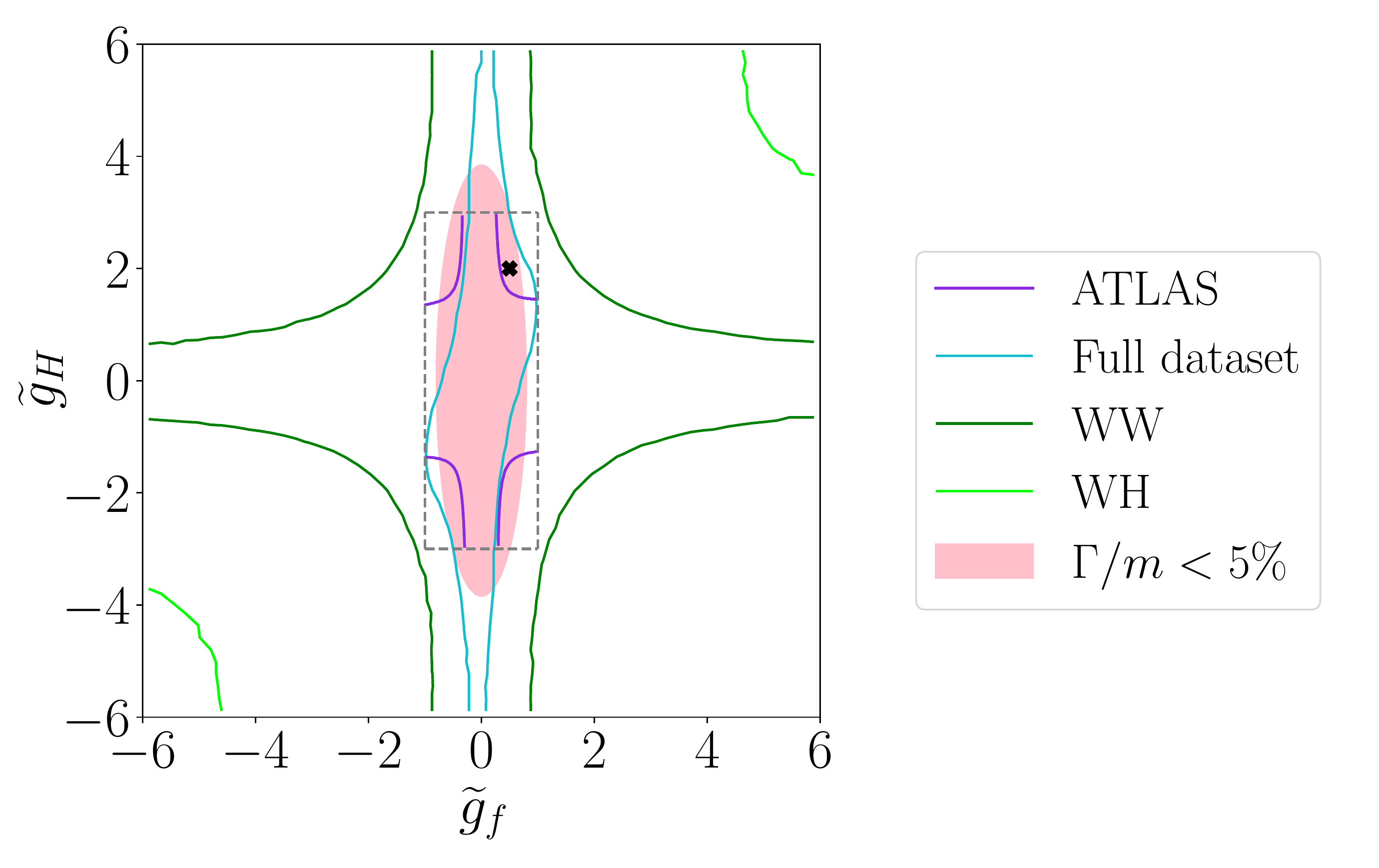}
  \caption{Left: $Z'$ prediction for $m_V=\unit[4]{TeV}$, $\gt_H=2$,
    $\gt_f=0.5$ (shown by a star in the right panel) for the $WW$
    search~\cite{Aad:2020ddw}, compared to the SMEFT prediction.
    Right: SMEFT limits ($\Delta\chi^2=5.991$) for $m_V=\unit[4]{TeV}$
    and profiled over the matching scale, for the $WW$ and $WH$
    distributions alone and the full dataset.  We also show the 95\%CL
    exclusion from the $WH$ resonance search~\cite{Aad:2020tps}. The
    gray box marks the ATLAS search region, the narrow-width is shaded
    in pink.}
  \label{fig:compare_to_atlas}
\end{figure}

A key question we would like to address in this work is whether a
global SMEFT analysis can be competitive with direct searches in
constraining a given UV model. Figure~\ref{fig:compare_to_atlas}
compares the constraints in the $(\gt_f, \gt_H)$ plane obtained in the
direct search of $WH$ resonances by ATLAS, Ref.~\cite{Aad:2020tps},
and from 2D SMEFT fits to different sets of observables.  In
particular, the light green line indicates the SMEFT constraints
obtained from the same distribution as in the direct search.  For all
lines in this plot, the heavy triplet mass is fixed to
$m_V=\unit[4]{TeV}$, the maximum value accessible by the resonance
search.  Strictly speaking, the direct and indirect constraints
extracted from the same measurement apply to complementary
regions of the parameter space: the former are valid for masses
$m_V\lesssim\unit[4]{TeV}$ and for narrow vector triplets within the
pink-shaded region of Fig~\ref{fig:compare_to_atlas}, while the latter
hold for $m_V\gg\unit[4]{TeV}$ irrespective of the resonance
width. Obviously, a comparison should be taken with a grain of salt.

Nevertheless, it can be instructive to examine the interplay between
the signals produced by a heavy resonance and by its corresponding
SMEFT approximation.  The left panel of
Fig.~\ref{fig:compare_to_atlas} shows the $m_{WW}$ resonant
distribution obtained for a benchmark point at $m_V=\unit[4]{TeV}$,
$\gt_H=2$, $\gt_f\equiv \gt_l=\gt_q=0.5$, compared to the ATLAS
measurement~\cite{Aad:2020tps} (black data points) and the SMEFT
signal matched to this benchmark model at dimension six.  This point
is indicated by a cross in Fig.~\ref{fig:compare_to_atlas} (right), and it is
excluded at 95\%CL by both the ATLAS $WH$ and $WW$ searches, but falls
within the $2\sigma$-allowed region of our SMEFT global analysis.
This discrepancy is obvious from the high-energy $m_{WW}$ tail, where
aside from the mass peak the dimension-6 SMEFT also misses the initial
rise of the distribution.  Among the Wilson coefficients that
contribute to $WW$ production, only
$f_W/\Lambda^2=\unit[0.28]{TeV^{-2}}$ takes a value above the permille
level, while $\fpWt=0$ because $\gt_q=\gt_l$.  This
results in SMEFT signals of only a few percent across the entire
$m_{WW}$ distribution, which are always well within the uncertainties.
It is worth pointing out that in such a situation the best place to
look for the SMEFT signal might not just be the bins where the energy
enhancement is largest, but rather those where the uncertainties are
smallest.

While not surprising, these conclusions do not extend to arbitrary BSM
scenarios. One characteristic of the case examined here is that the
resonance is narrow. As a consequence, the effect in $m_{WW}$ is only
visible close to $m_V$, where the SMEFT expansion immediately breaks
down. The situation improves when we include higher-dimensional
operators~\cite{Kilian:2015opv,Lang:2021hnd}. At dimension six, the
matching to our specific model suppresses all energy-enhanced SMEFT
contributions to $WW$ production, so the signal is under-estimated
across the e$m_{WW}$ distribution. This does not have to be the case
in other BSM models. For instance, it is possible that the dimension-6
approximation over-estimates the model predictions, in which case the
dimension-8 contributions need to be large and negative, and the
truncated SMEFT constraints appear more stringent than those from
direct searches.

Going beyond the comparison of resonance searches and SMEFT analyses
for one measurement, the true power of the SMEFT approach is that it
allows to combine a large number of different measurements.  This will
always improve the sensitivity of the SMEFT analyses and, on the other
hand, it allows to derive more general conclusions, by constraining
all model parameters simultaneously, as shown in
Fig.~\ref{fig:4tev_5param_fit}.  The light blue lines in
Fig.~\ref{fig:compare_to_atlas} show the constraints from a
2-parameter SMEFT fit to the entire dataset employed in this
work. Consistent with the discussions above, these limits are
dominated by EWPO, for which the SMEFT expansion is valid. In
particular, the constraint on $\gt_f$ is dominated by the leptonic
component $\gt_l$, which in turn is mostly associated to the
$f_{LLLL}$ Wilson coefficient.  Comparing these limits to those from
the ATLAS $WH$-search, we find that the latter are slightly stronger
for $|\gt_H|\gtrsim 1$ (with the caveat that they are only valid in
the narrow width regime), while the former dominate for
$|\gt_H|\lesssim 1$. Here, the $WH$ search has an unconstrained
direction along the $\gt_H=0$ axis, that is broken by the EWPO in the
SMEFT fit~\cite{delAguila:2010mx}.

\subsubsection*{Heavy vector results}

\begin{figure}[t]
 \centering
 \includegraphics[width=.24\textwidth]{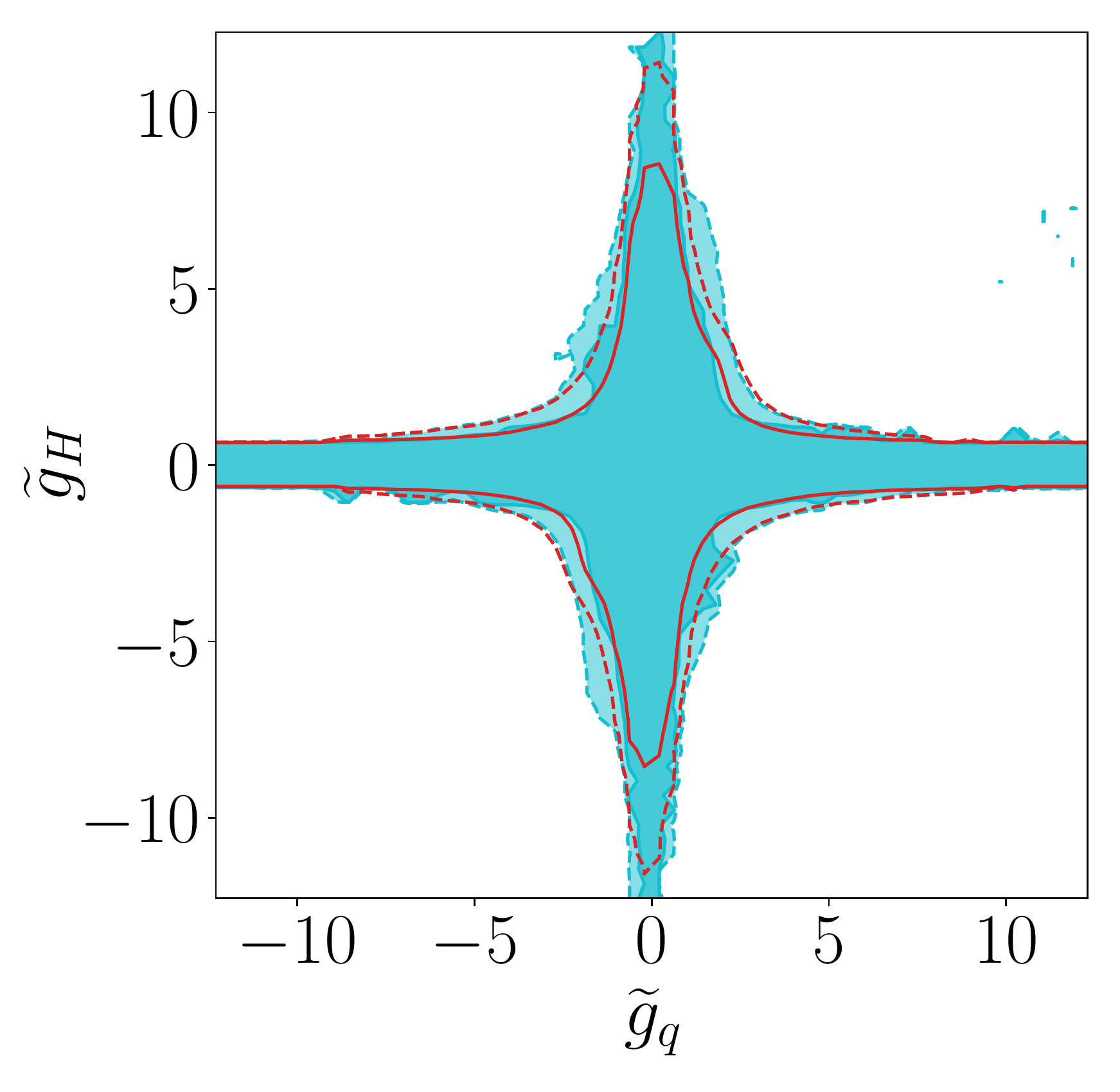}
 \includegraphics[width=.24\textwidth]{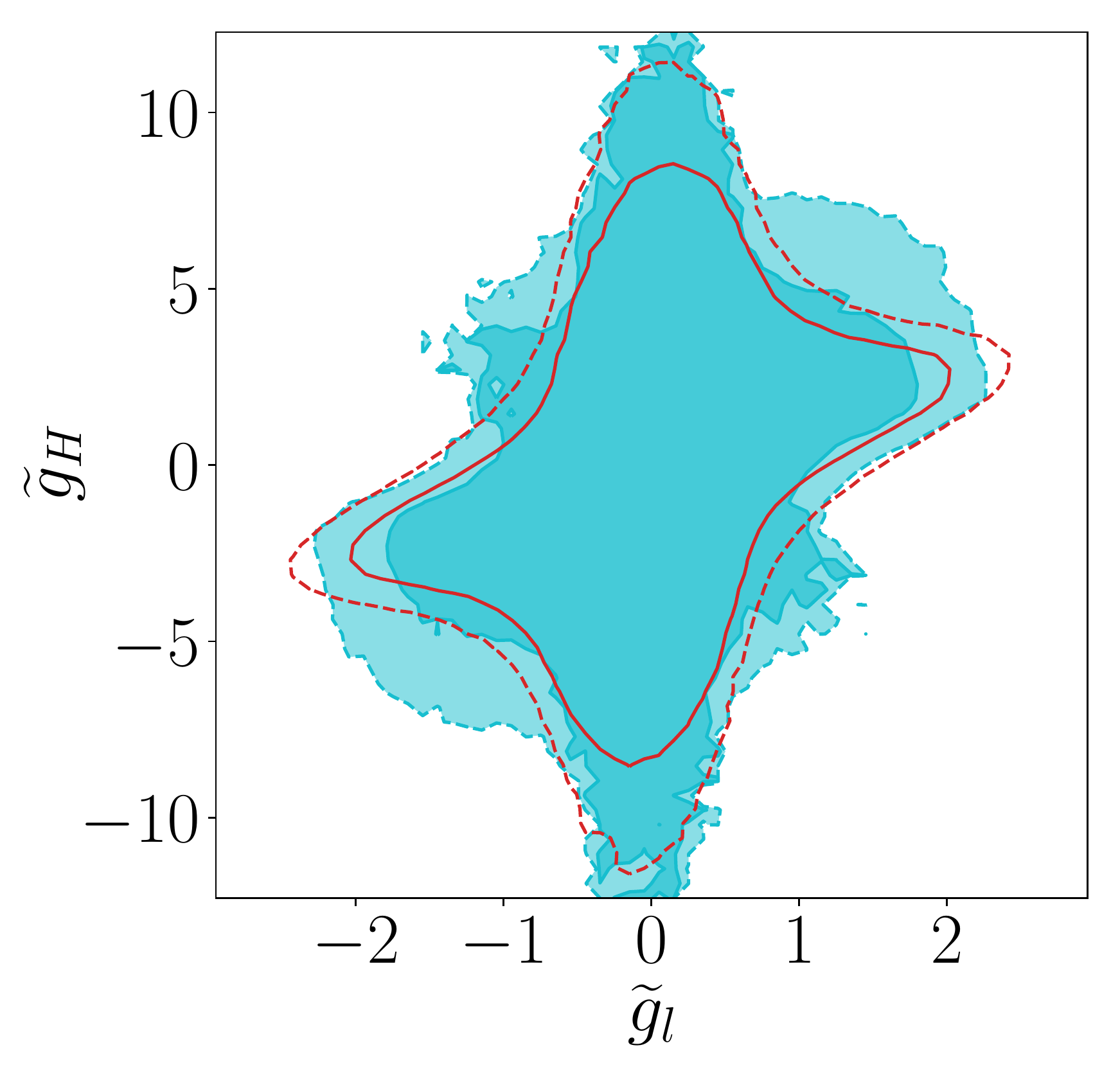}
 \includegraphics[width=.24\textwidth]{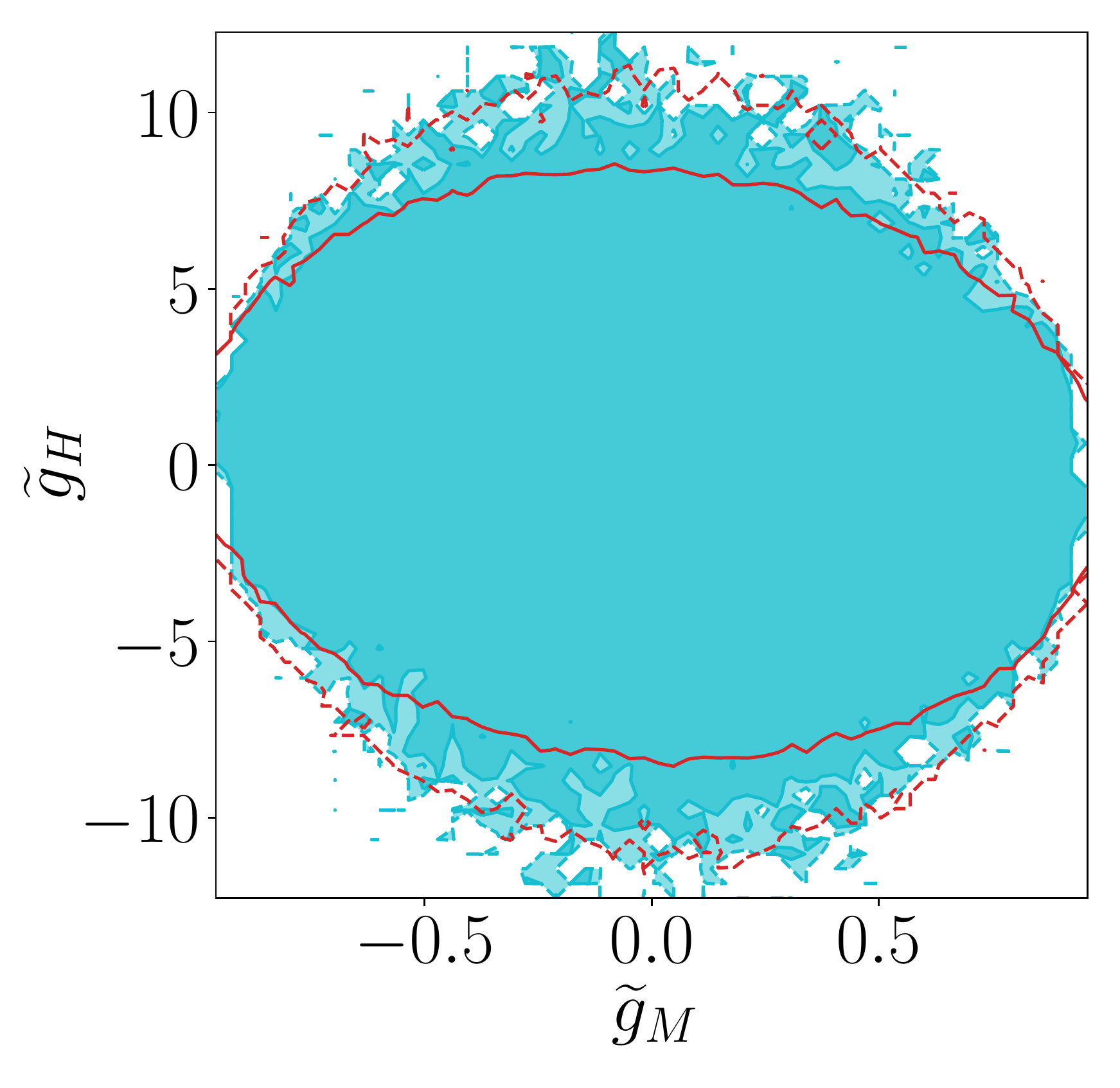}
 \includegraphics[width=.24\textwidth]{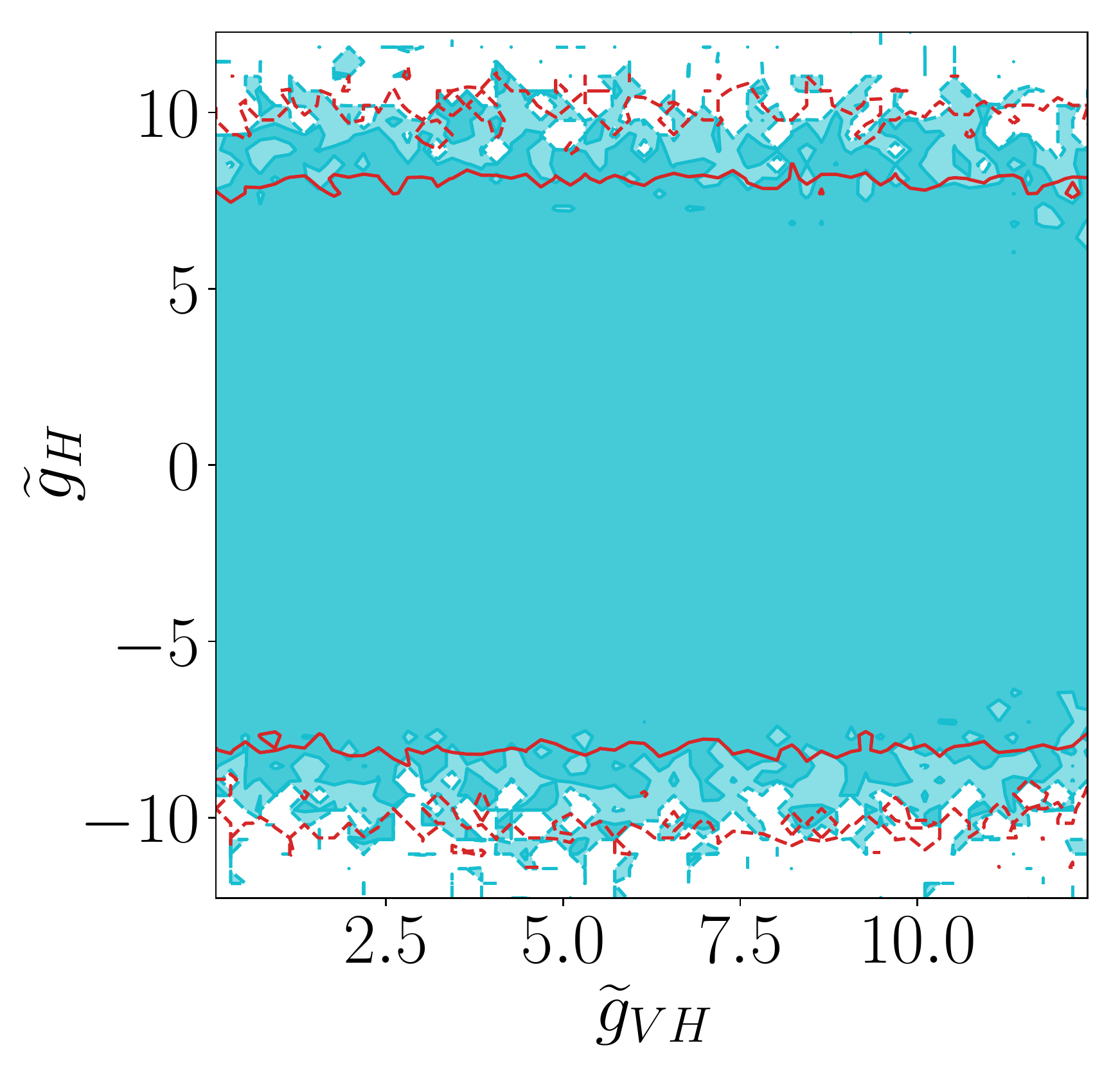}\\
 \includegraphics[width=.24\textwidth]{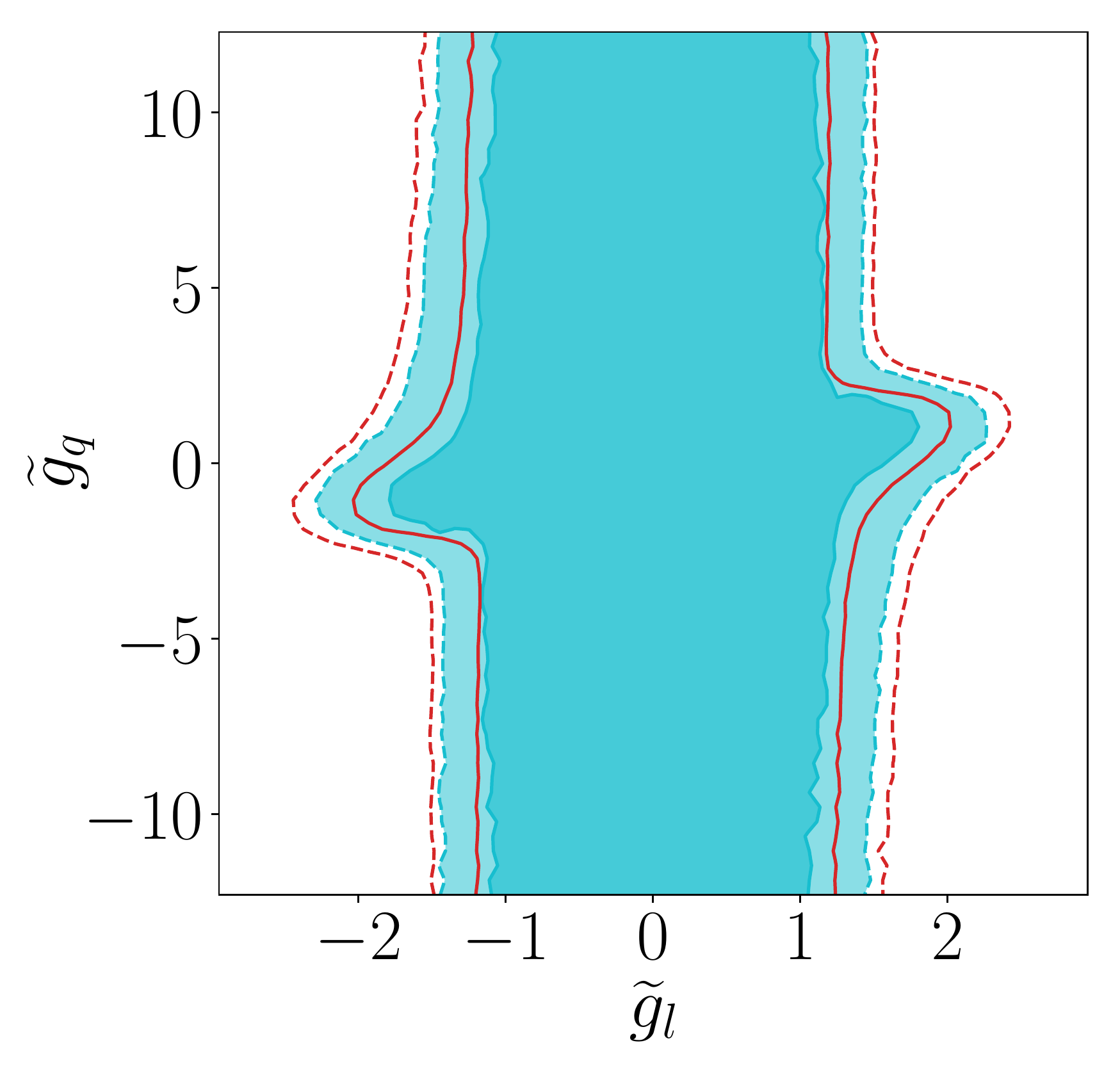}
 \includegraphics[width=.24\textwidth]{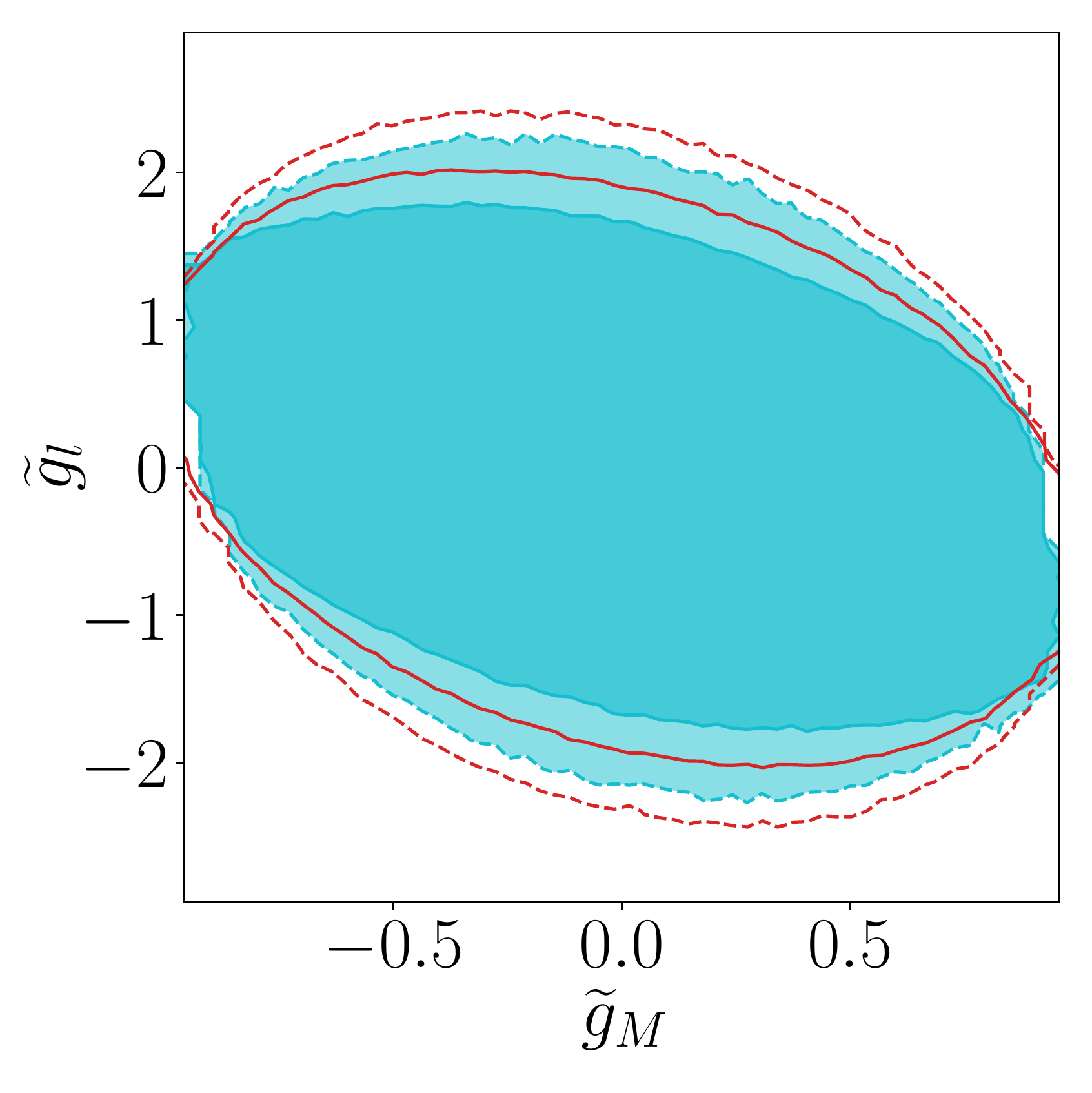}
 \includegraphics[width=.24\textwidth]{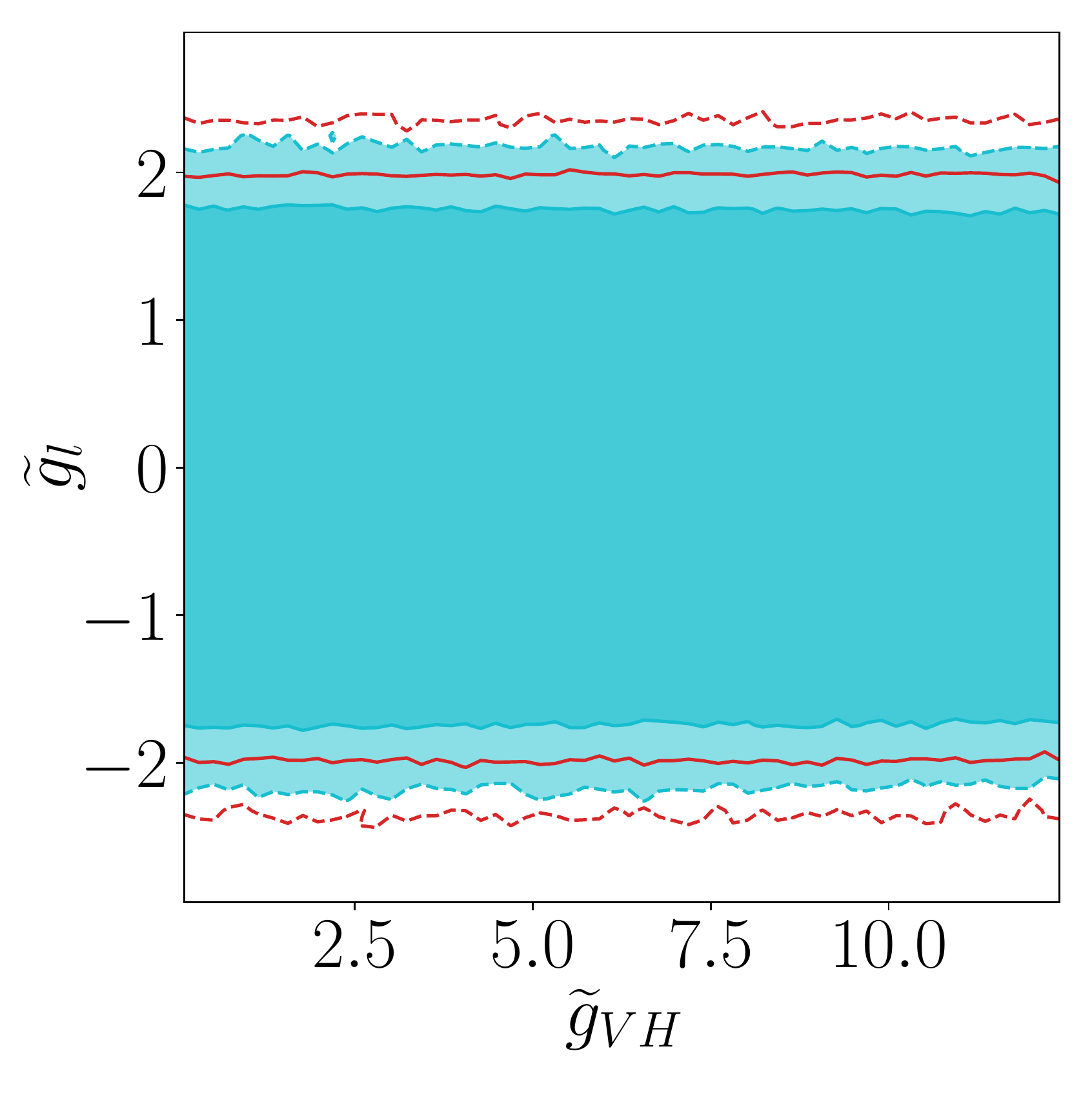}
 \caption{5-parameter fit to the full data set for the model
   parameters in Eq.\eqref{eq:lagrangian1}.  Each panel shows profiled
   $\Delta\chi^2=2.3$ (solid) and $\Delta\chi^2=5.991$ (dashed)
   contours. Red curves correspond to tree-level matching, the light
   blue region to 1-loop matching, profiled over three  $\gt$ parameters
   plus the matching scale $\qm = \unit[500]{GeV}~...~m_V$.  The
   panels for $\gt_M - \gt_q$ $\gt_{VH} -\gt_q$ and $\gt_{VH}-\gt_M$
   are not shown as they are unconstrained in the explored ranges.}
 \label{fig:8tev_5param_fit}
\end{figure}

One of the main motivations for the SMEFT formalism is that it allows
us to derive constraints on new particles with masses beyond the reach
of direct searches.  In this spirit, we can extend our SMEFT
constraints on the $\gt$ parameters for a heavy triplet mass to
$m_V=\unit[8]{TeV}$. Now, the dimension-6 SMEFT approximation is valid
all over the kinematic measurements discussed above.  The corresponding
results in Fig.~\ref{fig:8tev_5param_fit} can be directly compared to
those in Fig.~\ref{fig:4tev_5param_fit} for $m_V=\unit[4]{TeV}$.  As
expected, all the bounds on the model parameters are weaker for
heavier values of $m_V$ (see also Fig.~\ref{fig:dec_tilde}).  However,
a notable feature is that the limits do not simply scale with a factor
proportional to $m_V$, as one would naively expect from the SMEFT
analysis at dimension six.  The reason is that the matching
expressions that relate the Wilson coefficients to the model
parameters are generally non-trivial and do not scale universally with
$(\gt_i/m_V)$, as can be seen for instance in
Eq.\eqref{eq:fphi2_1}. Moreover, as we are considering a BSM state
that is not a singlet under $SU(2)$, the EW gauge coupling $g_2$
contributes to the matching independently of the $\gt$ parameters. The
result is that the degeneracy between $\gt_i$ and $m_V$ is largely
broken in the matching, leading to a complex likelihood structure that
changes significantly with $m_V$.

\section{Conclusions}
\label{sec:conclusions}

We have presented a global analysis of a Standard Model extension with
a gauge-triplet vector resonance in terms of the dimension-6 SMEFT
Lagrangian. We have performed a global \sfitter\ analysis including
electroweak precision observables, Higgs and di-boson measurements as
well as resonance searches at the LHC, and have compared our results
with limits obtained from direct searches.  To relate the full model
and the SMEFT we have employed one-loop matching with a focus on the
theory uncertainties from the choice of the matching scale.

First, we have shown that the theory uncertainty due to the choice of
the matching scale can have a large effect on the global analysis. In
particular, the bounds on the coupling of the new vector to the
SM-Higgs are significantly weakened once we profile over a variable
matching scale, illustrating how all theory uncertainties need to be
taken into account at least once we translate SMEFT results back into
models.

Comparing the SMEFT results with direct searches reveals an intriguing
complementarity. Direct and SMEFT searches are reliable in
different parameter regions; while direct searches are sensitive
to narrow resonances with kinematically accessible masses, SMEFT
searches apply to energies sufficiently below the resonance mass. The
SMEFT analysis can be sensitive to the onset of the resonance, but a
reliable description of this region requires a tower of
higher-dimensional operators. Specifically for the vector-triplet
model, the SMEFT model for the high-energy tail of kinematic
distributions turned out less sensitive than the resonance search, and
therefore provided conservative constraints.  On the other hand, the
SMEFT analysis can probe vector masses beyond the reach of resonance
searches. Here, we found that the one-loop matching dampens the
sensitivity decrease of the SMEFT analysis compared to the naively
expected scaling.

While SMEFT analyses cannot replace model-specific searches for new
physics, they add valuable constraints from a large variety of
measurements and are sensitive to new physics scales beyond the reach
of resonance searches. Only this complementarity of direct and
indirect searches allows us to make best use of current and future LHC
data.

\begin{center} \textbf{Acknowledgments} \end{center}

We would like to thank Anke Biek\"otter, Anja Butter, Tyler Corbett
and David Lopez-Val for extensive support with \sfitter\ and In\^es
Ochoa for her help with the $VH$ ATLAS analysis and for updating the
corresponding HEPData repository.  MK and BS are grateful to Alexander
Voigt for discussions on one-loop matching. We acknowledge support by
the state of Baden-Württemberg through bwHPC and the German Research
Foundation (DFG) through grant no INST 39/963-1 FUGG (bwForCluster
NEMO). BS was partially supported by the German Research
Foundation(DFG) by the grant STU 615/2-1.  The research of everyone is
supported by the Deutsche Forschungsgemeinschaft (DFG, German Research
Foundation) under grant 396021762 -- TRR~257 \textsl{Particle Physics
  Phenomenology after the Higgs Discovery}.

\appendix

\section{Appendix}
\subsection{Operator basis}
\label{app:basis}

\begin{table}[b!]
\renewcommand{\arraystretch}{1.5}
\begin{tabular}{*2{>{$}r<{$}@{ = }>{$}l<{$}}}
\toprule
\mathcal{O}_{G G} & \phi^{\dagger} \phi G_{\mu \nu}^{a} G^{a \mu \nu}  
& 
\mathcal{O}_{B W} & \phi^{\dagger} \hat{B}_{\mu \nu} \hat{W}^{\mu \nu} \phi \
\\
\mathcal{O}_{B B} & \phi^{\dagger} \hat{B}_{\mu \nu} \hat{B}^{\mu \nu} \phi 
&
\mathcal{O}_{W W} & \phi^{\dagger} \hat{W}_{\mu \nu} \hat{W}^{\mu \nu} \phi 
\\ 
\mathcal{O}_{B} & \left(D_{\mu} \phi\right)^{\dagger} \hat{B}^{\mu \nu}\left(D_{\nu} \phi\right)
&
\mathcal{O}_{W} & \left(D_{\mu} \phi\right)^{\dagger} \hat{W}^{\mu \nu}\left(D_{\nu} \phi\right) 
\\
\mathcal{O}_{W W W} & \operatorname{Tr}\left(\hat{W}_{\mu \nu} \hat{W}^{\nu \rho} \hat{W}_{\rho}^{\:\: \mu}\right) 

\\\hline
\mathcal{O}_{\phi 1} & \left(D_{\mu} \phi\right)^{\dagger} \phi \phi^{\dagger}\left(D^{\mu} \phi\right) 
&
\mathcal{O}_{\phi 2} & \frac{1}{2} \partial^{\mu}\left(\phi^{\dagger} \phi\right) \partial_{\mu}\left(\phi^{\dagger} \phi\right) 
\\\hline

\mathcal{O}_{b} & (\phi^{\dagger} \phi)\, \bar{q}_{3} \phi d_{3} 
& 

\mathcal{O}_{\tau} & (\phi^{\dagger} \phi)\, \bar{l}_{3} \phi e_{3} 
\\
\mathcal{O}_{t} & (\phi^{\dagger} \phi)\, \bar{q}_{3} \widetilde{\phi} u_{3} 
\\
\mathcal{O}_{L L L L} & \left(\bar{l}_{1} \gamma_{\mu} l_{2}\right)\left(\bar{l}_{2} \gamma^{\mu} l_{1}\right) 
& 
\mathcal{O}_{\phi e} & (\phi^{\dagger}i \overleftrightarrow{D}_{\mu}  \phi)\left(\bar{e}_{i} \gamma^{\mu} e_{j}\right) \delta^{ij}
\\
\mathcal{O}_{\phi d} & (\phi^{\dagger}i \overleftrightarrow{D}_{\mu} \phi)\left(\bar{d}_{i} \gamma^{\mu} d_{j}\right)  \delta^{ij}
&
\mathcal{O}_{\phi u} & (\phi^{\dagger}i \overleftrightarrow{D}_{\mu}  \phi)\left(\bar{u}_{i} \gamma^{\mu} u_{j}\right)  \delta^{ij}
\\
\mathcal{O}_{\phi Q}^{(1)} & (\phi^{\dagger}i \overleftrightarrow{D}_{\mu}  \phi)\left(\bar{q}_{i} \gamma^{\mu} q_{j}\right)  \delta^{ij}
& 
\mathcal{O}_{\phi Q}^{(3)} & (\phi^{\dagger}i \stackrel{\longleftrightarrow}{D^A_\mu} \phi)\left(\bar{q}_{i} \gamma^{\mu} t^{A} q_{j}\right)  \delta^{ij}
\\
\bottomrule
\end{tabular}
\caption{Basis of dimension-6 SMEFT operators adopted in our global
  analysis. Flavor indices are denoted by $i,j$ and are implicitly
  contracted when repeated.}
\label{tab:smeft_basis}
\end{table}

We consider the dimension-6 SMEFT Lagrangian
\begin{align}
\mathcal{L}_{\text {SMEFT}} \supset 
&-\frac{\alpha_{s}}{8 \pi} \frac{f_{G G}}{\Lambda^{2}} \mathcal{O}_{G G}
+\frac{f_{W W}}{\Lambda^{2}} \mathcal{O}_{W W}
+\frac{f_{B B}}{\Lambda^{2}} \mathcal{O}_{B B} 
+\frac{f_{B W}}{\Lambda^{2}} \mathcal{O}_{B W}
+\frac{f_{W}}{\Lambda^{2}} \mathcal{O}_{W}
+\frac{f_{B}}{\Lambda^{2}} \mathcal{O}_{B}
\notag \\
&+\frac{f_{W W W}}{\Lambda^{2}} \mathcal{O}_{W W W} 
+\frac{f_{\phi 1}}{\Lambda^{2}} \mathcal{O}_{\phi 1}
+\frac{f_{\phi 2}}{\Lambda^{2}} \mathcal{O}_{\phi 2}
+\frac{f_{\tau} m_{\tau}}{v \Lambda^{2}} \mathcal{O}_{\tau}
+\frac{f_{b} m_{b}}{v \Lambda^{2}} \mathcal{O}_{b}
+\frac{f_{t} m_{t}}{v \Lambda^{2}} \mathcal{O}_{t}+ 
\notag \\
&
+\frac{f_{L L L L}}{\Lambda^{2}} \mathcal{O}_{L L L L} 
+\frac{f_{\phi e}}{\Lambda^{2}} \mathcal{O}_{\phi e}
+\frac{f_{\phi d}}{\Lambda^{2}} \mathcal{O}_{\phi d}
+\frac{f_{\phi u}}{\Lambda^{2}} \mathcal{O}_{\phi u}
+\frac{f_{\phi Q}^{(1)}}{\Lambda^{2}} \mathcal{O}_{\phi Q}^{(1)}
+\frac{f_{\phi Q}^{(3)}}{\Lambda^{2}} \mathcal{O}_{\phi Q}^{(3)} \, ,
\end{align}
where the Wilson coefficients are denoted by $f_i$.  We use the
dimension-6 operator basis of Ref.~\cite{Biekotter:2018rhp}, which is
based on the HISZ set~\cite{Hagiwara:1993ck} and defined in
Tab.~\ref{tab:smeft_basis}. We adopt the `+' convention for the
covariant derivatives, e.g.  $D_\mu\phi=(\partial_\mu + ig' B_\mu/2
+ig t^A W^A_\mu)\phi$, where $t^A = \sigma^A/2$ are the $SU(2)$
generators and $\sigma^A$ the Pauli matrices. We have also defined
$(\phi^\dag i\overleftrightarrow{D}_\mu \phi) = i
\phi^{\dagger}(D_\mu\phi) - i (D_\mu\phi^\dag)\phi$ , $(\phi^\dag
i\overleftrightarrow{D}_\mu^I \phi) = i \phi^{\dagger}t^A(D_\mu\phi) -
i (D_\mu\phi^\dag)t^A\phi$ and the dual Higgs field
$\widetilde{\phi}=i \sigma^2\phi^\star$.  The field strengths are
normalized as $\hat{B}_{\mu \nu}=i g' B_{\mu \nu}/2$ and $\hat{W}_{\mu
  \nu}=i g t^A W_{\mu \nu}^A$.  Finally, the operators $\O_{\phi
  Q}^{(1), (3)}, \O_{\phi u}, \O_{\phi d}, \O_{\phi e}$ are defined in
a $U(3)^5$-invariant flavor structure, while for $\O_{LLLL}$ we only
retain the $(1221)$ contraction, that is relevant for the definition
of the Fermi constant, and for $\O_{b}, \O_{t}, \O_{\tau}$ we only
consider the 3rd fermion generation. The latter choice is justified
considering that, in a $U(3)^5$-symmetric scenario, these operators
are weighted by a Yukawa coupling insertion, that acts as a
suppression for the first two families.

The matching to the UV models described in Sec.~\ref{sec:basics_match}
is automated for the Warsaw basis of SMEFT
operators~\cite{Grzadkowski:2010es}, in the general flavor case. The
results obtained are provided on github at \cite{code} and we give
explicit expressions for the tree-level matching in
Appendix~\ref{app:matchingtree}.  In order to interface them to
\sfitter, the matching results are mapped onto the basis of
Tab.~\ref{tab:smeft_basis}.  In the following we denote the operators
in the Warsaw basis, defined as in Ref.~\cite{Grzadkowski:2010es}, by
$Q_k$ and the associated Wilson coefficients by $C_k$, such that the
SMEFT Lagrangian in this basis has the form
\begin{align}
\mathcal{L}_{\rm Warsaw} \supset \frac{1}{\Lambda^2}\sum_k \sum_{ij} C_{k,ij}\,Q_{k,ij}\,,
\end{align}
where $k$ runs over the operators labels and $i,j$ are flavor indices,
that are present for fermionic operators.  The relations between the
two operator bases are
\begin{align}
\label{Orelations_first}
 \O_{GG} &= Q_{\phi G}\,,
 &
 \O_{WWW} &= \frac{g^3}{4}Q_W\,,
 \notag \\
 \O_{BB} &=- \frac{g^{\prime 2}}{4} Q_{\phi B}\,,
 &
 \O_{WW} &=- \frac{g^{2}}{4} Q_{\phi W}\,,
 &
 \O_{BW} &=- \frac{gg^{\prime}}{4} Q_{\phi WB}\,,&
 \notag \\
 \O_{\phi1} &= Q_{\phi D}\,,
 &
 \O_{\phi2} &= -\frac12 Q_{\phi\square}\,,
 &
 \O_\phi &= Q_\phi\,,
 \notag \\
 \O_{\tau} &= Q_{e\phi,33}\,,
 &
 \O_{t} &=  Q_{u\phi,33}\,,
 &
 \O_{b} &=  Q_{d\phi,33}\,,
 \notag \\
 \O_{\phi e} &= Q_{\phi e,ij}\, \delta^{ij}\,,
 &
 \O_{\phi u} &= Q_{\phi u,ij}\, \delta^{ij}\,,
 &
 \O_{\phi d} &= Q_{\phi d,ij}\, \delta^{ij}\,,
 \notag \\
 \O_{\phi Q}^{(1)} &= Q_{\phi q,ij}^{(1)} \, \delta^{ij}\,,
 &
 \O_{\phi Q}^{(3)} &= \frac{1}{4}Q_{\phi q,ij}^{(3)}\, \delta^{ij}\,,
 &
 \O_{LLLL} &= Q_{ll,1221} \; ,
\end{align}
and
\begin{align}
   \O_W  &= \frac{g^2}{8}Q_{\phi W} + \frac{g'g}{8}Q_{\phi WB}
 -\frac{3g^2}{8} Q_{\phi\square} + \frac{g^2m_h^2}{4} (\phi^\dag \phi)^2 - \frac{g^2\lambda}{2} Q_\phi  
 \notag \\ 
 &
 -\frac{g^2}{4}\left[ 
(Y_e)_{ij}Q_{e\phi,ij} + (Y_u)_{ij} Q_{u\phi,ij} + (Y_d)_{ij} Q_{d\phi,ij} 
+\hc\right]
- \frac{g^2}{8}\left(Q_{\phi q,ij}^{(3)}+Q_{\phi l,ij}^{(3)}\right)\delta^{ij}
\notag \\
\O_B &= \frac{g^{\prime2}}{8}Q_{\phi B} + \frac{gg'}{8}Q_{\phi WB}
-
\frac{g^{\prime2}}{2}Q_{\phi D} - \frac{g^{\prime2}}{8}Q_{\phi\square} 
 \notag \\
&
-\frac{g^{\prime2}}{4}\left(\frac16 Q_{\phi q,ij}^{(1)}-\frac12Q_{\phi l,ij}^{(1)}
+\frac23 Q_{\phi u,ij} - \frac13 Q_{\phi d,ij} - Q_{\phi e,ij}
 \right)\delta^{ij}\; ,
 \label{Orelations_last}
\end{align}
where all repeated flavor indices are implicitly summed over, and
$\lambda$ is the quartic coupling in the Higgs potential, normalised
such that
\begin{align}
 V(\phi) = -\frac{m_h^2}{2} \phi^\dagger \phi + \frac{\lambda}{2}(\phi^\dagger\phi)^2 \;.
\end{align}
As the vector triplet model we are interested in is defined in a
flavor-symmetric limit, after the matching procedure the Wilson
coefficients of the Warsaw basis operators $Q_{\phi e,\phi u,\phi d}$
and $Q_{\phi l,\phi q}^{(1),(3)}$ will have the form
\begin{align}
  C_{\phi \psi, ij} &= \bar C_{\phi \psi}\, \delta_{ij} \; ,
\end{align}
while 
\begin{align}
 C_{ll,ijkl} = \bar C_{ll} \delta_{ij}\delta_{kl} + \bar C_{ll}^\prime \delta_{il}\delta_{kj} \; .
\end{align}
Using this notation, the mapping in terms of Wilson coefficients is
\begin{align}
 f_B &= \frac{8}{g^{\prime 2}} \bar C_{\phi l}^{(1)}
 &
 -\frac{\alpha_s}{8\pi}f_{GG} &= C_{\phi G}
 \notag \\
 f_W &= -\frac{8}{g^2} \bar C_{\phi l}^{(3)}
 &
 f_{WWW} &= \frac{4}{g^3} C_{W}
 \notag \\
 f_{BB} &= -\frac{4}{g^{\prime 2}}\left[C_{\phi B} -  \bar C_{\phi l}^{(1)}\right]
  &
f_{\phi1} &= C_{\phi D} + 4 \bar C_{\phi l}^{(1)}
 \notag \\
 f_{WW} &= -\frac{4}{g^{2}}\left[C_{\phi W} +  \bar C_{\phi l}^{(3)}\right]
 &
 f_{\phi2} &= -2 C_{\phi\square} - 2 \bar C_{\phi l}^{(1)} +6 \bar C_{\phi l}^{(3)}
 \notag \\
 f_{BW} &= 4\left[-\frac{C_{\phi WB}}{gg'} -\frac{\bar C_{\phi l}^{(3)}}{g^2}+  \frac{\bar C_{\phi l}^{(1)}}{g^{\prime2}}\right]
 \hspace*{-5mm}
& f_{\phi} &= C_{\phi} - 4 \lambda \bar C_{\phi l}^{(3)}
\end{align}
and for the fermionic ones
\begin{align}
\frac{m_\tau}{v} f_{\tau} &= C_{e \phi,33} -2 (Y_e)_{33} \bar C_{\phi l}^{(3)} \qquad
 &
 f_{\phi e} &= \bar C_{\phi e} - 2\bar C_{\phi l}^{(1)}
 \notag \\
\frac{m_t}{v} f_{t} &= C_{u \phi,33} -2 (Y_u)_{33} \bar C_{\phi l}^{(3)}
 &
 f_{\phi u} &=  \bar C_{\phi u} +\frac 43 \bar C_{\phi l}^{(1)}
 \notag \\
 \frac{m_b}{v}f_{b} &= C_{d \phi,33} -2 (Y_d)_{33} \bar C_{\phi l}^{(3)}
 &
 f_{\phi d} &=\bar C_{\phi d} -\frac23 \bar C_{\phi l}^{(1)}
 \notag \\
 \fpWs &= \bar C_{\phi q}^{(1)} + \frac13 \bar C_{\phi l}^{(1)}
 &
 \fpWt &=4  \left[\bar C_{\phi q}^{(3)}- \bar C_{\phi l}^{(3)}\right]
 \notag \\
 f_{LLLL} &= \bar C_{ll}^\prime \;.
\end{align}
In addition, the Higgs quartic coupling gets redefined as 
\begin{align}
 \lambda_{\rm HISZ} = \lambda_{\rm Warsaw} +  \frac{4m_h^2}{\Lambda^2}\,\bar C_{\phi l}^{(3)} \;.
\end{align}
This translates into corrections to the cubic and quartic Higgs
self-couplings, which do not contribute to any of the observables in
our fit.

\subsection{Matching expressions at tree-level}\label{app:matchingtree}

Matching the heavy vector triplet model defined in
Section~\ref{sec:basics_model} at tree level onto the Warsaw basis, we
obtain
\begin{align}
 C_{\phi\square} &= -\frac{3}{8}\frac{(\gt_H+g_2 \gt_M)^2}{\tilde m_V^2}
 \notag \\
 C_{\phi l,ij}^{(3)} = \bar C_{\phi l}^{(3)} \delta_{ij} &= -\frac{1}{4}\frac{(\gt_l+g_2\gt_M)(\gt_H+g_2\gt_M)}{\tilde m_V^2}\delta_{ij}
 \notag \\
 C_{\phi Q,ij}^{(3)} = \bar C_{\phi q}^{(3)} \delta_{ij} &= -\frac{1}{4}\frac{(\gt_q+g_2\gt_M)(\gt_H+g_2\gt_M)}{\tilde m_V^2} \delta_{ij}
 \notag \\
 C_{ll,ijkl}= \bar C_{ll} \delta_{ij}\delta_{kl} + \bar C_{ll}^\prime \delta_{il}\delta_{kj} &= \frac{1}{8}\frac{(\gt_l+g_2\gt_M)^2}{\tilde m_V^2} \left(\delta_{ij}\delta_{kl}-2\delta_{il}\delta_{kj}\right)
 \notag \\
 C_{f\phi, ij} &= -\frac{(Y_f)_{ij}}{4}\frac{(\gt_H+g_2 \gt_M)^2}{\tilde m_V^2}
\qqquad \text{($f=e,u,d$).}
\end{align}
These results were also derived e.g. in
Refs.~\cite{Low:2009di,Biekoetter:2014jwa,Brehmer:2015rna,deBlas:2017xtg}.
The full expressions for the 1-loop matching are derived here for the
first time and are provided at Ref.~\cite{code}.

\end{fmffile}
\bibliography{literature}

\end{document}